\documentclass[aps,prx,twocolumn,preprintnumbers,amsmath,amssymb,superscriptaddress]{revtex4-2}

\usepackage[utf8]{inputenc}
\usepackage{graphicx}
\usepackage{booktabs}
\usepackage{dcolumn}
\usepackage{bm}
\usepackage{amssymb}
\usepackage{amsmath}
\usepackage{mathrsfs}
\usepackage{epigraph}
\usepackage{braket}
\usepackage{gensymb}
\usepackage{lmodern}
\usepackage{tipa}
\usepackage{bbold}
\usepackage{esint}
\usepackage{mathdots}
\usepackage{xcolor}
\usepackage{appendix}
\usepackage{natbib}
\usepackage{multirow}
\usepackage[colorlinks=true, linkcolor=blue, urlcolor=blue, citecolor=blue]{hyperref}

\begin{document}

\title{Particle-hole asymmetry and a fluctuating pair-density-wave state in a lightly doped Mott insulator on the square lattice}

\author{Hao-Xin Wang}
\affiliation{Department of Physics, The Chinese University of Hong Kong, Sha Tin, New Territories, Hong Kong, China}

\author{Zheng-Yuan Yue}
\affiliation{Department of Physics, The Chinese University of Hong Kong, Sha Tin, New Territories, Hong Kong, China}

\author{Wen-Yuan Liu}
\email{wyliu@zju.edu.cn}
\affiliation{Institute for Advanced Study in Physics and School of Physics, and Zhejiang Key Laboratory of Micro-nano Quantum Chips and Quantum Control, Zhejiang University, Hangzhou 310027, China}

\author{Zheng-Cheng Gu}
 \email{zcgu@phy.cuhk.edu.hk}
\affiliation{Department of Physics, The Chinese University of Hong Kong, Sha Tin, New Territories, Hong Kong, China}

\date{November 10, 2026}
\begin{abstract}
Whether a lightly doped Mott insulator on the square lattice supports robust $d$-wave superconductivity remains a central question in strongly correlated physics and in the microscopic theory of cuprate superconductivity. Experiments and numerical studies reveal a pronounced particle--hole asymmetry, pointing to a delicate competition among $d$-wave superconductivity, pair-density-wave (PDW), stripe, charge-density-wave (CDW), and antiferromagnetic correlations. Here we combine large-scale density-matrix renormalization group (DMRG) and finite-size projected entangled pair states (PEPS) to study the square-lattice $t$-$J$ model at low doping, $\delta \leq 0.1$, while tuning the next-nearest-neighbor hopping $t'$ from the electron-like ($t'>0$) to the hole-like ($t'<0$) regime. We uncover a pronounced particle-hole-asymmetric reorganization of the low-energy state. The electron-like side supports robust, nearly uniform $d$-wave pairing, whereas decreasing $t'$ induces a period-two plaquette modulation and ultimately produces, on the hole-like side, a charge-dominated state with a strongly suppressed static pairing amplitude. This evolution is supported by DMRG calculations on open strips and PEPS calculations on open square lattices up to $16\times16$ sites that preserve $C_4$ symmetry at the Hamiltonian level. {On these squares,} both methods find that the pair correlations of the hole-like state select the plaquette momenta $(\pi,0)$ and $(0,\pi)$ and decay within a few lattice spacings, while the {dominant charge-density} modulation forms a bidirectional checkerboard whose period extrapolates to about $4a_0$, close to that observed in lightly doped Bi-2201. These observations establish a fluctuating plaquette PDW in the lightly doped hole-like regime, in which local pairs carry finite momentum but {lack} phase coherence. Our results {firmly establish} particle-hole asymmetry as a central organizing principle for competing pairing tendencies in the lightly doped $t$-$J$ model. The suppression of uniform pairing on the hole-like side further motivates investigating whether material-specific processes (such as extended Cu-O molecular-orbital dynamics beyond a strict Zhang-Rice singlet description, correlated hopping, and lattice coupling) enhance pair mobility and global superconducting coherence.

\end{abstract}

\maketitle

\section{Introduction}
\subsection{Overview}

Four decades after the discovery of high-$T_c$ cuprates, the microscopic mechanism underlying $d$-wave superconductivity in these materials remains under debate. The doped-Mott-insulator picture is widely regarded as a framework for capturing their essential physics~\cite{Anderson1987,PALee-RMP2006,Kivelson-Review2015}. In particular, motivated by the Zhang-Rice singlet construction~\cite{Zhang1988}, the square-lattice $t$-$J$ model provides a standard minimal setting for investigating whether strong local correlations can generate the intertwined orders associated with the cuprates, including $d$-wave superconductivity, charge order, spin order, and possibly pair-density-wave (PDW) order~\cite{Kivelson-Review2015,Scalapino-RMP2012,PALee-RMP2006,
Fradkin2015RMP,Arovas-ARCMP2022}.
Despite extensive work, {however, }the lightly doped regime
($\delta \leq 0.1$) of the two-dimensional $t$-$J$ model still lacks a settled qualitative picture. Major advances in density-matrix renormalization group (DMRG), quantum Monte Carlo, and tensor-network methods have substantially {advanced} the numerical understanding of these models~\cite{SimonsCollaboration2020,SimonsCollaboration2017,
SWZhang2024Coexistence,Chen2025PNAS,HongChen2019Science,
Shoushu2021Robust,White2021PNAS,YiFan2024Ground,Shengtao2022Pairing,
Hongchen2023EightLeg,Luxin2024Emergent,Corboz2011Stripes,
Corboz2014Competing,Jhengwei2021Study,Qiaoyi2023Tangent,
Xiao2023Temperature,Sinha2022Finite,zhang2025finite,
Wenyuan2025Accurate,Zhang2025Frustration,Dawei2024Phase}.
This progress brings two unresolved questions into sharper focus: how the superconducting tendency changes between the electron-like and hole-like regimes, and whether the low-energy pairing sector on the hole-like side acquires PDW character rather than remaining predominantly uniform $d$-wave.

The first question concerns the particle-hole asymmetry of
superconductivity. A natural benchmark for a minimal cuprate model is whether it reproduces the empirical tendency for superconductivity to be more robust in hole-doped than in electron-doped cuprates~\cite{PALee-RMP2006,Scalapino-RMP2012,Arovas-ARCMP2022,
Armitage2010ElectronDoped}.
In single-band descriptions, this asymmetry is commonly encoded
through the next-nearest-neighbor hopping $t'$. Existing numerical studies, however, report contrasting trends. Some find stronger or more robust superconductivity on the hole-doped side~\cite{SWZhang2024Coexistence}, whereas others find weaker pairing there~\cite{Chen2025PNAS,Shengtao2022Pairing,Dawei2024Phase,
Li2026FluctuatingPDW}, or a regime in which stripe order dominates and superconductivity is strongly suppressed~\cite{White2021PNAS,YiFan2024Ground,Jiang2025CompetitionCDWSC}. This issue tests the range of validity of the minimal description: a robust discrepancy with the observed asymmetry would motivate examining whether additional microscopic ingredients are required beyond the Hubbard or $t$-$J$ framework. These contrasting results therefore motivate a renewed study of the lightly doped regime using complementary large-scale numerical methods.

The second question concerns PDW order, in which the superconducting order parameter carries finite momentum and breaks translational symmetry in addition to the global $U(1)$
symmetry~\cite{Fradkin2015RMP,Agterberg:20}. Theoretically, PDW has been analyzed in several contexts: as a spatially modulated pairing state, including the Amperean-pairing mechanism~\cite{LeeAmperean2014} and the striped-superconductor scenario~\cite{Berg2009Striped}; as finite-momentum pairing correlations and order in one-dimensional and ladder models~\cite{Berg2010,Jaefari2012}; and as fluctuating, vestigial, or quantum-disordered pairing sectors that can remain relevant even when static PDW order is weak or
absent~\cite{Setty2023FluctuatingPDW,Berg2009Charge4e,
Dai2020QuantumDisorderedPDW}.
Experimentally, signatures consistent with PDW order or strong PDW fluctuations have been reported in several cuprate and iron-based systems~\cite{Hamidian:16,Edkins:19,DuZ:20,Ruan:18,LiX:21,
Liu2023PDW_Nature,lee2023pairdensity,ZhangYaoArXiv2024}.

Numerically, recent infinite projected entangled-pair state (iPEPS) calculations have identified several competing PDW states in the square-lattice $t$-$J$ model~\cite{Zheng2025CompetingPDW}, while DMRG studies
have reported PDW-like signatures in specific geometries or
grand-canonical formulations~\cite{xu2025stripespairdensitywave,Chen2025PNAS}.

{Finite-momentum pairing also}
offers a possible route {toward} understanding the microscopic origin of the pseudogap, one of the central unresolved problems of cuprate superconductivity. A key question is whether the hole-doped pseudogap is primarily a thermally disordered precursor of uniform $d$-wave superconductivity, or reflects a distinct correlated quantum state that can persist toward zero temperature when superconducting coherence is absent or suppressed.

Experimentally, the hole-doped pseudogap suppresses low-energy
spectral weight predominantly in the antinodal region below a
crossover temperature $T^{*}$, producing apparent Fermi arcs in
angle-resolved photoemission spectroscopy rather than a complete large Fermi surface~\cite{Kivelson-Review2015,Hashimoto2014Pseudogap}.
{More recent experiments point instead to small closed hole pockets. Laser ARPES on underdoped Bi-2201 resolved hole pockets that coexist with the arcs and are absent in overdoped samples~\cite{Meng2009Pockets}. In the inner CuO$_2$ planes of multilayer cuprates, which are shielded from disorder, closed hole pockets centered at $(\pi/2,\pi/2)$ appear at hole densities down to $p\approx0.007$, coexist with antiferromagnetic order, and develop a $d$-wave-like pairing gap~\cite{Kunisada2020Pockets,Chen2026PersistentPockets,Luo2026MultiplePockets}. In the pseudogap phase of HgBa$_2$CuO$_{4+\delta}$, the Yamaji effect in the $c$-axis magnetoresistance provides transport evidence for closed pockets, each occupying about $1.3\%$ of the Brillouin zone~\cite{Chan2025Yamaji}.}
{Experiments} on lightly doped Bi-2201 provide particularly direct
motivation for the low-doping regime considered here. Scanning
tunneling measurements on $p=0.03$ and $0.07$ samples revealed
doping-induced in-gap spectral weight and, with increasing doping, pseudogap-like low-energy suppression accompanied by checkerboard charge order~\cite{Cai2016LightlyDopedCuprates}.
Subsequent measurements on $p=0.08$ and $0.11$ samples found closely related local spectra across the insulator--superconductor transition, suggesting that global phase coherence emerges as charge-ordered plaquettes associated with local pairing proliferate spatially~\cite{Ye2023LocalPairing}. {Recent local shot-noise spectroscopy on underdoped Bi$_2$Sr$_2$CaCu$_2$O$_{8+\delta}$ resolves the charge of the tunneling carriers and finds that pairing sets in at the local pseudogap energy, inferring that pairs already exist on the pseudogap energy scale, and superconductivity is limited by their phase coherence~\cite{Niu2026PseudogapPairing}.}

Theoretically, a strongly momentum-dependent spectral gap without global superconducting coherence is not accounted for by a homogeneous weak-coupling BCS description alone. Among pairing-based explanations, PDW scenarios provide a natural connection between spectral reconstruction, charge modulation, and the loss of superconducting coherence. In the Amperean-pairing scenario, spinons on the same side of the Fermi surface pair at finite momentum; the resulting subsidiary charge order and reconstructed quasiparticle spectrum offer a unified account of Fermi arcs, their termination, and the relation between the arc-tip spanning vector and the charge-order wave vector~\cite{LeeAmperean2014}. A complementary construction quantum-disorders a PDW into a pseudogap metal containing small fermionic pockets and a bosonic Mott state of charge-$2e$ pairs~\cite{Dai2020QuantumDisorderedPDW}.
More exotically, the fractionalized Fermi liquid (FL$^*$) theory {proposes} that the spins of the parent Mott insulator form a spin liquid, and the doped holes form small Fermi pockets whose total area is set by the hole density $p$ rather than by $1+p$, without breaking any symmetry~\cite{Senthil2003FLstar,Punk2015DimerPseudogap}. The electron spectral weight resides mainly on the inner side of each pocket, so photoemission sees an arc, as in the phenomenological Yang-Rice-Zhang ansatz~\cite{YRZ2006Phenomenological}; FL$^*$ pockets of the predicted area also reproduce the Yamaji peak observed in HgBa$_2$CuO$_{4+\delta}$~\cite{Zhao2026Yamaji,Bonetti2025FLstarReview}.

Addressing this question requires numerical methods that can resolve both competing low-energy states and their two-dimensional spatial structure. DMRG has been a principal tool for these strongly correlated problems, but its access to two-dimensional physics is largely through long narrow cylinders or strips. This leaves open whether the inferred superconducting and PDW tendencies persist on wider, more isotropic geometries, and whether the pairing modulation is unidirectional or bidirectional when the finite Hamiltonian
preserves the lattice $C_4$ symmetry. PEPS provides a natural
complement because its tensor-network entanglement structure is adapted to two dimensions. Applications to doped fermion systems have nevertheless been restricted by the high cost of contraction and optimization. Recent finite-PEPS algorithms combining Monte Carlo sampling with variational optimization substantially extend the accessible bond dimensions and system
sizes~\cite{Wenyuan2017Gradient,He2019Gradient,Wenyuan2021Accurate,
Wenyuan2025Accurate}.

In this work, we combine DMRG and finite PEPS to investigate the lightly doped square-lattice $t$-$J$ model at $\delta\leq0.1$, tuning $t'$ from the electron-like ($t'>0$) through zero to the hole-like ($t'<0$) regime. DMRG calculations on open strips and
$8\times8$ squares are complemented by systematic PEPS calculations on $12\times12$ and $16\times16$ squares. The open square geometries preserve $C_4$ symmetry at the Hamiltonian level without geometrically favoring either lattice direction. This approach allows us to test whether the suppression of uniform $d$-wave superconductivity on the hole-like side is accompanied by the emergence of plaquette-PDW correlations.  Identifying this low-energy pairing structure provides a microscopic test of PDW-based explanations of the strongly underdoped pseudogap, beyond simply determining whether the model supports a uniform superconducting condensate.

\subsection{Model and summary of main results}\label{subsec:model}
We study the square-lattice $t$-$J$ model, whose Hamiltonian reads
\begin{equation}
    \begin{aligned}
        H_{tJ}=&-\sum_{\{i j\}, \sigma} t_{i j}\left(\hat{c}_{i, \sigma}^{\dagger} \hat{c}_{j, \sigma}+\text{H.c.}\right) \\
&+\sum_{\langle i j\rangle} J \left(\hat{\mathbf{S}}_{i} \cdot \hat{\mathbf{S}}_{j}-\frac{1}{4} \hat{n}_{i} \hat{n}_{j}\right),
    \end{aligned}
    \label{eq:hamiltonian}
\end{equation}
where $\hat{c}_{i\sigma}^\dagger$ creates an electron with spin $\sigma=\uparrow,\downarrow$ on site $i$, $\hat{n}_{i\sigma}=\hat{c}_{i\sigma}^\dagger \hat{c}_{i\sigma}$, $\hat{n}_i=\sum_\sigma \hat{n}_{i\sigma}$, and $\hat{\mathbf S}_i$ is the spin operator. The local Hilbert space is constrained by the no-double-occupancy condition. In the grand-canonical calculations we work with $K_\mu=H_{tJ}-\mu\hat N$, where $\hat N=\sum_i \hat n_i$ and $\mu$ is the chemical potential, and denote the grand-canonical energy of a state by $\Omega(\mu)=\langle K_\mu\rangle=\langle H_{tJ}\rangle-\mu\langle\hat N\rangle$. We include nearest-neighbor (NN) and next-nearest-neighbor (NNN) hoppings, $t_{ij}=t,t'$, together with NN superexchange $J$, and set $t=1$ throughout. The system size is $N=L_x\times L_y$, and the hole doping is $\delta=1-\frac{1}{N}\sum_i \langle n_i\rangle$. Throughout the manuscript we report the local hole density $\langle n_{i,h}\rangle \equiv 1-\langle n_i\rangle$, the local spin component $\langle S_i^z\rangle$, and the equal-time spin correlation relative to a fixed reference site $i_{\rm ref}$, $F_i\equiv\langle \mathbf{S}_{i_{\rm ref}}\cdot\mathbf{S}_i\rangle$. We equivalently write $F(\mathbf r)\equiv F_i$ with $\mathbf r=\mathbf r_i-\mathbf r_{i_{\rm ref}}$. We define the unnormalized singlet-pair annihilation operator on a bond $b=(i,i+a)$ as $\Delta_a(i)=\hat c_{i+a,\downarrow}\hat c_{i,\uparrow}-\hat c_{i+a,\uparrow}\hat c_{i,\downarrow}$, and write $\hat{\Delta}_a(i)=\Delta_a(i)/\sqrt{2}$ for the normalized operator. We use the singlet pair-pair correlation
\begin{equation}\label{eq: pairing correlation}
    P_s^{ab}(r)=\langle\Delta_a^\dagger(i)\Delta_b(i+r)\rangle,
\end{equation}
to diagnose superconducting (SC) off-diagonal long-range order, and, in the grand-canonical calculations only, the local singlet-pairing amplitude
\begin{equation}
    \Phi_a(i)\equiv {\langle\hat{\Delta}_a^\dagger(i)\rangle},
\end{equation}
on bond $a$ to diagnose the pairing symmetry and patterns.

All calculations are performed on the square lattice with open boundary conditions (OBC), using both elongated strips and $L\times L$ squares. Unlike the cylinder geometries commonly used in DMRG, which are open along their length but periodic around their circumference, an $L\times L$ square preserves the $C_4$ symmetry of the finite Hamiltonian. This distinction is particularly relevant in the doped $t$-$J$ model, where competing stripe, charge-density-wave (CDW), and $d$-wave pairing tendencies may respond differently to the finite geometry.

\begin{figure*}[t]
    \centering
    \includegraphics[width=0.92\textwidth]{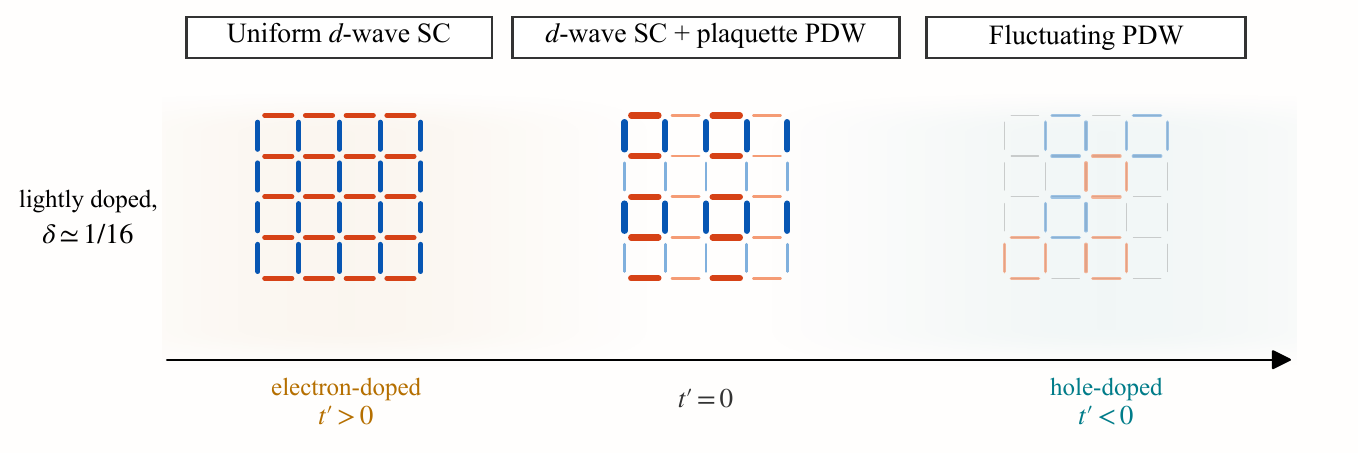}
    \caption{Schematic summary of the physical properties in the lightly doped $t$-$J$ model at $\delta\approx1/16$ as the NNN hopping $t'$ is tuned from $t'>0$ (the electron-like regime) to $t'<0$ (the hole-like regime). For $t'>0$, the dominant tendency is a predominantly uniform $d$-wave superconducting state. Near $t'=0$, a period-2 plaquette modulation along both lattice directions develops on top of the uniform $d$-wave background. For {sufficiently negative $t'$}, the uniform component is strongly suppressed and the low-energy pairing sector is short-ranged and modulated, which we refer to as a fluctuating PDW. This schematic summarizes the pairing-evolution trends inferred from the combined finite-size DMRG and PEPS results.}
    \label{fig:pairing_sector_summary}
\end{figure*}

In a square-lattice single-band model, a particle–hole transformation reverses the sign of the NNN hopping \(t'\), whereas the corresponding sign change of the NN hopping \(t\) can be removed by a sublattice gauge transformation. The ratio \(t'/t\) therefore provides the simplest parameterization of particle–hole asymmetry between the electron- and hole-doped regimes.

Band-structure estimates for hole-doped cuprates give $t'/t<0$ in the electron representation, with representative magnitudes $|t'/t|\sim0.15$ for La$_{2-x}$Sr$_x$CuO$_4$ (LSCO), $\sim0.25$ for HgBa$_2$CuO$_4$, and $\sim0.3$ for YBa$_2$Cu$_3$O$_{7-\delta}$ (YBCO), Bi$_2$Sr$_2$CaCu$_2$O$_{8+\delta}$ (Bi-2212), and Tl-based compounds~\cite{Pavarini2001,Andersen1995,Hirayama2019}. By the same particle--hole transformation, electron doping at a given $t'$ is equivalent to hole doping at $-t'$: with the cuprate sign $t'/t<0$, hole-doped cuprates correspond to hole doping with $t'<0$, and electron-doped cuprates to hole doping with $t'>0$. We perform calculations at light hole doping and use \(t'>0\) and \(t'<0\) as effective representations of electron- and hole-doped cuprates, respectively,
with \(t'=0\) providing a theoretically natural reference point for tracking the evolution between the two regimes. The values quoted above are one-electron band parameters, which is what the bare $t'$ of a lattice model represents. They should be distinguished from an effective $t'$ obtained by fitting a tight-binding contour to the Fermi surface measured in photoemission, which already contains the renormalization by electronic correlations. So $t'$ here is an input parameter.
In the DMRG strip calculations, we extend the parameter range to $|t'|=0.5$, reaching the upper end of the realistic cuprate range and enabling a systematic characterization of the $t'$ dependence of the ground state.

At a hole-doping level of $\delta\sim 1/8$, many Hubbard and {$t$-$J$} studies have found stripes and competing SC orders~\cite{SimonsCollaboration2017,Corboz2011Stripes,Corboz2014Competing,Chen2025PNAS,Hongchen2023EightLeg,Luxin2024Emergent,Zheng2025CompetingPDW}, and at very low doping, phase separation may become relevant~\cite{Kivelson-HQLin-PRL1990,Corboz2014Competing}. Here we focus on the lightly doped regime around $\delta\sim 1/16$ at $J=0.4$, with $t'$ ranging from $-0.5$ to $0.2$.

Figure~\ref{fig:pairing_sector_summary} summarizes the evolution of the ground-state pairing structure inferred from our DMRG and PEPS calculations as $t'$ is varied. At $t'=0.2$, the ground state exhibits a predominantly uniform $d$-wave pairing structure. Near $t'=0$, a period-2 plaquette modulation develops on top of this uniform $d$-wave background. For sufficiently negative $t'$, charge order becomes dominant, while {pairing is strongly suppressed and acquires a finite correlation length, with} a spatially modulated pattern that shows well-defined maxima in momentum space in the equal-time pair correlations. On the $8\times8$, $12\times12$, and $16\times16$ squares at $t'=-0.2$, these maxima lie at the plaquette momenta $(\pi,0)$ on horizontal and $(0,\pi)$ on vertical bonds, with the same momentum structure in DMRG and PEPS. The correlations  decay within a few lattice spacings on every lattice, {which means that} the pairs lack a common phase (Sec.~\ref{sec:fluctuating_pdw}). These local pairs are intertwined with a bidirectional checkerboard charge modulation whose period grows with lattice size and extrapolates to about $4a_0$, close to the period observed in lightly doped Bi-2201~\cite{Cai2016LightlyDopedCuprates,Ye2023LocalPairing} (Sec.~\ref{subsec:cdw_momentum}). In the electron momentum distribution, the near-nodal ridge stays sharp and close to the band Fermi surface, whereas the antinodal ridge weakens and flattens, a nodal--antinodal contrast {reminiscent of} the pseudogap (Sec.~\ref{subsec:disc_evolution}). These results motivate our description of this charge-dominated state as a fluctuating PDW.

\section{Numerical methods: DMRG and finite PEPS}
\label{sec:method}

We combine DMRG and finite PEPS in complementary geometries. DMRG is most accurate on narrow lattices, and we use it on six- and eight-leg lattices. Because the number of states DMRG must keep grows exponentially with the lattice width, wider squares are beyond its reach {for accurate simulation}. There we use finite PEPS, the two-dimensional extension of the matrix-product state, on squares of up to $16\times16$ sites. On the $6\times6$ and $8\times8$ squares, where the two methods overlap, we benchmark PEPS against DMRG.

\subsection{DMRG calculations}
\label{subsec:dmrg_method}
All DMRG results presented in this work were obtained using the standard two-site algorithm on open lattices at fixed hole number and $S^z_{\rm tot}=0$, exploiting $U(1)_N\times U(1)_{S^z}$ symmetry~\cite{Schollwock2011}. We denote the number of kept states in DMRG by $m$, and the truncation error by $\epsilon$.  The values of $m$ and $\epsilon$ for all DMRG calculations in this work are collected in Table~\ref{tab:dmrg_runs} (Appendix~\ref{app:dmrg_runs}). For the strip and $8\times8$ calculations at $\delta\le1/16$ on which our conclusions rest, $\epsilon\lesssim3\times10^{-5}$. Unless stated otherwise, DMRG results are shown at the largest $m$ reached, without extrapolation in $m$. {With a parallel, GPU-accelerated implementation we keep up to $m=48000$ states.}

\subsection{Finite fermionic PEPS and variational Monte Carlo optimization}
\label{subsec:peps_vmc}

Although PEPS has a natural area-law entanglement structure for 2D systems, its {computation} on finite systems is a challenging task. Conventional double-layer contraction PEPS methods with bond dimension $D$ {have a cost that scales} as high as $O(D^{10})$~\cite{Wenyuan2017Gradient}, {which severely limits the accessible $D$}. Here we adopt {variational Monte Carlo (VMC), in which the sum over the physical degrees of freedom of the PEPS is replaced by Monte Carlo sampling}~\cite{Wenyuan2017Gradient,He2019Gradient,Wenyuan2021Accurate,liu2024tensor}. {Physical quantities are evaluated by importance sampling of configurations, and the dominant cost, which scales as $O(D^6)$, is the contraction of single-layer tensor networks to obtain wave-function amplitudes.} This approach has been successfully applied to address the nature of various frustrated spin systems~\cite{liu2022gapless,liu2022emergence,liu2024emergent,liu2024j1j2j3,liu2024quantum} and has also been extended to fermionic models~\cite{Wenyuan2025Accurate}.

{For the $t$-$J$ model we use a fermionic PEPS in the Grassmann representation~\cite{gu2010grassmann,Ignacio2010Fermionic,Wenyuan2025Accurate}. Each tensor index carries a fermion-parity grading, and the signs from fermionic anticommutation are handled locally in each tensor operation, so a fermionic network is contracted with the same algorithms as a bosonic one. The amplitude $\psi_{\boldsymbol{\alpha}}(x)=\langle x|\psi(\boldsymbol{\alpha})\rangle$ of a fermion configuration $\ket x$ is obtained by boundary-MPS contraction of the open lattice. {For the calculations on the $12\times12$ and $16\times16$ lattices we use a boundary-MPS bond dimension $\chi=5D$--$7D$ to accommodate the sizable entanglement of the fermionic $t$-$J$ PEPS. This is significantly larger than the $\chi=2D$--$3D$ that gives converged} observables in earlier studies of frustrated spin models~\cite{liu2022gapless,liu2022emergence,liu2024emergent,liu2024j1j2j3,liu2024quantum}.}

Throughout this work, the ground state is optimized by stochastic reconfiguration (SR)~\cite{Sorella1998SR,Sorella2001SR,BeccaSorella2017,Vieijra2021DirectSampling,Wenyuan2025Accurate}. SR follows from the time-dependent variational principle. Each SR step is a short imaginary-time step projected onto the PEPS manifold, whose metric is the covariance matrix of the logarithmic derivatives $\partial_{\alpha_i}\ln\psi_{\boldsymbol{\alpha}}(x)$ with respect to the tensor elements $\alpha_i$. For fermionic PEPS, the $\mathbb{Z}_2$-graded signs in these derivatives are fixed by treating the tensors and their variations as Grassmann-valued objects~\cite{gu2010grassmann}. We use the step size $\eta=0.1$ throughout. Larger steps destabilize the optimization, and smaller ones slow it down.

\subsection{PEPS initial-state construction}
\label{subsec:Initialization}

{Because the lightly doped $t$-$J$ model has competing low-energy states, we start the SR optimization from several physically motivated initial states}~\cite{Corboz2014Competing,Zheng2025CompetingPDW}. Recent iPEPS calculations in the lightly doped regime ($\delta\le0.1$) identified several low-energy PDW states~\cite{Zheng2025CompetingPDW}. The closest competitor of the uniform $d$-wave state is the plaquette PDW, denoted P2 in Ref.~\cite{Zheng2025CompetingPDW}: its bond pairing has a $2\times2$ unit cell, preserves $C_4$ symmetry, and changes sign under lattice translations. Both iPEPS states were optimized at $t'=0$. They have a uniform hole density and an antiferromagnetic spin background, and they serve as the parent states of our finite-PEPS initial states.

To build a finite initial PEPS, we tile the open lattice with the four tensors of the $2\times2$ iPEPS unit cell and reduce every virtual bond that points out of the lattice to dimension one by a singular value decomposition. The resulting state carries the iPEPS pairing pattern in the bulk; all tensors are then optimized by SR. Because each iPEPS already approximates the corresponding state in the thermodynamic limit, this construction speeds up convergence considerably, and the subsequent optimization tests whether the state remains stable on the finite open lattice.

{For different initialization states which sharply different in pairing momenta, SR can either drive them to the same state, or leave each in its own local minimum.} In the second case the optimized states are compared at the same $\mu$ through $\Omega(\mu)$ (Sec.~\ref{subsec:symmetry_ensembles}). The lower state is our variational estimate of the ground state, and the higher one is metastable within the PEPS manifold at that $D$.

Earlier PEPS-VMC studies instead built initial states from finite-size simple-update states, sometimes with weak temporary pinning fields that steer the optimization toward a chosen charge or spin pattern~\cite{Jiang2008SimpleUpdate,Wenyuan2017Gradient,Wenyuan2021Accurate,Wenyuan2025Accurate}. We use such initial states for the $6\times6$ benchmark (Sec.~\ref{sec:bench_peps_dmrg}), where the optimized PEPS agrees closely with DMRG, and for the stripe at $\delta=1/8$ (Appendix~\ref{app:hole8_benchmark}).

\subsection{Symmetry sectors, ensembles, and Monte Carlo sampling }
\label{subsec:symmetry_ensembles}

In PEPS-VMC the global quantum numbers can be fixed at two levels: by the symmetry of the tensors and by the Monte Carlo sampler. For example, a $\mathbb{Z}_2^{\rm f}$-symmetric PEPS, which conserves only fermion parity, can be optimized at fixed total hole number $N_h$ and spin projection $S^z_{\rm tot}$ by sampling only that sector; this is equivalent to optimizing the projected state $\hat P_{N_h,S^z_{\rm tot}}|\psi\rangle$.

{Canonical ensemble calculations restrict the sampling to fixed $N_h$ and $S^z_{\rm tot}=0$. They are used for the benchmarks (Sec.~\ref{sec:bench_peps_dmrg}), so that PEPS and DMRG are compared in the same $U(1)_N\times U(1)_{S^z}$ sector. Grand-canonical ensemble calculations are used on the $12\times12$ and $16\times16$ lattices. There the tensors inherit $\mathbb{Z}_2^{\rm f}\times U(1)_{S^z}$ symmetry from the iPEPS parent states, and the sampling covers all particle numbers this symmetry allows. In the latter case, SR minimizes $\Omega(\mu)$ at a fixed chemical potential, and the hole concentration {becomes} an output.} {Because the particle number is not fixed, the pairing amplitude $\Phi_a(i)$ can be measured directly, as in grand-canonical DMRG~\cite{White2021PNAS,Chen2025PNAS}. Its sign and magnitude on each bond can {be} used to diagnose the pairing symmetry and spatial modulation. We also measure the pair--pair correlations to {diagnose superconducting off-diagonal long-range order}.}

{Configurations are generated by local Monte Carlo updates. The usual update acts on a nearest-neighbor two-site window and has a low acceptance rate in the $t$-$J$ model. We instead update windows of three contiguous sites: a sweep visits every such window along both lattice directions, shifting it by one site at each step, and successive windows overlap on two sites. Within a window, the candidate configurations are those compatible with the conserved quantum numbers, and the new one is selected with the Suwa--Todo algorithm, which satisfies global balance without imposing detailed balance~\cite{SuwaTodo2010}. In the cases tested, the three-site update roughly doubles the acceptance rate at small extra cost; in representative $12\times 12$ and $16\times 16$ systems the acceptance rate is $0.4$--$0.5$.}

{The SR optimization typically {uses} $1.28\times10^4$--$5.12\times10^4$ total samples per SR iteration, with $50$--$500$ warm-up sweeps {in each Markov chain at the start of an optimization}. After convergence, observables are computed using $5.12\times10^4$--$1.92\times 10^5$ Monte Carlo samples.  Statistical uncertainties are estimated by standard binning analysis from the Monte Carlo data. }

\section{Evolution of pairing patterns with \texorpdfstring{$t'$}{t'} on long strips}

\label{sec:dmrg_t2_evolution}
In this section, we present the DMRG results for {six- and eight-leg strips with open boundaries along both the $x$ and $y$ directions, at fixed $J=0.4$ and hole doping $\delta=1/16$}.
\subsection{6-leg systems}

\begin{figure}[htbp]
    \centering
    \includegraphics[width=0.8\linewidth]{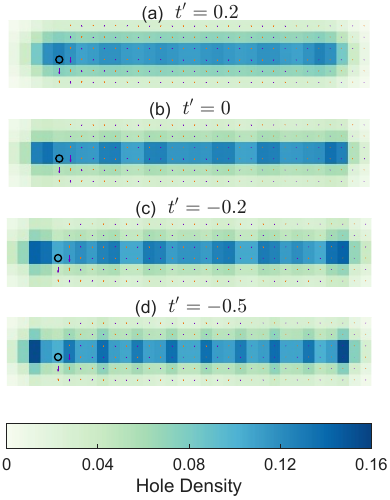}
    \caption{Hole density (colormap) and spin correlations (arrows) on six-leg open strips at $\delta=1/16$ and $J=0.4t$, with the reference site marked by a black circle. {The} spin-correlation arrows share a common scale across the four panels, with the longest arrow corresponding to $|F_i|=0.274$ in the $t'=-0.2$ panel.}
    \label{fig:6leg_density}
\end{figure}

We begin with six-leg strips, i.e., $L_y=6$. Figure~\ref{fig:6leg_density} presents the hole density $\langle n_{i,h}\rangle$ together with the spin correlations $F_i$ for four representative values $t'=0.2,\,0,\,-0.2,\,-0.5$. In all cases, the hole density is concentrated in the interior and suppressed near the boundaries.
The charge profile is primarily modulated along $\hat x$, and the modulation becomes progressively stronger as $t'$ is reduced from $0.2$ to $-0.5$. The $t'=0.2$ panel is closest to a nearly uniform stripe, while the $t'=-0.5$ panel shows the strongest charge modulation. This trend matches earlier DMRG cylinder studies of lightly doped $t$-$J$ and Hubbard models~\cite{Shoushu2021Robust,YiFan2024Ground,Chen2025PNAS}, where quasi-long-range stripe order is observed on the electron-like side and is enhanced into long-range CDW order on the hole-like side. The spin correlations remain predominantly antiferromagnetic. For $t'=0.2$, the displayed correlations follow the N\'eel sign pattern throughout the strip. For other $t'$, we do not clearly resolve antiphase domain walls of the type reported near $\delta=1/8$, where the staggered spin pattern reverses sign across hole-rich stripes~\cite{White1998Stripe}. Along the strip, the spin correlations decay exponentially for all four values of $t'$, and the trends described in this subsection persist on shorter $6\times16$ strips (Appendix~\ref{app:dmrg_obc_continuations}).

\begin{figure}[htbp]
    \centering
    \includegraphics[width=\linewidth]{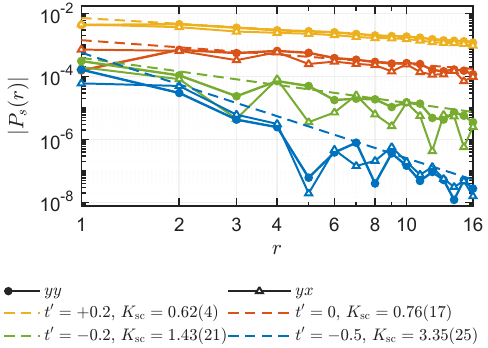}
    \caption{ Absolute values of singlet-pairing correlations in six-leg systems at $\delta=1/16$ in a log-log plot.  $r=x-x_{\rm ref}$ is the longitudinal separation. The reference bonds are vertical bonds. Circles show $yy$ correlations with vertical target bonds in the middle row of the 6-leg lattice; triangles show $yx$ correlations with horizontal target bonds on the lower middle row of the lattice.  Dashed lines are power-law fits to the envelope of the correlations; the legend reports the fitted $K_{\rm sc}$ and its regression standard error.}
    \label{fig:6leg_sc}
\end{figure}

We now turn to the singlet pair-pair correlation $P_s^{ab}(r)$ (Fig.~\ref{fig:6leg_sc}) with $r$ the horizontal separation. We typically fix the target bonds in the central rows. To characterize the decay we fit the envelope of the SC correlations to $A_{\rm sc}r^{-K_{\rm sc}}$. We interpret $K_{\rm sc}$ within the Luttinger-liquid framework routinely used for quasi-1D systems~\cite{White2002FriedelLL}: $K_{\rm sc}<2$ implies a divergent singlet-pairing susceptibility as $T\to0$, and $K_{\rm sc}<1$ further makes pairing the dominant instability. {As $t'$ is reduced, the exponents (Fig.~\ref{fig:6leg_sc}) cross all three regimes: pairing is dominant at $t'=0.2$ and $0$ [$K_{\rm sc}=0.62(4)$ and $0.76(17)$], divergent at $t'=-0.2$ [$K_{\rm sc}=1.43(21)$], and short-ranged at $t'=-0.5$ [$K_{\rm sc}=3.35(25)$].  Note that at $t'=-0.5$ the power-law fit cannot be distinguished from an exponential one, and a large $K_{\rm sc}$ signals fast, short-ranged decay of pairing.}

\begin{figure}
    \centering
    \includegraphics[width=\linewidth]{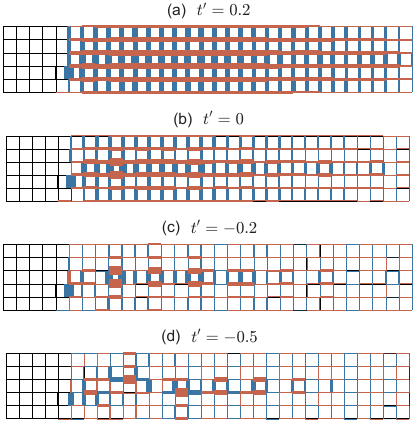}
    \caption{Two-dimensional view of normalized singlet-pairing correlations in six-leg systems at $\delta=1/16$ with different $t'$.  We visualize $P_s^{ab}(r)\,r^{K_{\rm sc}}/A_{\rm sc}$ using the fit parameters from the fixed central target bonds in Fig.~\ref{fig:6leg_sc}.  Each panel is scaled independently, so only the sign structure and relative modulation within a panel are meaningful. Blue and red bonds denote positive and negative correlations, respectively.}
    \label{fig:6leg_sc_2d}
\end{figure}

{To examine the pairing symmetry and spatial modulations, we plot in
Fig.~\ref{fig:6leg_sc_2d} the normalized correlations $P_s^{ab}(r)\,r^{K_{\rm sc}}/A_{\rm sc}$ using the fit parameters from
Fig.~\ref{fig:6leg_sc}, so that the sign structure and modulation remain visible across the whole strip.}
At $t'=0.2$, the correlations exhibit a uniform $d$-wave sign structure, with opposite signs on horizontal and vertical target bonds, consistent with previous cylindrical DMRG calculations~\cite{Shoushu2021Robust}. At $t'=0$, this sign structure persists, while the correlation amplitudes develop visible modulations along both lattice directions. The correlations on vertical target bonds show a period-two strong--weak alternation along $y$, while the longitudinal modulation follows the charge-density modulation.
At $t'=-0.2$, the modulation becomes more pronounced relative to the uniform background. For $t'=-0.5$, the pairing sign structure becomes patchy and the uniform $d$-wave component is no longer identifiable. Away from the reference bond, the pairing pattern approximately repeats every three lattice spacings along $x$ by inspection, coinciding with the charge modulation period.

\subsection{8-leg systems}
\begin{figure}[htbp]
    \centering
    \includegraphics[width=\linewidth]{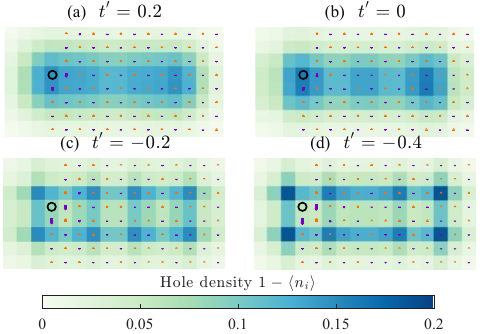}
    \caption{Hole density and spin correlations in $8\times 16$ systems at $\delta=1/16$ and $J=0.4t$, with different values of NNN hopping $t'$. The hole density $\langle n_{i,h}\rangle$ is shown by a common colormap, and the spin correlations are shown by arrows on a common scale across all panels; the reference site is marked by a black circle. The longest arrow corresponds to $F_i=-0.380$ in the $t'=-0.4$ panel.}
    \label{fig:8leg_density}
\end{figure}

 We perform DMRG calculations on $8\times16$ systems for $t'=0.2$, $0$, $-0.2$, and $-0.4$.  As in the six-leg systems, the hole density is suppressed near the  boundaries and concentrated in the interior. On the wider strips, the charge profile develops a more pronounced two-dimensional structure as $t'$ is reduced [Fig.~\ref{fig:8leg_density}]. At $t'=0.2$, the central density profile remains relatively smooth. For negative $t'$, a $4\times4$-like charge motif becomes apparent, with enhanced hole density surrounding a region of lower density. This structure is most clearly visible at $t'=-0.4$.

The spin correlations remain predominantly antiferromagnetic. At $t'=0.2$, the spin correlations follow the N\'eel sign pattern, whereas deviations appear for $t'\leq0$. The regions with reversed staggered spin correlations do not form a clear, regular array of antiphase domain walls associated with the charge modulation.

\begin{figure}[t]
    \centering
    \includegraphics[width=\linewidth]{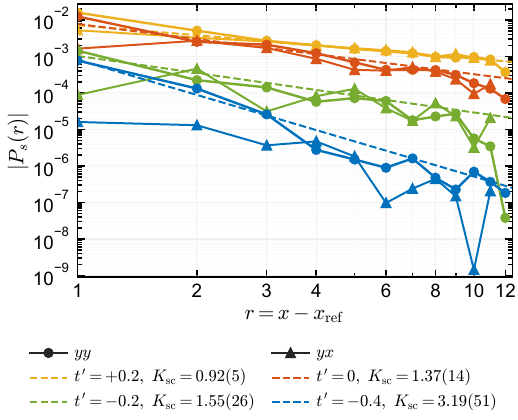}
    \caption{Singlet-pairing correlations in eight-leg systems at $\delta=1/16$. Plotting conventions follow Fig.~\ref{fig:6leg_sc}. The $yy$ channel uses vertical target bonds connecting the two central rows, while the $yx$ channel uses horizontal target bonds on the upper of these rows. The correlations at all four values of $t'$ are extrapolated quadratically to zero truncation error.}
    \label{fig:8leg_sc}
\end{figure}

Figure~\ref{fig:8leg_sc} shows the singlet-pairing correlations extrapolated to the zero-truncation-error limit, evaluated on fixed central target bonds as in the six-leg systems.
Power-law fits to the correlation envelopes give effective decay exponents $K_{\rm sc}=0.92(5)$, $1.37(14)$, $1.55(26)$, and $3.19(51)$ for $t'=0.2$, $0$, $-0.2$, and $-0.4$, respectively. The same monotonic suppression of pairing as $t'$ is reduced remains visible. In these fits, the first three exponents satisfy the Luttinger-liquid criterion $K_{\rm sc}<2$ for a divergent pairing susceptibility introduced above. Again, at $t'=-0.4$ the fit cannot distinguish algebraic from exponential decay, as the fitting range is too short, and the exponent primarily measures the decay rate; nevertheless, the fast decay of the pairing correlations at $t'=-0.4$ is unambiguous.

\begin{figure}[htbp]
    \centering
    \includegraphics[width=\linewidth]{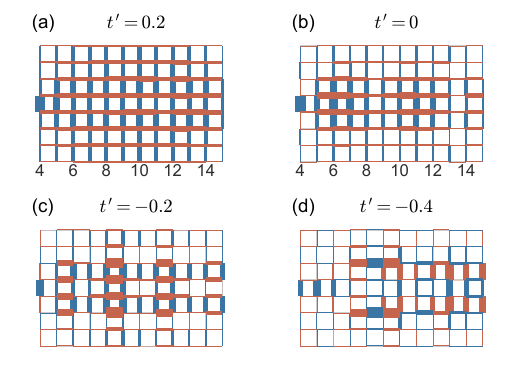}
    \caption{Two-dimensional view of normalized singlet-pairing correlations in eight-leg systems at $\delta=1/16$. We visualize the same zero-truncation-error correlations as in Fig.~\ref{fig:8leg_sc}, normalized as $P_s^{ab}(r)\,r^{K_{\rm sc}}/A_{\rm sc}$ using that figure's fit parameters. The reference bond lies in the middle of the rung at $x_0=3$. The top-row numbers give the zero-based lattice $x$ coordinate after cropping. Each panel for different $t'$ uses its own bond-width scale, so only the sign structure and relative modulation within a panel are meaningful. Blue and red bonds denote positive and negative correlations, respectively.}
    \label{fig:8leg_sc_2d}
\end{figure}

Figure~\ref{fig:8leg_sc_2d} shows the spatial structure of the singlet-pairing correlations after normalization by the fitted power-law envelope, following the same procedure as for the six-leg systems.
At $t'=0.2$, the correlations retain the uniform $d$-wave sign structure, with opposite signs on horizontal and vertical target bonds.
At $t'=0$, this sign structure persists, while the correlation amplitudes show more visible spatial modulations. The longitudinal modulation has a spacing comparable to that of the hole-density modulation in Fig.~\ref{fig:8leg_density}. The amplitude modulation becomes pronounced at $t'=-0.2$, producing a plaquette-like pattern while largely preserving the $d$-wave sign structure. At $t'=-0.4$, correlations on bonds of either orientation change sign across the strip, departing from the uniform $d$-wave sign structure.
The resulting pattern includes plaquettes whose four bounding bonds carry correlations of the same sign, with both positive and negative plaquettes visible.

Together, the six- and eight-leg results show that reducing $t'$ strengthens the charge modulation, suppresses the pairing correlations ($K_{\rm sc}$ rises from below 1 at $t'=0.2$ to above 3 at the most negative $t'$), and enhances their spatial modulation, with the uniform $d$-wave sign structure giving way to sign-changing patterns at the most negative $t'$ studied.

\section{\texorpdfstring{$L\times L$}{LxL} systems with \texorpdfstring{$C_4$}{C4} symmetry}
\label{sec:c4_square_clusters}

In this section, we study $L\times L$ systems that preserve the $C_4$ lattice symmetry. We ask whether the $t'$-driven {evolution} found on the strips persists on a square, and whether {$C_4$-symmetric systems reveal new features}. We first follow the evolution with $t'$ by DMRG on the open $8\times8$ lattice, then briefly benchmark finite PEPS against DMRG on open $6\times6$ and $8\times8$ lattices, and finally use PEPS to compare competing states on the $12\times12$ and $16\times16$ lattices.

\subsection{\texorpdfstring{Evolution with $t'$ on the $8\times8$ square}{Evolution with t' on the 8x8 square}}
\label{subsec:8x8_dmrg_scan}

\begin{figure}[t]
    \centering
    \includegraphics[width=\linewidth]{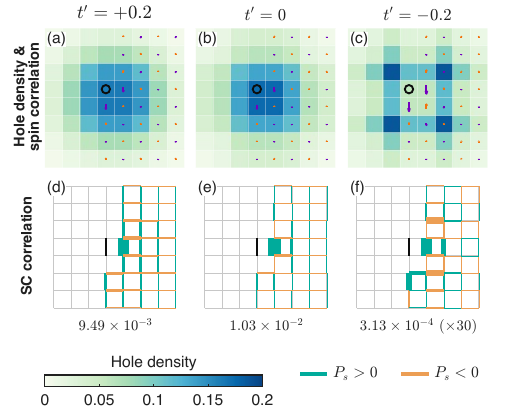}
    \caption{DMRG results on the open $8\times8$ lattice at doping $\delta=1/16$ and $J=0.4$ for $t'=0.2$, $0$, and $-0.2$ (columns). Upper row (a--c): hole density and equal-time spin correlation $F_i=\langle\mathbf S_{\rm ref}\cdot\mathbf S_i\rangle$ relative to the reference site (black circle); orange and purple arrows denote positive and negative $F_i$ on a common scale, with the longest arrow corresponding to $F_i=-0.345$ in the $t'=-0.2$ panel. Lower row (d--f): bond-resolved singlet-pairing correlation referenced to the black bond, with bond color denoting sign and bond width denoting magnitude. Bond widths use a common scale for the $t'=0.2$ and $0$ columns; the $t'=-0.2$ values are multiplied by 30 for visibility, and the numerical label below each lower-row panel gives the unscaled maximum $|P_s|$ for that column. The $t'=0$ column is the DMRG reference for the PEPS benchmark in Fig.~\ref{fig:benchmark}.}
    \label{fig:8x8_t2_scan_}
\end{figure}

We perform DMRG calculations on the open $8\times8$ lattice for $t'=0.2$, $0$, and $-0.2$, at the same $J=0.4$ and $\delta=1/16$ (four holes) as on the strips; the truncation errors are below $3.5\times10^{-5}$ (Appendix~\ref{app:dmrg_runs}).
The upper panels in Fig.~\ref{fig:8x8_t2_scan_} show the hole density together with spin correlations $F_i$ relative to the reference site marked by a black circle, and the lower panels show the bond-resolved singlet-pairing correlation $P_s$ relative to the reference bond marked in black. The $t'=0$ state also serves as the DMRG reference for the PEPS benchmark in Sec.~\ref{sec:bench_peps_dmrg}.

The charge distribution evolves with $t'$ as on the strips, but without selecting a direction [Fig.~\ref{fig:8x8_t2_scan_}(a--c)]. For $t'\ge0$ the holes concentrate near the center of the lattice and the density profile is smooth. At $t'=-0.2$ the central profile develops a pronounced bidirectional, star-like CDW: four hole-rich cells surround the center, holding roughly one hole each. None of the three states forms a unidirectional stripe, in contrast to the stripe phases found in long-cylinder DMRG at larger doping~\cite{Shoushu2021Robust,YiFan2024Ground}; on the same $8\times8$ square, stripes form only as $\delta$ approaches $1/8$ (Appendix~\ref{app:dmrg_8x8_doping_appendix}).

The pairing correlations change more abruptly than the charge profile. At $t'=0.2$ and $0$ [Fig.~\ref{fig:8x8_t2_scan_}(d,e)], they have comparable maxima, $9.5\times10^{-3}$ and $1.0\times10^{-2}$, and the uniform $d$-wave sign structure, with opposite signs on horizontal and vertical bonds. At $t'=-0.2$ [Fig.~\ref{fig:8x8_t2_scan_}(f)], the maximum $|P_s|$ drops to $3.1\times10^{-4}$, about $1/30$ of the $t'\ge0$ values, and the sign structure departs from the uniform $d$-wave pattern, forming extended plaquette-like groups of same-sign bonds. These groups do not display the strict $2\times2$ periodicity of the long-range plaquette-PDW state. This is {expected given the short-range pair correlations and the loss of} phase coherence of the state. {We return to this point in Sec.}~\ref{sec:fluctuating_pdw}.

The appearance of charge order and the loss of uniform pairing suggest that the low-energy electronic states themselves reorganize as $t'$ is lowered. To test this, we compute the spin-summed electron momentum distribution of the ground state,
\begin{equation}
    n(\mathbf k)=\frac{1}{N}\sum_{i,j,\sigma}
    e^{i\mathbf k\cdot(\mathbf r_i-\mathbf r_j)}
    \langle c^\dagger_{i\sigma}c_{j\sigma}\rangle,
    \label{eq:dmrg_single_particle_momentum}
\end{equation}
where $N=L_xL_y$ and the electron operators act in the no-double-occupancy Hilbert space. On an open lattice $\mathbf k$ is not a good quantum number; we treat it as a continuous variable, so that $n(\mathbf k)$ is a smooth function of momentum. Two exact properties set its scale. Because the $i=j$ terms are included, the Brillouin-zone average of $n(\mathbf k)$ equals the electron density $1-\delta$. Because double occupancy is excluded, $\{c_{i\sigma},c^\dagger_{i\sigma}\}=1-n_{i\bar\sigma}$ in the projected space, which bounds the distribution by $0\le n(\mathbf k)\le1+\delta$ instead of $0\le n(\mathbf k)\le2$ for unprojected electrons.

In a Fermi liquid, $n(\mathbf k)$ jumps on the Fermi surface, and Luttinger's theorem fixes the area enclosed by this surface to $a_{\rm FS}=(1-\delta)/2=15/32$ of the Brillouin zone per spin, irrespective of the interaction strength~\cite{Oshikawa2000Luttinger}. As a reference, we draw the Fermi surface of the unprojected band
\[
 \epsilon_0(\mathbf k)=-2t(\cos k_x+\cos k_y)
                    -4t'\cos k_x\cos k_y
\]
at the same bare $t'$ and density $15/16$. On these finite lattices the jump is smoothed out, so we locate it by the ridge of $|\nabla_{\mathbf k}n|$, i.e., the inflection line of $n(\mathbf k)$ (Appendix~\ref{app:momentum_ridge}). A step broadened symmetrically about $k_F$, as by a superconducting gap, keeps its inflection line at $k_F$, so the ridge marks the underlying Fermi surface also in a paired state. The ridge thus provides two independent tests {of the large-Fermi-surface assumption}: its shape can be compared with a band contour, and the area $a_{\rm ridge}$ that it encloses with the Luttinger count $a_{\rm FS}$.

\begin{figure*}[t]
    \centering
    \includegraphics[width=0.72\textwidth]{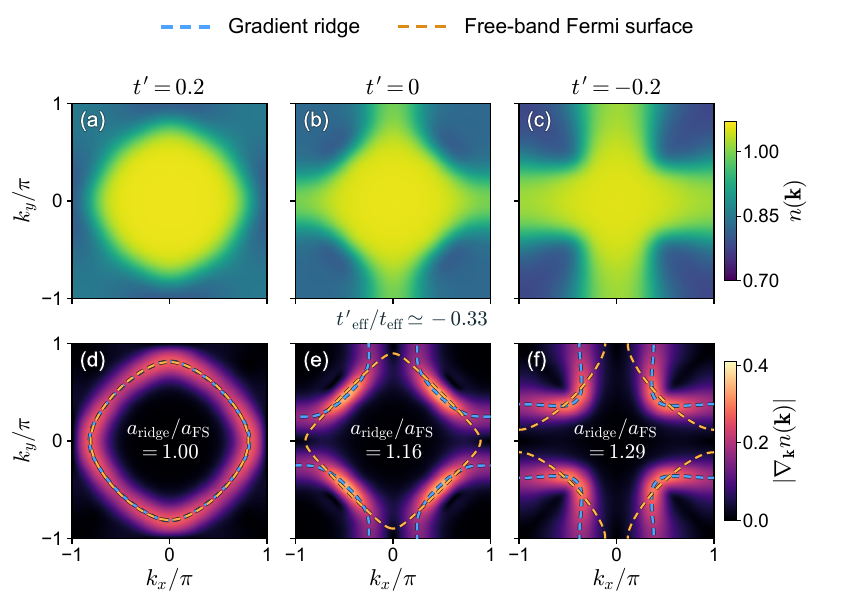}
    \caption{Electron momentum distribution on the open $8\times8$ square at $J=0.4$ and $\delta=1/16$ for $t'=0.2$, $0$, and $-0.2$ (columns). Upper row (a--c): $n(\mathbf k)$ of Eq.~\eqref{eq:dmrg_single_particle_momentum}; lower row (d--f): $|\nabla_{\mathbf k}n(\mathbf k)|$. Orange dashed curves show the Fermi surface of free fermions at the same $t'$ and density $15/16$, as a reference. Blue {dashed} curves trace the gradient ridge defined in the text. For $t'=0$, a band with $t'_{\rm eff}/t_{\rm eff}\simeq-0.33$ fits the ridge well, with the effective chemical potential also adjusted in the fit.}
    \label{fig:square_momentum_tprime}
\end{figure*}

The $t'=0.2$ state passes both tests [Fig.~\ref{fig:square_momentum_tprime}(d)]: the ridge coincides with the free-band contour and encloses $a_{\rm ridge}/a_{\rm FS}=1.00$, as expected for a large Fermi surface. For $t'\le0$ both tests fail. At $t'=0$ the ridge closes around $M=(\pi,\pi)$, i.e., it is hole-like, whereas the free-band contour at the same density is electron-like [Fig.~\ref{fig:square_momentum_tprime}(e)], and the deviation grows at $t'=-0.2$ [Fig.~\ref{fig:square_momentum_tprime}(f)]. The enclosed area exceeds the Luttinger count, with $a_{\rm ridge}/a_{\rm FS}=1.16$ at $t'=0$ and $1.29$ at $t'=-0.2$. The ridge is also not uniform along its length: near the nodal directions it stays sharp and close to the band contour, whereas near the antinodes it weakens and flattens. The eight-leg strip shows the same evolution over a wider range of $t'$, with $a_{\rm ridge}/a_{\rm FS}$ growing from $1.01$ at $t'=0.2$ to $1.32$ at $t'=-0.4$ (Appendix~\ref{app:ridge_tprime}, Fig.~\ref{fig:strip_momentum_tprime}), so this redistribution of momentum-space weight {is robust across different geometries}. On the square, it is already present at $t'=0$, where the pairing is still uniform $d$-wave, and thus precedes both the CDW and the collapse of uniform pairing at $t'=-0.2$.

The shape and the area of the ridge carry different information. The shape is well described by a renormalized band. Fitting the full ridge at $t'=0$ with a $t$-$t'$ contour, with the chemical potential also free, gives $t'_{\rm eff}/t_{\rm eff}\simeq-0.33$ on the square and $-0.35$ on the strip (Appendix~\ref{app:ridge_effective_fit}): correlations generate an appreciable negative diagonal hopping even though none is present in the Hamiltonian. ARPES on the cuprates shows a deformation in the same direction. Yoshida \emph{et al.}~\cite{Yoshida2006UnderlyingFS} fitted the underlying Fermi surface of LSCO with a $t$-$t'$-$t''$ band and found that $|t'_{\rm eff}/t_{\rm eff}|$ decreases from $0.20$ at $x=0.03$ to $0.12$ at $x=0.30$. Taking the overdoped, Fermi-liquid-like state as the closest to the bare band, this trend indicates that correlations make the effective $t'$ more negative as the Mott insulator is approached, although doping-dependent structural changes may also contribute.

The area, in contrast, cannot be accounted for by any large Fermi surface. The best-fit band at $t'=0$ holds $n_{\rm fit}\simeq1.09$ electrons per site on the square and $1.10$ on the strip, instead of the actual $15/16$. Read as a large Fermi surface, the ridge would therefore imply a negative hole doping, $\delta_{\rm eff}\simeq-0.09$, which is impossible in the $t$-$J$ model. Constraining the fitted band to the actual density keeps $t'_{\rm eff}/t_{\rm eff}$ negative, but the contour then no longer follows the ridge (Appendix~\ref{app:ridge_effective_fit}). The excess area is thus not a renormalization of a large Fermi surface; it signals that the large-Fermi-surface picture itself breaks down. The same apparent violation is observed experimentally: the large-Fermi-surface count that reproduces the nominal doping in LSCO gives negative hole concentrations in lightly doped Na-CCOC~\cite{Yoshida2006UnderlyingFS}.

One possibility is that the low-energy states form small Fermi pockets near the nodes rather than a large Fermi surface. In the pocket picture of the underdoped cuprates~\cite{YRZ2006Phenomenological,Punk2015DimerPseudogap,Bonetti2025FLstarReview}, the antinodal states are gapped and each nodal pocket carries most of its spectral weight on one side, with the weight on the other side strongly suppressed. The sharp near-nodal segments of the ridge could then belong to such pockets, and the area enclosed by the ridge would no longer count the carriers, which removes the apparent conflict with the Luttinger count. The weakened, flattened antinodal ridge is consistent with an antinodal gap, and at $t'=-0.2$ the bidirectional charge modulation offers an additional scattering channel that can reconstruct the antinodal states (Sec.~\ref{subsec:cdw_momentum}). The $8\times8$ lattice is too small to resolve the weakly weighted parts of such pockets or to fix their position and size; this requires the frequency-resolved spectral function on larger lattices.

In summary, the $8\times8$ square reproduces the $t'$-driven changes found on the strips in a $C_4$-symmetric form. As $t'$ is lowered, the uniform $d$-wave correlations collapse, a bidirectional CDW develops, and the momentum distribution departs from a large Fermi surface, most strongly near the antinodes, consistent with small Fermi pockets near the nodes. The lattice is small, however, and its open boundaries strongly shape the charge distribution. Because the cost of DMRG grows exponentially with the lattice width, we turn to finite PEPS to reach larger squares and to determine which competing state has the lowest energy. We first benchmark PEPS against DMRG and then compare the competing states on the $12\times12$ and $16\times16$ lattices.

\subsection{Benchmark of finite PEPS against DMRG}
\label{sec:bench_peps_dmrg}

\begin{table*}[t]
    \caption{Energy benchmark for the $t$-$J$ model at $t'=0$ and $S^z_{\rm tot}=0$.
    $N_h$ is the number of holes, $m$ the DMRG kept-state dimension, and $D$ the PEPS bond dimension.
    Both energies are total energies in units of $t=1$. Parentheses give Monte Carlo standard errors in the last quoted digit.
    The relative error is $(E_{\rm PEPS}-E_{\rm DMRG})/|E_{\rm DMRG}|$, evaluated using the finite-$m$ DMRG reference listed for each group.}
    \label{tab:benchmark_energy}
    \centering
    \normalsize
    \setlength{\tabcolsep}{8pt}
    \renewcommand{\arraystretch}{1.1}
    \begin{tabular}{ccccrcrr}
    \toprule
    Lattice & $N_h$ & $J/t$ & $m$ & $E_{\rm DMRG}/t$ & $D$ & $E_{\rm PEPS}/t$ & Relative error \\
    \midrule
$6\times6$ & 2 & 0.3 & $20,000$ & $-14.820062$ & 8 & $-14.757(5)$ & $4.245\times10^{-3}$ \\
& & & & & 12 & $-14.806(2)$ & $9.22\times10^{-4}$ \\
& & & & & 16 & $-14.8165(5)$ & $2.38\times10^{-4}$ \\
    \addlinespace
$8\times8$ & 2 & 0.4 & $38,000$ & $-30.327531$ & 8 & $-30.268(2)$ & $1.976\times10^{-3}$ \\
& & & & & 12 & $-30.302(1)$ & $8.26\times10^{-4}$ \\
& & & & & 16 & $-30.3140(5)$ & $4.45\times10^{-4}$ \\
    \addlinespace
$8\times8$ & 4 & 0.4 & $40,000$ & $-33.534191$ & 10 & $-33.400(2)$ & $4.002\times10^{-3}$ \\
    & & & & & 12 & $-33.4856(8)$ & $1.449\times10^{-3}$ \\
    & & & & & 16 & $-33.5067(7)$ & $8.186\times10^{-4}$ \\
    \bottomrule
    \end{tabular}
\end{table*}

We benchmark finite PEPS against DMRG on the same open $6\times6$ and $8\times8$ lattices {with the same parameters},
comparing energies, hole densities, spin correlations, and singlet pair--pair correlations. Table~\ref{tab:benchmark_energy} summarizes the energy comparisons for three representative cases, using the DMRG energy at the largest $m$
in each case as the reference.

At $D=12$, the bond dimension used for most of the results below, the PEPS energies agree with the DMRG references to within $0.15\%$ in all three $t'=0$ benchmarks. The relative energy differences decrease monotonically with $D$ in these cases, reaching $0.082\%$ at $D=16$ for the four-hole $8\times8$ lattice. See Appendix~\ref{app:hole8_benchmark} for a separate benchmark
at $\delta=1/8$.

\begin{figure}[tbp]
    \centering
    \includegraphics[width=\linewidth]{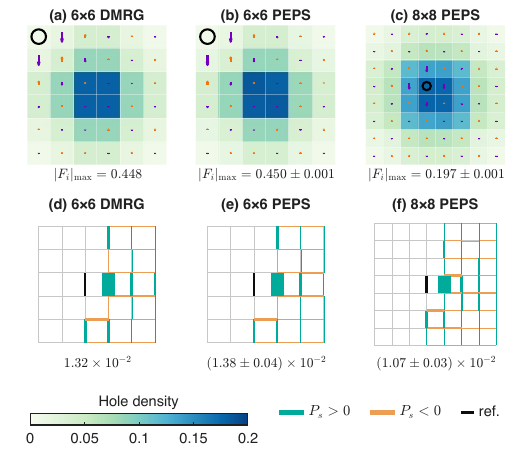}
    \caption{DMRG--PEPS benchmark on open square lattices at $t'=0$. The upper row shows the hole density and equal-time spin correlation $F_i$ for (a) DMRG with $m=20000$ and (b) PEPS with $D=16$ on the $6\times6$ lattice with two holes at $J/t=0.3$, and for (c) PEPS with $D=12$ on the $8\times8$ lattice with four holes at $J/t=0.4$; the black circle marks the reference site. The lower row (d--f) shows the corresponding bond-resolved singlet-pairing correlations on the target bonds measured by DMRG. Bond color denotes the sign of $P_s$, bond width denotes $|P_s|$, and the black bond marks the reference bond. The labels below the panels give $\max|F_i|$ (upper row) and $\max|P_s|$ (lower row), with the Monte Carlo standard error for PEPS. In (a,b) the arrows are normalized to the largest $|F_i|$ of each panel, and (d,e) share one bond-width scale. The corresponding $8\times8$ DMRG results ($m=40000$) form the $t'=0$ column of Fig.~\ref{fig:8x8_t2_scan_}; panels (c,f) use the same arrow, bond-width, and color scales as that figure. In (f), the only bonds whose sign differs from DMRG are two vertical bonds in the fifth column, where DMRG gives $P_s\approx-7.27\times10^{-5}$ and the PEPS values are consistent with zero within two Monte Carlo standard errors.}
    \label{fig:benchmark}
\end{figure}

The hole density, spin {correlations}, and SC correlations obtained by DMRG and PEPS for the $6\times6$ two-hole and $8\times8$ four-hole systems are shown in Fig.~\ref{fig:benchmark}; the $8\times8$ DMRG results are the $t'=0$ column of Fig.~\ref{fig:8x8_t2_scan_}, so panels (c,f) of Fig.~\ref{fig:benchmark} are to be compared with panels (b,e) of Fig.~\ref{fig:8x8_t2_scan_}. Both methods give the same spatial pattern of charge modulation, with peak hole densities differing by less than $3\%$, and the {residual} local moments $\langle S_i^z\rangle$ are small. With the same reference site in both methods, the equal-time spin correlator $F_i$ has the same sign at every site compared except
one on the $6\times6$ lattice, where DMRG gives $|F_i|<10^{-3}$.

On both lattices, PEPS and DMRG give the same predominantly $d$-wave sign pattern for the singlet pair--pair correlations. On the target bonds common to both calculations, the magnitudes of $P_s^{ab}$ agree across bonds, with the largest magnitudes differing by less than $5\%$.

These comparisons establish that finite PEPS accurately captures the charge profiles, spin-correlation sign patterns, and singlet pair--pair correlations on the benchmark lattices. We therefore use finite PEPS, primarily at $D=12$, to compute these observables on the $12\times 12$ and $16\times 16$ lattices.

\subsection{Competing states on large square systems: a finite PEPS simulation}
\label{subsec:12x12}

We now return to the reorganization found by DMRG on the $8\times8$ square in Sec.~\ref{subsec:8x8_dmrg_scan}, where charge order strengthens and the pairing pattern changes sharply on the negative-$t'$ side. To investigate this reorganization on a larger square, we study the open $12\times12$ lattice using finite PEPS with bond dimensions $D=10$ and $12$. Every tensor of the finite PEPS is independent, so the charge, spin, and pairing patterns relax freely on all 144 sites, without the small unit cells and uniform-density constraint of the iPEPS studies~\cite{Yue2024FluctuatingPDW,Zheng2025CompetingPDW}, while the open square preserves the $C_4$ symmetry that cylinders break. At each $t'$, we start from two states, one with uniform $d$-wave pairing and the other with a plaquette-PDW pattern, constructed as described in Sec.~\ref{subsec:Initialization}. For each bond dimension $D$, we use the same pair of initial states at all values of $t'${. This {examines} how the Hamiltonian changes their relative stability.} The calculations are grand canonical (Sec.~\ref{subsec:symmetry_ensembles}): the two states are compared at the same $t'$, chemical potential $\mu$, and $D$, using $\Omega(\mu)$. Their mean hole densities can relax independently and remain within $\delta\leq0.1$.

SR optimization approximates imaginary-time evolution under $K_\mu$ projected onto the finite-$D$ PEPS manifold (Sec.~\ref{subsec:peps_vmc}); we run $200$--$500$ steps at $\eta=0.1$. During the optimization the density profile, the antiferromagnetic background, and the prescribed pairing pattern are all free to evolve, so an initial ordering tendency survives only if it lowers the energy.  At both $D$ and over the full optimization, the two initial states relax into distinct local minima of the variational energy. We denote the SR-optimized states obtained from the uniform $d$-wave and the plaquette-PDW initial states by $|\psi_d\rangle$ and $|\psi_{\rm pPDW}\rangle$, respectively; the subscripts refer only to the initialization, and the physical character of each state is determined from its measured observables.

\begin{figure*}[t]
    \centering
    \includegraphics[width=\textwidth]{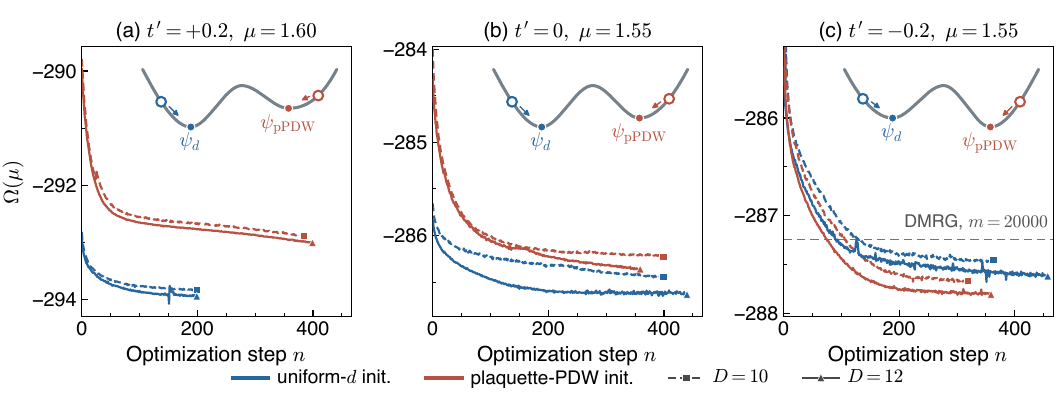}
    \caption{Grand-canonical energy evolution during stochastic reconfiguration (SR) optimization of PEPS on $12\times12$ systems at $J=0.4$, comparing uniform $d$-wave and plaquette-PDW initial states at $D=10$ and $D=12$. The vertical axis shows the variational grand-canonical energy $\Omega(\mu)$. SR approximates imaginary-time evolution within the finite-$D$ PEPS variational space. The horizontal axis counts SR optimization steps $n$, tracking the relaxation of the wave function toward lower energy.  Faint bands show one standard error (too small to be visible). The horizontal gray dashed line in panel (c) marks an auxiliary fixed-particle-number DMRG calculation with $N_h=12$.
Blue and red identify the uniform-$d$ and plaquette-PDW initializations;
dashed lines with squares and solid lines with triangles denote $D=10$ and
$12$, respectively. Insets schematically illustrate the optimized basins
and their energetic ordering. Open and filled circles denote initial and
optimized states; arrows indicate SR relaxation.}
\label{fig:12x12_energy_competition}
\end{figure*}

\begin{table}[t]
\caption{
Grand-canonical PEPS results on the $12\times12$ lattice at $J=0.4$. Blocks (a,b) identify the initial pairing patterns. Each entry gives the total grand-canonical energy $\Omega$ (upper line, in units of $t$) and mean hole density $\delta$ (lower line). Bold energies are lower within the pair of initializations at the same $t',\mu,D$, where both results are available. Parentheses give uncertainties in the last digits.}
\label{tab:peps_runs}
\small
\setlength{\tabcolsep}{3pt}
\begin{ruledtabular}
\begin{tabular}{c c c c}
$D$ & $t'=+0.2$ & $t'=0$ & $t'=-0.2$ \\
 & $\mu=1.60$ & $\mu=1.55$ & $\mu=1.55$ \\
\hline
\multicolumn{4}{c}{(a) Uniform $d$-wave initialization} \\[2pt]
$10$ & \shortstack{$\boldsymbol{-293.830(2)}$\\$0.08812(7)$} & \shortstack{$\boldsymbol{-286.446(2)}$\\$0.06277(4)$} & \shortstack{$-287.471(3)$\\$0.06714(6)$} \\[4pt]
$12$ & \shortstack{$\boldsymbol{-293.928(3)}$\\$0.09656(7)$} & \shortstack{$\boldsymbol{-286.620(2)}$\\$0.07272(4)$} & \shortstack{$-287.566(3)$\\$0.07357(6)$} \\[4pt]
\hline
\multicolumn{4}{c}{(b) Plaquette-PDW initialization} \\[2pt]
$10$ & \shortstack{$-292.891(4)$\\$0.02794(8)$} & \shortstack{$-286.21(3)$\\$0.0337(1)$} & \shortstack{$\boldsymbol{-287.668(3)}$\\$0.08325(6)$} \\[4pt]
$12$ & \shortstack{$-293.005(1)$\\$0.04071(3)$} & \shortstack{$-286.364(2)$\\$0.05685(3)$} & \shortstack{$\boldsymbol{-287.800(3)}$\\$0.08343(7)$} \\[4pt]
\end{tabular}
\end{ruledtabular}
\end{table}

Figure~\ref{fig:12x12_energy_competition} shows the energy relaxation, and Table~\ref{tab:peps_runs} summarizes the optimized energies $\Omega_d$ and $\Omega_{\rm pPDW}$ and hole densities. At both $D$, $|\psi_d\rangle$ has the lower energy at $t'=+0.2$ and $0$, whereas $|\psi_{\rm pPDW}\rangle$ is lower at $t'=-0.2$, by $0.20t$ ($D=10$) and $0.23t$ ($D=12$). The plaquette-PDW states start higher in energy than the $d$-wave states [Fig.~\ref{fig:12x12_energy_competition}(c)], so the reversal reflects a strong energetic preference for $|\psi_{\rm pPDW}\rangle$ at $t'=-0.2$. {At $t'=0$, where iPEPS does not resolve the energetic ordering, the finite PEPS distinguishes the two states clearly: $|\psi_d\rangle$ lies lower} by $0.23t$ ($D=10$) and $0.257t$ ($D=12$). The energetic ordering reverses between $t'=0$ and $-0.2$, two points computed at the same chemical potential $\mu=1.55$. Within this variational comparison, we identify the lower-energy state as the ground state. The fixed-$N_h=12$ DMRG result shown in Fig.~\ref{fig:12x12_energy_competition}(c) lies above the PEPS energies and serves only as an auxiliary, unconverged energy reference (Appendix~\ref{app:dmrg_12x12_t2m02}).

The reversal of the energetic ordering between $t'=0$ and $-0.2$ shows that the ground state changes character in this interval. Below we characterize the lower-energy state at each $t'$, and we return to the relation between the two optimized states at the end of Sec.~\ref{subsubsec:12x12_evolution}.

\subsubsection{Evolution of charge, spin, and pairing with \texorpdfstring{$t'$}{t'}}
\label{subsubsec:12x12_evolution}

We now examine the real-space observables of the ground state {identified by} the above energy comparison at each $t'$: $|\psi_d\rangle$ for $t'=+0.2$ and $0$, and $|\psi_{\rm pPDW}\rangle$ for $t'=-0.2$ (Table~\ref{tab:peps_runs}). The metastable states are discussed in Appendix~\ref{app:metastable_branches}. The mean hole densities of these states lie between $\delta\simeq0.06$ and $0.10$, and for each $t'$ the $D=10$ and $D=12$ results show the same qualitative features. Both initial states have a uniform hole density, so every charge modulation described below develops during the optimization.

\begin{figure}[htbp]
    \centering
    \includegraphics[width=\columnwidth]{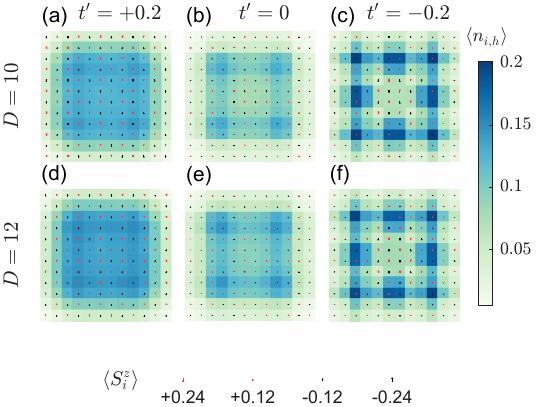}
    \caption{Hole density $\langle n_{i,h}\rangle$ (color) and local spin polarization $\langle S_i^z\rangle$ (arrows) of the PEPS ground states on the $12\times12$ lattice at $J=0.4$, for $D=10$ (upper row) and $D=12$ (lower row) and $t'=+0.2$, $0$, and $-0.2$ (columns). The states are $|\psi_d\rangle$ for $t'\ge0$ and $|\psi_{\rm pPDW}\rangle$ for $t'=-0.2$; their chemical potentials and mean hole densities are listed in Table~\ref{tab:peps_runs}. All panels share one color scale and one arrow scale.}
\label{fig:12x12_density_spin_triptych}
\end{figure}

\begin{figure}[t]
    \centering
    \includegraphics[width=\columnwidth]{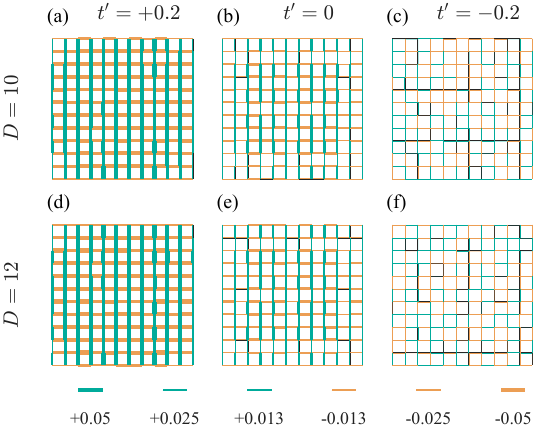}
    \caption{Singlet-pairing amplitude $\Phi_a(i)$ of the states in Fig.~\ref{fig:12x12_density_spin_triptych}. Bond color denotes the sign and bond width the magnitude of $\Phi_a$, on a common scale for all panels.}
    \label{fig:12x12_sc_triptych}
\end{figure}

\emph{Charge and spin.}---Figure~\ref{fig:12x12_density_spin_triptych} shows the hole density and the local spin polarization. At $t'=+0.2$, the holes occupy the interior of the lattice, the two outermost rings are nearly hole-free, and the density inside is almost uniform. The local spin polarization $\langle S^z_i\rangle$ shows an antiferromagnetic (AFM) pattern throughout the lattice, including the hole-rich interior.
At $t'=0$, the hole density is lower and only weakly modulated, with holes accumulating on a square ring about three sites from the edge, and the AFM pattern is strongly suppressed.
At $t'=-0.2$, the holes concentrate on a square frame of hole-rich islands on the third ring from the edge (hole density up to $0.20$), forming a bidirectional, $C_4$-symmetric CDW. The frame separates a central region of lower hole density ($\simeq0.07$--$0.10$) from the two outer rings. The AFM pattern is strongly nonuniform: it is strongest in the central region, where $|\langle S^z_i\rangle|$ reaches $0.22$ ($0.24$ for $D=10$), nearly vanishes on the frame ($\simeq0.03$), and remains weak in the outer rings ($\simeq0.04$--$0.06$) even though they are the least doped. Moreover, the staggered sign of $\langle S^z_i\rangle$ reverses across the frame, so the hole-rich frame acts as an antiphase domain wall of the AFM background---a closed, bidirectional analogue of the charge--spin arrangement in stripes~\cite{Zaanen-PRB1989,White1998Stripe}.
The metastable state at $t'=-0.2$, optimized from the uniform $d$-wave initialization, is also strongly charge modulated, but with a different $C_4$-symmetric pattern [Appendix~\ref{app:metastable_branches}, Fig.~\ref{fig:app_t2minus_dwave_density_sc}(a,b)], so strong charge modulation is a robust feature of the low-energy states at $t'=-0.2$. None of the states forms a unidirectional stripe. This sequence reproduces the trend of the $8\times8$ DMRG results (Fig.~\ref{fig:8x8_t2_scan_}): a smooth charge profile for $t'\ge0$ and a bidirectional arrangement of hole-rich cells at $t'=-0.2$. The comparison is qualitative: the $8\times8$ calculation is at fixed $\delta=1/16$, whereas here $\mu$ sets the doping. A finite $\langle S^z_i\rangle$ reflects the spin-rotation symmetry breaking of the variational state; we use it as a proxy for the antiferromagnetic correlations $F_i$ shown in the $8\times8$ panels.

\begin{table}[t]
\caption{ Pairing amplitudes of finite PEPS and parent iPEPS. Panels (a,b) use the $12\times12$ parameters of Table~\ref{tab:peps_runs}. (a) Maximum bond amplitude in the ground state; parentheses give the Monte Carlo standard error at the selected bond. (b) Uniform $d$-wave amplitude $\Phi_d$ and relative root-mean-square residual $R_\Phi/|\Phi_d|$ on the central $8\times8$ sites [Sec.~\ref{subsubsec:12x12_plaquette}], for the $d$-wave minimum, which is metastable at $t'=-0.2$. (c) Full-update uniform $d$-wave parent iPEPS at $J=0.4$, $t'=0$, and $\mu=1.568$. }
\label{tab:sc_summary}
\setlength{\tabcolsep}{5pt}
\begin{ruledtabular}
\begin{tabular}{lccc}
 & $t'=+0.2$ & $t'=0$ & $t'=-0.2$ \\
\hline
\multicolumn{4}{c}{(a) Ground state: $\max_b|\Phi_b|$ $(10^{-2})$} \\[2pt]
$D=10$ & 4.71(9) & 3.53(8) & 0.7(6) \\
$D=12$ & 4.9(1) & 3.24(7) & 0.5(3) \\[3pt]
\hline
\multicolumn{4}{c}{(b) States in the $d$-wave minimum} \\[2pt]
\multicolumn{4}{l}{$D=10$} \\
$|\Phi_d|$ $(10^{-2})$ & 4.10 & 2.64 & 1.25 \\
$R_\Phi/|\Phi_d|$ $(\%)$ & 5.2 & 20.5 & 31.8 \\[3pt]
\multicolumn{4}{l}{$D=12$} \\
$|\Phi_d|$ $(10^{-2})$ & 4.14 & 2.60 & 1.19 \\
$R_\Phi/|\Phi_d|$ $(\%)$ & 6.2 & 17.1 & 29.6 \\[3pt]
\hline
\multicolumn{4}{c}{(c) Parent uniform $d$-wave iPEPS} \\[2pt]
$D$ & $\delta$ & $\max_b|\Phi_b|$ $(10^{-2})$ & $|\Phi_d|$ $(10^{-2})$ \\
$10$ & 0.05828 & 3.12 & 3.01 \\
$12$ & 0.05894 & 2.91 & 2.78 \\
\end{tabular}
\end{ruledtabular}
\end{table}

\emph{Local pairing amplitude.}---The bond-resolved amplitude $\Phi_a(i)$ changes in both magnitude and spatial structure as $t'$ decreases (Fig.~\ref{fig:12x12_sc_triptych}). At $t'=+0.2$, it has a nearly uniform $d$-wave form, with opposite signs on horizontal and vertical bonds. At $t'=0$, the amplitude is moderately reduced, while a period-two modulation develops on top of the $d$-wave pattern; we analyze this modulation in Sec.~\ref{subsubsec:12x12_plaquette}. In the ground state at $t'=-0.2$, the pairing amplitude is strongly suppressed throughout the lattice, and the bond signs no longer follow the $d$-wave pattern. At $t'=-0.2$ the one-standard-error interval of $\Phi_a(i)$ includes zero on about one third of the bonds at both $D$, so the signs on these bonds are unresolved. The maximum amplitudes are summarized in Table~\ref{tab:sc_summary}(a).

This suppression is not a doping effect: the $t'=-0.2$ state ($\delta\simeq0.083$) and the $D=10$ state at $t'=+0.2$ ($\delta\simeq0.088$) have nearly the same hole density. A small $\Phi_a$ alone, however, does not establish the absence of pairing. On a finite lattice $\Phi_a\neq0$ requires the state to mix sectors of different particle number, and a state close to a particle-number eigenstate has $\Phi_a\approx0$ even when its pair correlations are appreciable. The pair--pair correlations and the particle-number distribution test this possibility.

\begin{figure*}[t]
    \centering
    \includegraphics[width=\textwidth]{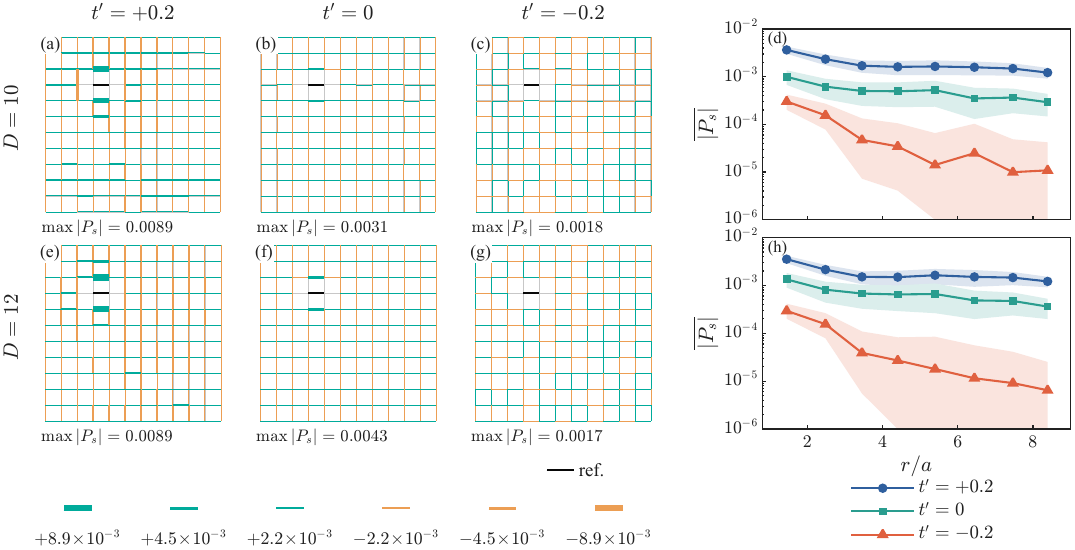}
    \caption{Singlet pair--pair correlations $P_s^{ab}$ of the states in Fig.~\ref{fig:12x12_density_spin_triptych}, for $D=10$ (upper row) and $D=12$ (lower row). Panels (a--c) and (e--g) show bond-resolved correlations for $t'=+0.2$, $0$, and $-0.2$, respectively. The black bond is the reference bond; bond color denotes the sign and bond width the magnitude of $P_s^{ab}$, on a common scale for all six maps. The label below each map gives $\max|P_s|$. Panels (d,h) compare $\overline{|P_s|}$ for the three values of $t'$. The bar denotes an equally weighted spatial average of absolute correlations within distance intervals of one lattice constant.  Shading gives 95\% Monte Carlo confidence intervals, {with a Bonferroni correction}.}
    \label{fig:12x12_sc_pair_corr_grid}
\end{figure*}

\emph{Pair--pair correlations.}---Figure~\ref{fig:12x12_sc_pair_corr_grid} shows the singlet pairing correlation $P_s^{ab}(r)$ measured from a fixed reference bond. In a grand-canonical state $P_s$ contains the disconnected part $\langle\Delta_a^\dagger\rangle\langle\Delta_b\rangle$, so at large distances it approaches $2\Phi_a\Phi_b$ whenever $\Phi$ is finite. Consistent with this, the correlations at $t'=+0.2$ and $0$ remain finite across the lattice with a $d$-wave sign structure, and their magnitude decreases from $t'=+0.2$ to $t'=0$ as $\Phi$ does. At $t'=-0.2$, the bond maps no longer exhibit the $d$-wave sign pattern. The distance dependence in panels (d,h) shows that the suppression of $\overline{|P_s|}$ relative to $t'\ge0$ becomes more pronounced away from the reference bond: the correlations fall strongly over the first few lattice spacings and remain much weaker at larger separations. At $t'=-0.2$ the pairing correlations on this lattice are predominantly short-ranged; at large separations they are too small for a precise Monte Carlo estimate.

\begin{figure}[htbp]
    \centering
    \includegraphics[width=0.98\columnwidth]{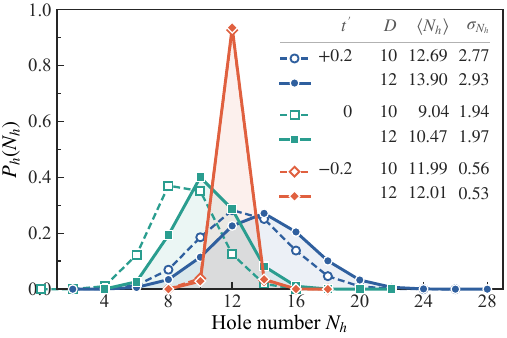}
    \caption{Hole-number distribution $P_h(N_h)$ of the $D=10$ (dashed lines, open symbols) and $D=12$ (solid lines, filled symbols) states in Fig.~\ref{fig:12x12_density_spin_triptych}. The legend gives the mean hole number and standard deviation $\sigma_{N_h}$ of each distribution. Only even $N_h$ appear because fermion parity is conserved.}
    \label{fig:12x12_particle_number_distribution}
\end{figure}

\emph{Particle-number distribution.}---The hole-number distribution \(P_h(N_h)\) makes this distinction more explicit [Fig.~\ref{fig:12x12_particle_number_distribution}]. For \(t'=+0.2\) and \(0\), \(P_h(N_h)\) has a broad, approximately Gaussian envelope, as expected for a phase-coherent BCS state with large particle-number fluctuations. By contrast, at \(t'=-0.2\), more than \(90\%\) of the weight resides in the \(N_h=12\) sector. The optimized state is therefore nearly a particle-number eigenstate with \(\delta=1/12\). Since the SC amplitude \(\Phi_a=\langle \hat\Delta_a^\dagger\rangle\) requires coherence between sectors whose particle numbers differ by two, this near locking of \(N_h\) naturally accounts for the strongly suppressed \(\Phi_a\). Because \(N_h=12\) is one hole per row and column, the locking could also be a commensuration effect of the \(12\times12\) lattice; the $16\times16$ lattice, where the locked sector has no such relation to the lattice, tests this possibility (Sec.~\ref{subsec:16x16}).

The higher-energy state at each $t'$ shows that this reversal is a crossing between two competing states. At $t'=0$, $|\psi_{\rm pPDW}\rangle$ already shows the hole-rich square frame that characterizes the $t'=-0.2$ ground state [Appendix~\ref{app:metastable_t20}, Fig.~\ref{fig:app_t20_pdw_density_sc}(a,b)]. At $t'=-0.2$, $|\psi_d\rangle$ still has the $d$-wave sign on every bond, like the ground states at $t'\ge0$ [Appendix~\ref{app:t2minus_dwave_branch}, Fig.~\ref{fig:app_t2minus_dwave_density_sc}(c,d)].
 If each local minimum evolves smoothly with $t'$, their grand-canonical energies at fixed $\mu=1.55$ must cross between $t'=0$ and $-0.2$, as sketched in the insets of Fig.~\ref{fig:12x12_energy_competition}.

\begin{figure*}[t]
    \centering
    \includegraphics[width=0.72\textwidth]{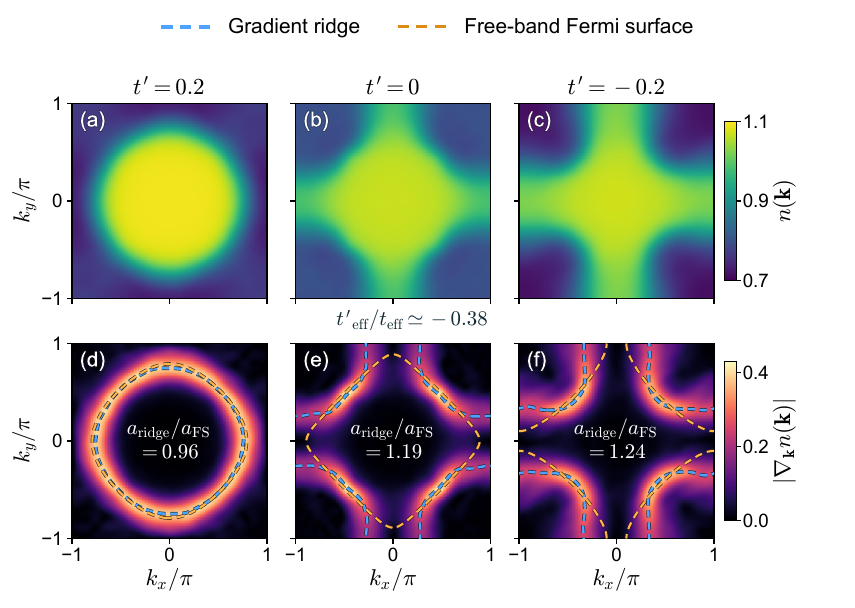}
    \caption{{Electron momentum distribution of ground states on the $12\times12$ {square} at $J=0.4$. {They are obtained from the $D=12$ PEPS states, with doping levels} $\delta\simeq0.0966$, $0.0724$, and $0.0835$. Upper row (a--c): $n(\mathbf k)$ of Eq.~\eqref{eq:dmrg_single_particle_momentum}; lower row (d--f): $|\nabla_{\mathbf k}n(\mathbf k)|$. Blue dashed curves trace the gradient ridge and orange dashed curves show the free-band Fermi surface at each electron density $1-\delta$.  At $t'=0$, the geometric band-contour fit gives $t'_{\rm eff}/t_{\rm eff}\simeq-0.38$.}}
    \label{fig:12x12_momentum_tprime}
\end{figure*}

\emph{Electron momentum distribution.}{---We also compute the electron momentum distribution of the ground states on the $12\times12$ lattice. We present the $D=12$ results in} Fig.~\ref{fig:12x12_momentum_tprime} {. The $D=10$ results are consistent with them. The pattern agrees qualitatively with that on the $8\times8$ lattice. At $t'=0.2$, the ridge of $|\nabla_{\mathbf k}n(\mathbf k)|$ encloses the $\Gamma$ point and nearly follows the free-band Fermi surface. From $t'=0$ on, the ridge encloses the $M$ point, and its enclosed area is significantly larger than that of the free-band Fermi surface at the same doping. The ridge is sharp near the nodal directions, whereas it smears and flattens near the antinodes.}

In summary, the $12\times12$ PEPS results reproduce the trend seen {in} the $8\times8$ DMRG calculation. As $t'$ decreases from $+0.2$ to $-0.2$, the ground state evolves from a nearly uniform $d$-wave superconductor with a smooth density profile, through a $d$-wave state carrying a period-2 plaquette modulation at $t'=0$, to a number-locked state with a bidirectional CDW and a strongly suppressed, spatially irregular pairing amplitude. The next subsection analyzes the plaquette modulation at $t'=0$. The pairing at $t'=-0.2$ is analyzed in Sec.~\ref{sec:fluctuating_pdw}.

\begin{figure}[htbp]
    \centering
    \includegraphics[width=0.98\columnwidth]{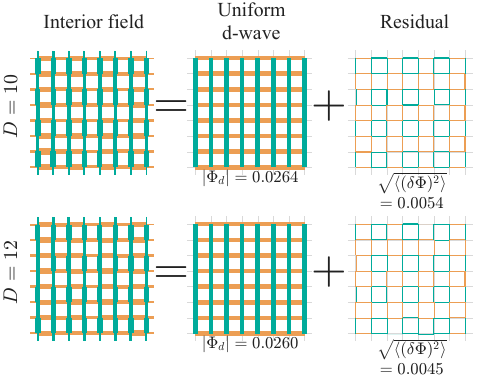}
    \caption{Decomposition of representative $12\times12$ bond-resolved singlet-pairing amplitudes at $t'=0$ for $D=10$ (top row) and $D=12$ (bottom row), after omitting the outer two bond rings. In each row, the left panel shows the remaining interior amplitude, the middle panel shows the uniform $d$-wave background extracted from the interior average, and the right panel shows the residual amplitude after subtraction. Bond widths use a common normalization across both rows. The amplitudes at all three values of $t'$ are summarized in Table~\ref{tab:sc_summary}.}
    \label{fig:12x12_sc_decomposition}
\end{figure}

\subsubsection{Plaquette modulation of the \texorpdfstring{$d$}{d}-wave pairing at \texorpdfstring{$t'=0$}{t'=0}}
\label{subsubsec:12x12_plaquette}
At $t'=0$, the period-two modulation coexists with a larger uniform $d$-wave component. To distinguish the two, we decompose the bond amplitude into its uniform part and the remaining spatial variation (Fig.~\ref{fig:12x12_sc_decomposition}). We use the central $8\times8$ sites, comprising 112 nearest-neighbor bonds, to exclude the two outermost rings where the low hole density suppresses pairing. With $\eta_b=+1$ ($-1$) on horizontal (vertical) bonds, the uniform component is $\Phi_d=N_b^{-1}\sum_b\eta_b\Phi_b$, and the residual is $\delta\Phi_b=\Phi_b-\eta_b\Phi_d$. Its root-mean-square magnitude, $R_\Phi=\langle(\delta\Phi)^2\rangle^{1/2}$, measures the total deviation from the uniform $d$-wave form. At $t'=0$, $R_\Phi$ is $17\%$--$21\%$ of $|\Phi_d|$. The residual amplitude, $4.5$--$5.4\times10^{-3}$, is well above the typical per-bond Monte Carlo standard error of $7.5\times10^{-4}$; we also extract the uniform extended-$s$ component $N_b^{-1}\sum_b\Phi_b$, which is below $3\times10^{-4}$.

The residual in Fig.~\ref{fig:12x12_sc_decomposition} is predominantly a plaquette modulation: its sign alternates between neighboring horizontal bonds along each row and between neighboring vertical bonds along each column. The two bond orientations have the same modulation amplitude, preserving $C_4$ symmetry, with wave vectors $\mathbf Q_x=(\pi,0)$ and $\mathbf Q_y=(0,\pi)$, respectively. This spatial structure is approximately described by
\begin{equation}
\Phi_b\simeq\eta_b\left[\Phi_d+\Phi_P\,(-1)^{\mathbf r_b\cdot\hat{\mathbf e}_b}\right],
\label{eq:plaquette_modulation}
\end{equation}
where $\Phi_P$ is the plaquette-modulation amplitude, $\mathbf r_b$ is the left (lower) site of bond $b$, and $\hat{\mathbf e}_b$ is its direction.
The $\Phi_P$ term has the pattern of the P2 plaquette-PDW state of Ref.~\cite{Zheng2025CompetingPDW}, from which $|\psi_{\rm pPDW}\rangle$ is initialized: its magnitude is uniform over the bonds and only its sign varies, such that plaquettes whose four bonds carry the same sign alternate in sign and form a checkerboard.
To this accuracy, the ground state at $t'=0$ is the uniform $d$-wave state with an admixture of the competing P2 state. Since $|\Phi_P|<|\Phi_d|$, the admixture changes the magnitude but not the sign of the pairing amplitude: in the total field, the strong plaquettes, whose four bonds are all enhanced, form a $2\times2$ superlattice (one plaquette in four).

On an OBC lattice the edges fix the phase of such a modulation---here the strong plaquettes include the four corner plaquettes---so its presence alone does not distinguish a bulk tendency from a boundary-induced oscillation; its spatial profile does. At $t'=0$ the relative contrast between neighboring strong and weak bonds is $0.15$--$0.22$ both at the center of the lattice and near the edge of the analysis window, i.e., the modulation does not decay into the bulk. In contrast, at $t'=+0.2$, whose optimization starts from the same uniform $d$-wave initial state, a much weaker modulation with the same phase is present near the edges but significantly decreases toward the center. This indicates that the $t'=0$ modulation is a bulk feature of the optimized state, whereas at $t'=+0.2$ it is induced by the edges.

A natural alternative is that the modulation is a by-product of uniform \(d\)-wave pairing coexisting with the N\'eel background at the antiferromagnetic wave vector \(Q_{\rm AF}\). Symmetry excludes this. To see why, let \(\Delta_{d,0}\) denote the uniform \(d\)-wave singlet component and let \(\Delta_{\alpha,P}\) denote a finite-momentum singlet-pair component with the wave vector \(P\) of the observed plaquette modulation. In a translation-invariant Landau expansion, a term linear in \(\Delta_{\alpha,P}\) must have the form $\Delta_{\alpha,P}^{*}\Delta_{d,0} O_P+\mathrm{c.c.},$
where \(O_P\) is a neutral spin-singlet composite carrying the same momentum \(P\) and the appropriate lattice form factor. For a single commensurate N\'eel field \(\mathbf M_{Q_{\rm AF}}\) with \(Q_{\rm AF}=(\pi,\pi)\), the lowest spin-rotation-invariant composite \(\mathbf M_{Q_{\rm AF}}\cdot\mathbf M_{Q_{\rm AF}}\) carries momentum \(2Q_{\rm AF}\equiv 0\), and therefore only renormalizes or mixes uniform pairing channels. The familiar cubic invariant \(\Delta_{d,0}^{*}\mathbf M_{Q_{\rm AF}}\cdot\mathbf \Pi_{-Q_{\rm AF}}+\mathrm{c.c.}\) instead induces a triplet \(\pi\)-pair component, not the bond-resolved singlet modulation plotted here. Neither invariant generates a singlet component at \((\pi,0)\) or \((0,\pi)\), so the plaquette modulation is not a by-product of N\'eel order.

For comparison, Table~\ref{tab:sc_summary}(c) lists the pairing amplitudes of the full-update parent iPEPS used to construct the uniform $d$-wave initial states. These amplitudes are measured using the corner transfer matrix renormalization group with environment dimension $\chi=100$, and the uniform pairing component $|\Phi_d|$ is averaged over the $2\times2$ unit cell.
The parent states are nearly uniform, with $R_\Phi/|\Phi_d|=2.6\%$ ($3.7\%$), much smaller than the corresponding values of $20.5\%$ ($17.1\%$) in the finite-size states at $t'=0$.

Within the $d$-wave minimum, the uniform pairing component weakens steadily as $t'$ decreases, while the spatial modulation becomes larger relative to it [Table~\ref{tab:sc_summary}(b)]. From $t'=+0.2$ to $0$, the residual grows in absolute magnitude and develops the plaquette structure described above. At $t'=-0.2$, where this minimum is metastable, the residual decreases, but the uniform component falls more rapidly, so that $R_\Phi/|\Phi_d|$ reaches about $30\%$. Consequently, negative $t'$ enhances spatial modulation relative to uniform pairing even within the $d$-wave minimum. This trend is consistent with the energy comparison, in which the state from the plaquette-PDW initialization becomes the ground state at $t'=-0.2$ (Table~\ref{tab:peps_runs}). We next test this ground state on a larger lattice.

\begin{samepage}
\subsection{\texorpdfstring{The ground state at negative $t'$ on the $16\times16$ lattice}{The ground state at negative t' on the 16x16 lattice}}
\label{subsec:16x16}

\end{samepage}

To examine whether the energetic advantage of the plaquette-PDW-initialized state at $L=12$ is a finite-size effect, and whether its modulated pairing persists, we extend the PEPS calculation to the $16\times16$ lattice at the same parameters, $t'=-0.2$, $J=0.4$, and $\mu=1.55$, again starting from both the uniform $d$-wave and the plaquette-PDW initial states.
At both $D=10$ and $D=12$, the state optimized from the plaquette-PDW initialization has the lower energy, $\Omega=-513.355(5)$ and $-513.551(3)$, respectively. Its hole density, $\delta=0.085847(8)$ at $D=10$ and $0.086024(7)$ at $D=12$, is close to that of the $12\times12$ state ($\delta\simeq0.083$).

\begin{figure*}[t]
    \centering
    \includegraphics[width=\textwidth]{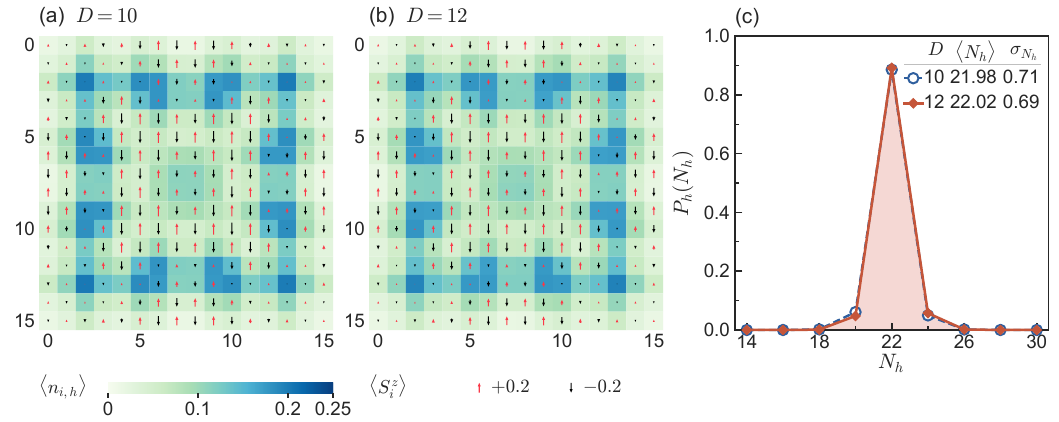}
    \caption{Hole-density and spin patterns (a,b) and hole-number distribution (c) of the PEPS states optimized from the plaquette-PDW initial state on the $16\times16$ lattice at $t'=-0.2$, $J=0.4$, and $\mu=1.55$, for $D=10$ (a) and $D=12$ (b). In (c), the legend gives the mean hole number $\langle N_h\rangle$ and its standard deviation $\sigma_{N_h}$; Monte Carlo standard errors are smaller than the markers.}
    \label{fig:16x16_density_spin_sc}
    \label{fig:16x16_particle_number_distribution}
\end{figure*}

The charge and spin structure of the $12\times12$ state reappears on the larger lattice [Fig.~\ref{fig:16x16_density_spin_sc}(a,b)]. The holes again form a square frame of hole-rich regions, which are now broader and include several islands of roughly $2\times2$ sites. The staggered spin moment is weaker on the frame than in the less-doped interior, some neighboring spins near the hole-rich regions are aligned parallel, and the staggered pattern reverses sign across the frame.

{The hole-number distribution again remains concentrated in a single sector [Fig.~\ref{fig:16x16_density_spin_sc}(c)], with most of its weight at $N_h=22$.}
This result removes one of the explanations left open on the $12\times12$ lattice (Sec.~\ref{subsec:12x12}). There the locked sector, $N_h=12$, corresponds to one hole per row and per column, a filling commensurate with the lattice. On the $16\times16$ lattice the locked sector, $N_h=22$, has no such relation to the lattice. The near locking follows the hole density $\delta \simeq 0.085$, not the commensuration of the cluster. As on the $12\times12$ lattice, the near locking of the particle number is accompanied by an almost vanishing pair field, here with no spatially regular pattern. We analyze the pairing correlations together with the other lattices in Sec.~\ref{sec:fluctuating_pdw}.

\section{\texorpdfstring{Fluctuating plaquette PDW at negative $t'$}{Plaquette pairing without phase coherence at negative t'}}
\label{sec:fluctuating_pdw}

On the hole-like side, uniform $d$-wave pairing is strongly suppressed on every lattice we study, and the PEPS ground states are nearly particle-number eigenstates. In such a state the pair field vanishes whether or not the holes are paired, so the suppression admits two physical pictures: either the holes are unpaired and form a normal metal or a charge-ordered insulator, or they still bind into local pairs that lack a common phase. The pair correlations discriminate between the two. For independent electrons they reduce to products of one-electron propagators, so any weight beyond this product is a two-particle correlation, and its momentum identifies the pairing channel. In Sec.~\ref{subsec:dmrg_strip_pair_fourier} we resolve the pair correlations in momentum space on the three $C_4$-symmetric lattices and separate them from this one-electron background. In Sec.~\ref{subsec:cdw_momentum} we identify the charge modulation that accompanies the pairing and its imprint on the momentum distribution.

\subsection{Momentum-resolved pair correlations}
\label{subsec:dmrg_strip_pair_fourier}

On the hole-like side the pair correlations are weak and irregular in real space on every lattice we study. On the $8\times8$ square at $t'=-0.2$, the DMRG correlations are about 30 times weaker than at $t'\geq0$ and have lost the uniform $d$-wave sign structure [Fig.~\ref{fig:8x8_t2_scan_}(f)]. On the $12\times12$ lattice, the PEPS correlations likewise lose the $d$-wave sign structure and fall strongly over the first few lattice spacings [Fig.~\ref{fig:12x12_sc_pair_corr_grid}(c,g)]; on the $16\times16$ lattice they drop by an order of magnitude within about four lattice spacings and vary nonmonotonically with separation. Two features obscure any pattern: the correlations survive over only a few lattice spacings, and their magnitude follows the local hole density. A Fourier transform overcomes both, because it accumulates the correlations of many pairs of bonds and sorts them by wave vector. We treat the two bond orientations separately, since a modulated pairing pattern can carry different wave vectors on horizontal and vertical bonds.

{From the full bond--bond correlation matrix, for} bonds of orientation $a=x,y$ we sum over all pairs of bonds,
\begin{equation}
    \tilde P_s^{aa}(\mathbf q)
    =\frac{1}{N_a}
      \sum_{i,j}' P_s^{aa}(i,j)
      e^{i\mathbf q\cdot(\mathbf r_i-\mathbf r_j)}.
    \label{eq:dmrg_strip_pair_fourier}
\end{equation}
Here, $P_s^{aa}(i,j)$ is the correlation of Eq.~\eqref{eq: pairing correlation} between bonds of orientation $a$ starting at sites $i$ and $j$, $\mathbf r_i$ is the position of site $i$, $N_a$ is the number of such bonds, and the prime restricts the sum to pairs of bonds that share no site. This is the pair analog of a structure factor: a maximum at $\mathbf q$ means that the pair correlations, accumulated over the lattice, oscillate with wave vector $\mathbf q$. We transform the signed correlations and display the modulus.

\begin{figure*}[t]
    \centering
    \includegraphics[width=\textwidth]{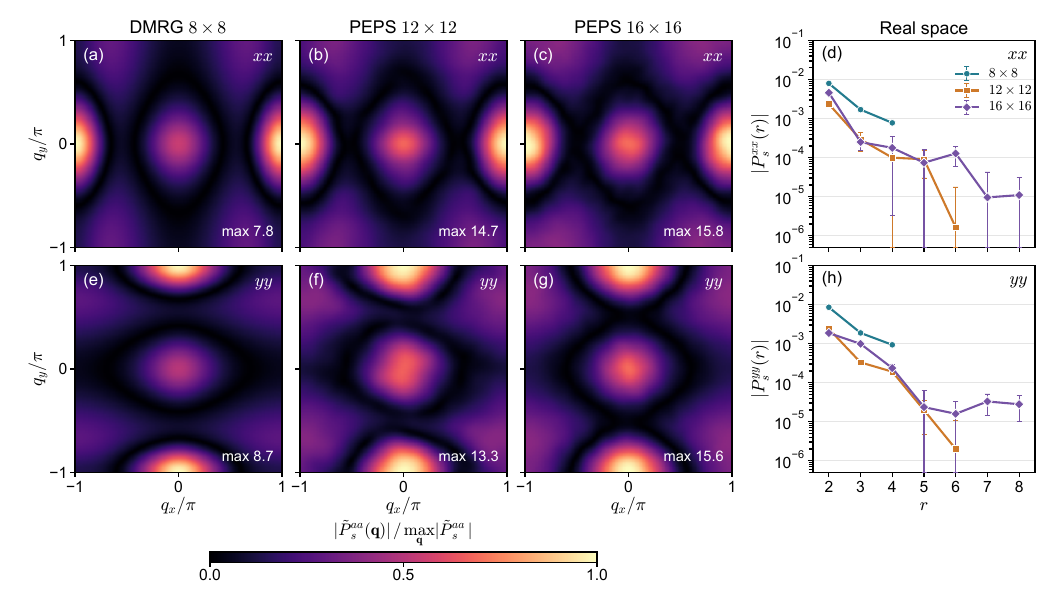}
    \caption{Singlet-pair correlations on $L\times L$ lattices at $t'=-0.2$ and $J=0.4$ in momentum space $|\tilde P_s^{aa}(\mathbf q)|$ of Eq.~\eqref{eq:dmrg_strip_pair_fourier} (a--c) and (e--g) and in real space (d,h). Upper row (a--d): correlation between horizontal bonds ($xx$); lower row (e--h): correlation between vertical bonds ($yy$). (a--c, e--g) Momentum space, in order of increasing size. (a,e) DMRG on the $8\times8$ square at $\delta=1/16$ and $m=30000$. (b,f) and (c,g) PEPS on the $12\times12$ ($D=12$) and $16\times16$ ($D=10$) lattices at $\mu=1.55$. (d,h) Real space $|P_s^{aa}(i_0,i_0+r\hat{\mathbf e}_a)|$, along the bond direction ($+x$ for $xx$, $+y$ for $yy$) for $2\le r\le L/2$, for the PEPS reference bonds above and for the DMRG references $(2,2)$--$(3,2)$ and $(2,2)$--$(2,3)$. PEPS error bars are one Monte Carlo standard error. }
    \label{fig:pair_momentum_squares}
\end{figure*}

The momentum maps in Fig.~\ref{fig:pair_momentum_squares}(a--c,e--g) show the result. On all three lattices and with both methods, $|\tilde P_s^{xx}|$ is largest at $\mathbf Q_x=(\pi,0)$ and $|\tilde P_s^{yy}|$ at $\mathbf Q_y=(0,\pi)$. The maxima are broad, as expected for correlations that extend over only a few lattice spacings. {The momentum-space pair correlations respect $C_4$ symmetry. On the $12\times12$ lattice we have also checked that the $D=10$ results agree with those at $D=12$.}

{The peak heights grow only slowly from the $8\times8$ to the $16\times16$ lattice, which signals short-range order. The real-space correlations from fixed reference bonds confirm this: along the bond direction} [Fig.~\ref{fig:pair_momentum_squares}(d,h)] they fall by one to two orders of magnitude within three or four lattice spacings on every lattice.

In real space, long-range order at these momenta, $\mathbf Q_x$ on horizontal and $\mathbf Q_y$ on vertical bonds, is the plaquette $d$-wave pattern: on each bond orientation the pair amplitude alternates in sign with period two along the bond direction. This pattern already appeared at $t'=0$ as the modulated part of the $d$-wave state [the $\Phi_P$ term of Eq.~\eqref{eq:plaquette_modulation}; Fig.~\ref{fig:12x12_sc_decomposition}]. Taken on its own, it reads
\begin{equation}
\Phi_b^{\rm P2}=\eta_b\,\Phi_P\,e^{i\mathbf Q_{\hat{\mathbf e}_b}\cdot\mathbf r_b},
\qquad \mathbf Q_{\hat{\mathbf x}}=\mathbf Q_x,\quad \mathbf Q_{\hat{\mathbf y}}=\mathbf Q_y,
\label{eq:plaquette_pdw}
\end{equation}
with $\eta_b$, $\mathbf r_b$, and $\hat{\mathbf e}_b$ as in Eq.~\eqref{eq:plaquette_modulation}. The X-point maxima identify a short-range version of this order. The leading real-space pairing modes of the DMRG correlation matrix show it directly: on both the $8\times8$ square and the $8\times16$ strip they are strongly modulated, and several are recognizably plaquette patterns (Appendix~\ref{app:pair_eigenmodes}).

The maps also carry a weaker, broad weight near $\mathbf q=0$. The X-point maximum exceeds it by a factor of about $1.5$--$2$ {on all three lattices}. The mixed-orientation correlations $P_s^{xy}$ and $P_s^{yx}$ (not shown) identify this weight as residual uniform $d$-wave pairing. On all three lattices $P_s^{xy}$ and $P_s^{yx}$ are {largest at} $\mathbf q=0$, where they are negative, the relative sign of $d$-wave pairing between horizontal and vertical bonds; {however,} their modulus there is about $30\%$ of the parallel-orientation maxima. At $\mathbf Q_x$ and $\mathbf Q_y$ they {almost vanish}. This is expected for an inversion-symmetric state: singlet bonds are even under exchange of their endpoints, so inversion-related pairs of perpendicular bonds contribute a factor $\cos(\mathbf q\cdot\mathbf d)$, and both components of the separation $\mathbf d$ of their centers are half-integers.

These maxima could in principle be a memory of the PEPS optimization, because the lower-energy PEPS states at $t'=-0.2$ are optimized from a plaquette-PDW initial state (Sec.~\ref{subsec:12x12}) and could retain its pattern in a local minimum of the variational energy. DMRG provides an independent test: it is initialized randomly, yet it gives the same momentum structure [Fig.~\ref{fig:pair_momentum_squares}(a,e)]. The agreement extends to the weaker features: besides the maxima at the X point and at $\Gamma$, both methods give weaker secondary maxima near the zone boundary. We conclude that the plaquette pairing is a property of the hole-like ground state, which the PEPS ground state captures. The eight-leg strip, whose two directions are inequivalent, selects the same momenta at $t'=-0.4$(Appendix~\ref{app:strip_pair_fourier}).

Short-range pair correlations do not by themselves signal pairing: even free electrons have nonzero pair correlations, given entirely by products of one-electron propagators and carrying the momentum structure of one-electron motion. To isolate the pairing, we split the DMRG pair correlation exactly into its normal Wick part, built from the one-electron density matrix of the same state, and a residual (Appendix~\ref{app:pair_wick_decomposition}). The Wick part contains everything the one-electron density matrix encodes, including any charge or spin modulation such as the checkerboard charge order of Sec.~\ref{subsec:cdw_momentum}. The residual is the connected two-particle part, which vanishes identically in any normal independent-electron state, ordered or not. On the $8\times8$ square the Wick part is nearly flat in momentum, whereas the residual is positive at the X points and largest there; on the eight-leg strip it is likewise positive at the X points. The plaquette momentum selection is therefore a {genuine }two-particle correlation, inherited neither from one-electron motion nor from the charge modulation.

Two methods and three $C_4$-symmetric lattices of increasing size thus select the same plaquette momenta in the pair correlations.  Beneath the irregular real-space signal lies organized plaquette pairing without long-range phase coherence. To our knowledge, this is the first identification of finite-momentum pairing in a lightly doped Mott model that resolves both plaquette momenta on $C_4$-symmetric lattices {and is reproduced by two independent, controlled methods}.

\subsection{\texorpdfstring{Checkerboard charge modulation and the $4a_0\times 4a_0$ pattern}{Checkerboard charge modulation and the 4a0 x 4a0 pattern}}
\label{subsec:cdw_momentum}

{The real-space charge-density profiles on the $8\times 8$, $12\times 12$, and $16\times 16$ lattices differ in detail. To identify the features they share, we analyze them in momentum space.}  We Fourier transform the hole-density profile $\langle n_{i,h}\rangle$ after subtracting the average hole density $\delta$:

\begin{equation}
 I_h(\mathbf q)=\left|\frac{1}{N}\sum_i\bigl(\langle n_{i,h}\rangle-\delta\bigr)
                   e^{-i\mathbf q\cdot\mathbf r_i}\right|^2.
 \label{eq:charge_profile_fourier}
\end{equation}

\begin{figure*}[t]
    \centering
    \includegraphics[width=\textwidth]{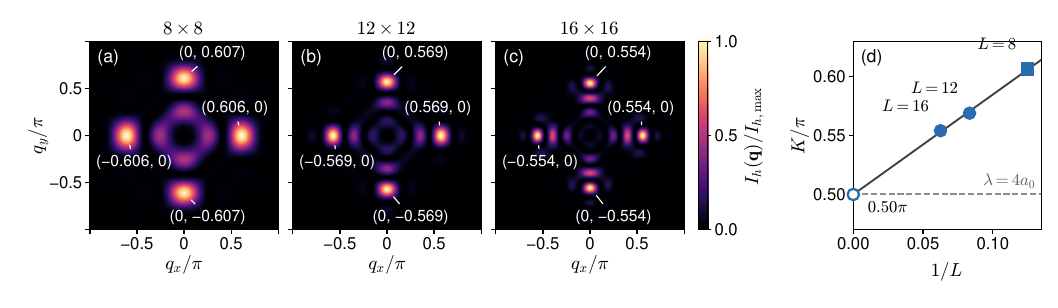}
    \caption{Elastic charge structure factor at $t'=-0.2$ and $J=0.4$: (a) $8\times8$ DMRG with $m=30000$ and four holes, (b) $12\times12$ PEPS with $D=12$, and (c) $16\times16$ PEPS with $D=10$; (b) and (c) are at $\mu=1.55$. In (a--c), colors show $I_h(\mathbf q)/I_{h,\max}$, the Fourier power of the mean-subtracted hole-density profile, Eq.~\eqref{eq:charge_profile_fourier}, normalized to its maximum over momentum separately in each panel. Labels give the axial peak coordinates, and the mean hole densities are $\delta=0.0625$, $0.0834$, and $0.0858$ in (a), (b), and (c), respectively. (d) Axis-averaged peak position $K=(K_x+K_y)/2$ versus $1/L$ for the three lattices in (a--c) (square, DMRG; circles, PEPS). The solid line is a linear fit in $1/L$, the open circle marks its extrapolation, and the dashed line marks $K=\pi/2$, i.e., $\lambda=4a_0$.}
    \label{fig:cdw_momentum}
\end{figure*}

Figure~\ref{fig:cdw_momentum} shows $I_h(\mathbf q)$ for $L\times L$ lattices with $L=8$, 12, and 16 at $t'=-0.2$. In all three lattices, four dominant peaks appear at $\pm\mathbf K_x$ and $\pm\mathbf K_y$, where $\mathbf K_x=(K_x,0)$, $\mathbf K_y=(0,K_y)$, and $K_x\simeq K_y\equiv K$. Their coordinates are given in the figure. These peaks correspond to the bidirectional modulation along the lattice axes seen in real space. As $L$ increases, the peaks move monotonically toward smaller $K$, and the corresponding modulation period $\lambda=2\pi/K$ grows from $3.30a_0$ to $3.51a_0$ and $3.61a_0$ for $L=8$, 12, and 16, respectively, where $a_0$ is the lattice spacing. {Because the $12\times12$ and $16\times16$ PEPS states are obtained at the same chemical potential, the drift of the CDW momenta between them, together with the small change of the mean hole density, is a finite-size effect. The doping change itself cannot account for the drift: more holes would shorten the modulation period and increase $K$, whereas the data show the opposite trend.}

This bidirectional modulation, whose period grows with system size, is reminiscent of the checkerboard charge order in lightly hole-doped cuprates. Scanning tunneling microscopy resolved an approximately $4a_0\times4a_0$ electronic checkerboard in lightly hole-doped Ca$_{2-x}$Na$_x$CuO$_2$Cl$_2$~\cite{Hanaguri2004Checkerboard}. In Bi-2201, a short-range checkerboard with period $\simeq4a_0$ appears already in the insulating regime at $p\simeq0.07$~\cite{Cai2016LightlyDopedCuprates}; subsequent measurements found periods of $3.9a_0$ and $4.0a_0$ at $p=0.08$ and $0.11$, respectively~\cite{Ye2023LocalPairing}.
The monotonic drift of the peaks and the experimental period lead us to conjecture that, as the lattice grows, the peaks move toward $(\pm\pi/2,0)$ and $(0,\pm\pi/2)$, corresponding to the $4a_0$ period observed experimentally. {Figure~}\ref{fig:cdw_momentum}(d) {presents a heuristic finite-size extrapolation of the modulation wave vector in $1/L$.}
A linear extrapolation of the peak position in $1/L$ gives $K\simeq0.50\pi$, i.e., $\lambda\simeq4.0a_0$ [Fig.~\ref{fig:cdw_momentum}(d)].

The checkerboard modulation provides a mechanism for the reconstruction of the electronic structure seen in $n(\mathbf k)$  [Fig.~\ref{fig:square_momentum_tprime}(f), and Fig.~\ref{fig:strip_momentum_tprime}(g,h)]. The charge modulation scatters electrons between $\mathbf k$ and $\mathbf k\pm\mathbf K_\alpha$, with $\alpha=x,y$. Hybridization with these translated states can shift dispersions and redistribute momentum-resolved spectral weight, thereby moving the steepest variation of $n(\mathbf k)$ away from the unreconstructed band contour.  Density-wave reconstruction has likewise been invoked to explain the displaced antinodal back-bending and particle--hole asymmetry observed by photoemission in Bi-2201~\cite{Hashimoto2010ParticleHole}. {Note, however, that the Fermi-surface reconstruction already appears at $t'=0$, where the hole density is nearly uniform; CDW scattering therefore cannot be its only origin.}

\section{Summary and Discussion}

\subsection{Evolution of the ground state with \texorpdfstring{$t'$}{t'}}
\label{subsec:disc_evolution}
Combining state-of-the-art finite-size DMRG and finite PEPS calculations on open strips and $C_4$-symmetric squares of up to $16\times16$ sites, we have tracked how the ground state of the lightly doped ($\delta \leq 0.1$) square-lattice $t$-$J$ model evolves as the next-nearest-neighbor hopping $t'$ is tuned from the electron-like ($t'/t>0$) through $t'/t=0$ to the hole-like ($t'/t<0$) side at fixed $J=0.4t$. Finite PEPS reproduces DMRG on the small squares and extends the calculation to lattice sizes that DMRG cannot reach. We identify three regimes, distinguished by the structure of the pairing and density modulation. All geometries pass through the same three regimes, but the values of $t'$ at which they enter the second and third regimes depend on the geometry.

On the electron-like side ($t'=0.2$), all lattices {show uniform $d$-wave pairing, a smooth hole density, and N\'eel antiferromagnetic spin correlations; the momentum-distribution ridge} follows the free-band Fermi surface and encloses the Luttinger area.

In the second regime, the $d$-wave sign structure persists, the pairing correlations weaken moderately, and the pairing amplitude acquires a spatial modulation on top of the uniform $d$-wave background. The squares  are in this regime near $t'=0$; the strips are in it at both $t'=0$ and $t'=-0.2$. On the six- and eight-leg strips at $t'=-0.2$, the DMRG pairing correlations still decay with $K_{\rm sc}<2$, and the amplitude modulation becomes plaquette-like while the $d$-wave sign structure is largely preserved [Figs.~\ref{fig:6leg_sc_2d} and~\ref{fig:8leg_sc_2d}]. The electronic structure is also reconstructed in this regime: on the $8\times8$ square and the eight-leg strip at $t'=0$, the momentum-distribution ridge closes around $(\pi,\pi)$, i.e., it bounds a hole-like Fermi surface; it encloses about $16\%$ more than the Luttinger area, and corresponds to an effective diagonal hopping $t'_{\rm eff}/t_{\rm eff}\simeq-0.33$. On the $C_4$-symmetric $12\times12$ lattice at $t'=0$, the PEPS optimization spontaneously develops a bulk, bidirectional period-2 plaquette modulation, superposed on a larger uniform $d$-wave background [Fig.~\ref{fig:12x12_sc_decomposition}].

In the third regime, the pair correlations become short-ranged and are dominated by the plaquette momenta $(\pi,0)$ and $(0,\pi)$. The squares enter it already at $t'=-0.2$; the strips enter it at $t'\le -0.4$. On the strips at these values, the pair correlations decay faster than $r^{-2}$ and change sign without the $d$-wave pattern [Figs.~\ref{fig:6leg_sc_2d} and~\ref{fig:8leg_sc_2d}]. On the $8\times8$ square at $t'=-0.2$, the largest pairing correlation falls to about $1/30$ of its $t'\ge0$ value, and the signs form plaquette-like groups of same-sign bonds instead of the uniform $d$-wave pattern. On the $12\times12$ and $16\times16$ squares, which only PEPS reaches, the state optimized from the plaquette-PDW initialization becomes the ground state at $t'=-0.2$. On the eight-leg strip and on all three squares, the equal-time pair structure factor peaks at the same plaquette momenta, $(\pi,0)$ for horizontal and $(0,\pi)$ for vertical bonds. The pair correlations, however, decay within a few lattice spacings, so the plaquette pairs lack superconducting phase coherence. The aspect ratio sets how the hole-rich regions and the antiferromagnetic domain walls fit within the open boundaries, and can thereby shift the relative energies of the competing states and the value of $t'$ at which the third regime sets in.

The states of the third regime also carry a bidirectional checkerboard charge modulation. On the squares its wave vectors are $(\pm K,0)$ and $(0,\pm K)$, with $K$ slightly larger than $\pi/2$; the eight-leg strip at $t'=-0.4$ shows a similar $4\times4$-like charge motif [Fig.~\ref{fig:8leg_density}]. As $L$ grows from 8 to 16, $K$ decreases and the period grows from $3.30a_0$ to $3.61a_0$; a heuristic linear extrapolation in $1/L$ gives a period close to $4a_0$, the period of the checkerboard observed by STM in lightly doped Bi-2201 (Sec.~\ref{subsec:cdw_momentum}). We also find that the checkerboard peak of $I_h(\mathbf q)$ decreases from $L=12$ to $L=16$, consistent with the short-range checkerboard seen by STM in lightly doped Bi-2201~\cite{Cai2016LightlyDopedCuprates}.

The electron momentum distribution at $t'\le 0$ shows a nodal--antinodal contrast. Near the nodal directions the ridge stays sharp, whereas near the antinodes it weakens and flattens. This is reminiscent of the cuprate pseudogap, in which low-energy spectral weight is suppressed predominantly near the antinodes~\cite{Hashimoto2014Pseudogap}.

The qualitative sequence is common to all systems and both methods: uniform $d$-wave $\to$ $d$-wave with plaquette modulation $\to$ fluctuating plaquette PDW intertwined with checkerboard charge order. The geometry shifts only the boundaries between these regimes. Because the uniform $d$-wave and fluctuating-PDW states break different symmetries, if both survive in the thermodynamic limit, they must be separated by a phase transition.

\subsection{Comparison with previous numerical studies}
\label{subsec:disc_comparison}

Recent ground-state studies near $t'=0$ agree that finite-momentum pairing competes closely with uniform $d$-wave pairing, but differ in the form it takes. The iPEPS simple-update study~\cite{Zheng2025CompetingPDW} finds the uniform $d$-wave and plaquette-PDW states too close in energy to rank within the thermodynamic-limit ansatz. On eight-leg cylinders at $\delta=0.1$--$0.2$, Chen, Haldane, and Sheng~\cite{Chen2025PNAS} find robust $d$-wave superconductivity at $t'=0$, possibly coexisting with a weak \emph{unidirectional} period-2 PDW whose momentum $(0,\pi)$ lies along the periodic direction. For $|t'|\gtrsim0.15t$ of either sign, the PDW and the charge modulation are suppressed and a uniform $d$-wave superconductor remains.

On the hole-like side, two recent Hubbard-model studies reach a different conclusion. Using DMRG on a $45^\circ$-rotated lattice at $U=12t$ and $t'=-0.3t$, Xu, Liu, and Jiang~\cite{xu2025stripespairdensitywave} find, near $\delta\simeq0.1$, a holon Wigner crystal in which bidirectional charge order coexists with short-range, spatially oscillating pair correlations, as on our open squares at $t'=-0.2$. With further doping these correlations develop into a PDW with a divergent susceptibility above $\delta\simeq0.12$, in the range of doping and $t'$ where Ref.~\cite{Chen2025PNAS} finds a uniform $d$-wave superconductor. Using MPO-based thermal tensor networks on width-6 cylinders at $U=8t$ and $t'=-0.2t$, Li, Qi, and Li~\cite{Li2026FluctuatingPDW} find $d$-wave superconductivity on the electron-doped side, whereas on the hole-doped side the zero-momentum pairing response is suppressed and the dominant pair fluctuation, formed by near-nodal electrons, carries a momentum near $(0,\pi)$ along the periodic direction.

These studies leave three questions open. (i) Over what range of $t'$ does the PDW compete with uniform $d$-wave pairing: does it include $t'=0$, and does it extend to the hole-like side? (ii) Is the PDW the leading pairing channel, as in Refs.~\cite{xu2025stripespairdensitywave,Li2026FluctuatingPDW}, or only a weak component on top of uniform $d$-wave superconductivity, as in Ref.~\cite{Chen2025PNAS}? (iii) Is the modulation unidirectional or bidirectional, and does it run along the lattice axes or along the diagonals? Cylinder calculations cannot settle (iii) on their own, because the periodic direction and the lattice orientation already break the $C_4$ symmetry; indeed, the unidirectional pair momenta of Refs.~\cite{Chen2025PNAS,Li2026FluctuatingPDW} both lie along the periodic direction.

Our calculations address these questions at light doping, $\delta\le0.1$. Combining DMRG and finite PEPS on open strips and $C_4$-symmetric squares of up to $16\times16$ sites, we find the following. (i) A period-2 plaquette modulation is already present at $t'=0$, superposed on a larger uniform $d$-wave background, consistent with the weak PDW of Ref.~\cite{Chen2025PNAS}. (ii) On the hole-like side the uniform component disappears, at $t'=-0.2$ on the squares and at $t'\le-0.4$ on the strips. The remaining pair correlations are short-ranged and dominated by the plaquette momenta. The PDW becomes the leading pairing channel but lacks phase coherence. (iii) Because the open square favors neither lattice direction, the modulation is free to break $C_4$ down to $C_2$ or to preserve it. {We find that it is} bidirectional, along the lattice axes, with period two in both directions: $(\pi,0)$ on horizontal and $(0,\pi)$ on vertical bonds. The eight-leg strip, whose two directions are inequivalent, selects the same momenta. This fluctuating PDW is accompanied by a checkerboard charge modulation whose period extrapolates to about $4a_0$.

\subsection{Outlook}
\label{subsec:disc_outlook}

The most direct extension concerns doping. Xu, Liu, and Jiang~\cite{xu2025stripespairdensitywave} find that the short-range, spatially oscillating pair correlations at $\delta\simeq0.1$ develop into a PDW with a divergent susceptibility above $\delta\simeq0.12$. The short-range plaquette PDW found here may likewise be the lightly doped precursor of {a long-range PDW}. Near $\delta=1/8$, where the dynamical decoupling of superconducting layers in stripe-ordered La$_{2-x}$Ba$_x$CuO$_4$ has been attributed to PDW order~\cite{Li2007LayerDecoupling,Berg2007LayerDecoupling,Berg2009Striped}, the question becomes whether a long-range PDW is a competing ground state of the $t$-$t'$-$J$ model.

A second direction concerns the electronic structure. On the lattices studied here, the momentum distribution already shows two features reported by photoemission on lightly doped cuprates: a large-Fermi-surface count that would imply a negative hole doping, as in lightly doped Na-CCOC, and a negative diagonal hopping generated by correlations, $t'_{\rm eff}/t_{\rm eff}\simeq-0.33$ at bare $t'=0$, in line with the growth of $|t'_{\rm eff}/t_{\rm eff}|$ toward low doping in LSCO~\cite{Yoshida2006UnderlyingFS} (Sec.~\ref{subsec:8x8_dmrg_scan}). Equal-time data cannot separate small nodal pockets from Fermi arcs; the frequency-resolved spectral function on larger lattices would. The correlation-generated $t'_{\rm eff}$ also matters for how the model is parameterized. A $t'$ fitted to the measured Fermi surface already contains this {correlation }contribution and therefore overstates the bare $|t'|$ that should enter the $t$-$J$ Hamiltonian. The hole-like regime of the cuprates may thus correspond to a smaller bare $|t'|$ than such fits suggest.

Third, the hole-like state bears on the anomalous metal of monolayer Bi-2201, which intervenes between the insulating and superconducting states of a single CuO$_2$ plane~\cite{Luo2026MonolayerCuprate}. The ingredients identified here {(local pairs that fail to establish a common phase, checkerboard charge order, and possibly a small fraction of gapless electrons near the nodes)} suggest that the anomalous metal arises from local pairing that survives the loss of superconducting coherence and remains coupled to mobile near-nodal carriers. Whether such a state conducts or freezes into a paired insulator {calls for further studies of compressibility} and transport. The scenario also motivates a search for finite-momentum pair correlations together with gapless near-nodal spectral weight in the anomalous metallic regime.

Finally, at the light dopings studied here, uniform $d$-wave pairing in the $t$-$J$ model is robust on the electron-like side and gives way on the hole-like side to phase-incoherent plaquette pairing, opposite to the empirical trend in the cuprates. If this asymmetry persists at the higher dopings where cuprate superconductivity is strongest, the minimal model lacks ingredients that give pairs mobility and global phase coherence on the hole-doped side, such as Cu--O molecular-orbital dynamics beyond a strict Zhang--Rice singlet~\cite{Emery1987ThreeBand,Ebrahimnejad2014HoleDynamics,Mai2021ThreeBandPairing}, correlated hopping~\cite{Hirsch1989BondCharge,Belinicher1996GeneralizedtJ,Batista1995CorrelatedHopping}, or coupling to the lattice; bond-modulating Su--Schrieffer--Heeger phonons, for example, produce robust $d$-wave superconductivity in DMRG studies of the Hubbard model~\cite{Wang2025SSHHubbard}. The hole-like state found here sets a specific target for such ingredients: the pairs already form, and what is missing is the coherence of their phases.

\section{Acknowledgment}
We thank Zheng-Yu Weng, Fuchun Zhang, Qianghua Wang, Yayu Wang, Yuanbo Zhang, Xingjiang Zhou, and Hai-Hu Wen for helpful discussions. This work is supported by funding from Hong Kong’s Research Grants Council (AoE/P-604/25-R, CRF C7015-24G, CRS HKU701/24).  W.-Y. Liu is supported by National Natural Science Foundation of China (Grant No.12534009).

\appendix

\section{\texorpdfstring{Additional DMRG results on six-leg strips}{Additional DMRG results on six-leg strips}}
\label{app:dmrg_obc_continuations}

\begin{figure}[htbp]
    \centering
    \includegraphics[width=\linewidth]{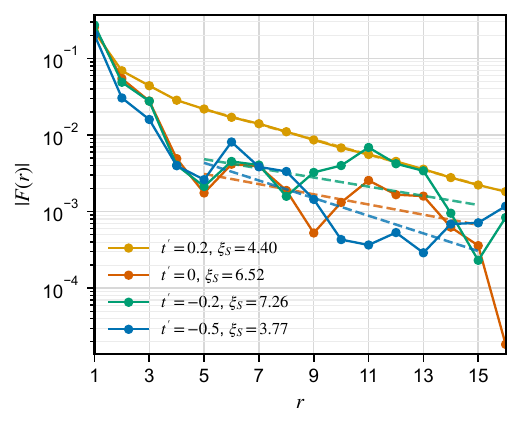}
    \caption{Equal-time spin correlations $|F(r)|$ on the open $6\times32$ strips of Fig.~\ref{fig:6leg_density} at $\delta=1/16$ and $J=0.4$, for target sites on the leg of the reference site; $r$ is the longitudinal separation. The four curves correspond to $t'=0.2$, $0$, $-0.2$, and $-0.5$. Dashed lines show exponential fits $|F(r)|=A e^{-r/\xi_S}$ over $5\le r\le15$, with the fitted $\xi_S$ values given in the legend.}
    \label{fig:app_6leg_spin_decay}
\end{figure}

Along the $6\times32$ strips of Fig.~\ref{fig:6leg_density}, the equal-time spin correlations decay exponentially for all four values of $t'$ (Fig.~\ref{fig:app_6leg_spin_decay}). Such short-ranged spin correlations are consistent with a spin gap, as expected for the Luther--Emery liquid that underlies the Luttinger-liquid interpretation of $K_{\rm sc}$ in Sec.~\ref{sec:dmrg_t2_evolution}. The decay separates $t'=0.2$ from $t'\le0$. At $t'=0.2$, $|F(r)|$ decreases smoothly with $\xi_S\simeq4.4$, consistent with the N\'eel sign pattern that persists across the strip in Fig.~\ref{fig:6leg_density}. For $t'\le0$, it drops about an order of magnitude below the $t'=0.2$ values by $r\simeq5$ and fluctuates non-monotonically beyond, accompanying the stronger longitudinal charge modulation at these values of $t'$. The fitted values $\xi_S\simeq3.8$--$7.3$ for $t'\le0$ describe only this weak, irregular tail and do not indicate longer-ranged magnetic correlations. Like uniform $d$-wave pairing, antiferromagnetic correlations are most robust on the electron-like side.

\begin{figure}[htbp]
    \centering
    \includegraphics[width=0.98\linewidth]{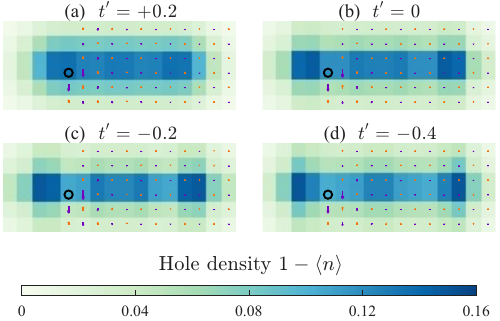}
    \caption{Hole density (color) and equal-time spin correlations $F_i$ (arrows) on open $6\times16$ strips at $\delta=1/16$ and $J=0.4$; the reference site is marked by a black circle. Panels are ordered left to right and top to bottom as $t'=0.2$, $0$, $-0.2$, and $-0.4$. All arrows use a common scale. For calibration, the $F_i$ value on the site immediately to the right of the reference site is $-0.215$, $-0.266$, $-0.286$, and $-0.238$ in panels (a)--(d), respectively.}
    \label{fig:app_dmrg_6x16}
\end{figure}

\begin{figure}[htbp]
    \centering
    \includegraphics[width=0.98\linewidth]{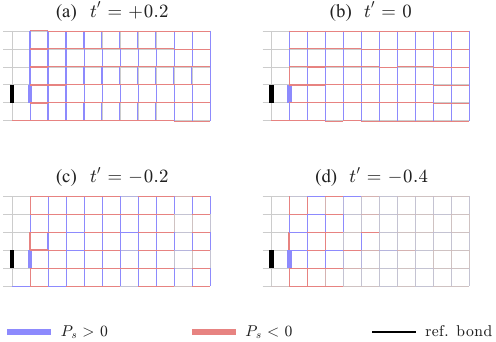}
    \caption{Bond-resolved singlet-pairing correlations $P_s$ for the $6\times16$ states of Fig.~\ref{fig:app_dmrg_6x16}. The black bond marks the reference vertical bond, and target bonds to its right are shown. Blue and red bonds denote positive and negative $P_s$, respectively. Each panel uses its own bond-width scale, so only the sign structure and relative modulation within a panel should be compared; the unscaled maximum $|P_s|$ values are $1.06\times10^{-2}$, $2.44\times10^{-3}$, $6.33\times10^{-4}$, and $5.31\times10^{-4}$ for panels (a)--(d), respectively.}
    \label{fig:app_dmrg_6x16_pairing}
\end{figure}

The six-leg trends of Sec.~\ref{sec:dmrg_t2_evolution} do not depend on the strip length: DMRG on shorter $6\times16$ strips at the same $J$ and $\delta$, for $t'=0.2$, $0$, $-0.2$, and $-0.4$, reproduces them (Figs.~\ref{fig:app_dmrg_6x16} and~\ref{fig:app_dmrg_6x16_pairing}). As $t'$ is reduced, the longitudinal density modulation strengthens while the spin correlations $F_i$ retain an antiferromagnetic background (Fig.~\ref{fig:app_dmrg_6x16}). The spin correlations develop local phase slips once $t'\le0$. The pairing correlations in Fig.~\ref{fig:app_dmrg_6x16_pairing} evolve as on the $6\times32$ strips (Fig.~\ref{fig:6leg_sc_2d}). At $t'=0.2$ they carry the uniform $d$-wave sign structure across the displayed strip, and at $t'=0$ this sign structure persists with a visible amplitude modulation. For $t'<0$ they fall off within a few lattice spacings of the reference bond, and bonds violating the $d$-wave sign appear, first locally next to the reference bond at $t'=-0.2$ and then throughout the weak correlations at $t'=-0.4$. Over the same range, the largest $|P_s|$, found next to the reference bond, decreases by a factor of about 20.

\section{\texorpdfstring{Momentum-distribution ridge: extraction and evolution with $t'$}{Momentum-distribution ridge: extraction and evolution with t'}}
\label{app:momentum_ridge}

\subsection{Ridge extraction and controls}
\label{app:ridge_extraction}

On an open $L_x\times L_y$ lattice, Eq.~\eqref{eq:dmrg_single_particle_momentum} is a finite Fourier polynomial in $\mathbf k$, with displacements bounded by the lattice size. Its gradient $\nabla_{\mathbf k}n$ and Hessian $H_n$ are therefore evaluated analytically from the measured $\langle c^\dagger_{i\sigma}c_{j\sigma}\rangle$, without interpolation. With $\hat{\mathbf g}=\nabla_{\mathbf k}n/|\nabla_{\mathbf k}n|$, the ridge is the set of points where $|\nabla_{\mathbf k}n|$ is a local maximum along $\hat{\mathbf g}$. Stationarity requires $\hat{\mathbf g}^{T}H_n\hat{\mathbf g}=0$: the second derivative of $n$ along its own gradient vanishes, so the ridge is the inflection line of $n(\mathbf k)$. We check that the second derivative of $|\nabla_{\mathbf k}n|$ along $\hat{\mathbf g}$ is negative at every extracted point. For an occupation step broadened symmetrically about $k_F$, whether by temperature or by a gap as in $n_{\mathbf k}=1-\xi_{\mathbf k}/E_{\mathbf k}$, the inflection lies at $k_F$.

The ridge closes around $M=(\pi,\pi)$ on the momentum torus. We parameterize it by 2048 rays from $M$ and, on each ray, take the root of the stationarity condition nearest the dominant gradient maximum. The hole-side area is $\frac12\oint r(\theta)^2\,d\theta$, and $a_{\rm ridge}$ is its complement in the Brillouin zone. No smoothing, symmetry averaging, density threshold, or adjustment toward a target area enters. Applied to the exact bulk band contour, the same integration recovers $15/32$ to within $10^{-9}$.

Two alternative conventions test the sensitivity to the definition of the ridge: the maximum of $|\nabla_{\mathbf k}n|$ along each ray from $M$, and the maximum of the radial derivative $\partial_r n$ along each ray. They differ from the primary definition only where the ridge is not normal to the ray. For the DMRG states they give $a_{\rm ridge}=0.6132$ and $0.6062$ on the $8\times16$ strip and $0.5994$ and $0.5912$ on the $8\times8$ square, compared with $0.6172$ and $0.6055$ from the primary definition. The excess over $a_{\rm FS}$ thus lies between $26\%$ and $32\%$ for every convention. For unprojected free electrons on the same open lattices, all three conventions give $0.4769$--$0.4773$ and $0.4740$--$0.4745$, within $2\%$ of $a_{\rm FS}$. The spread among conventions, at most $0.015$ of the zone, is an order of magnitude smaller than the difference of $0.13$--$0.14$ between the interacting and free ridges.

\subsection{\texorpdfstring{Ridge tracing across the $t'$ scans}{Ridge tracing across the t' scans}}
\label{app:ridge_tprime}

\begin{figure*}[t]
    \centering
    \includegraphics[width=0.92\textwidth]{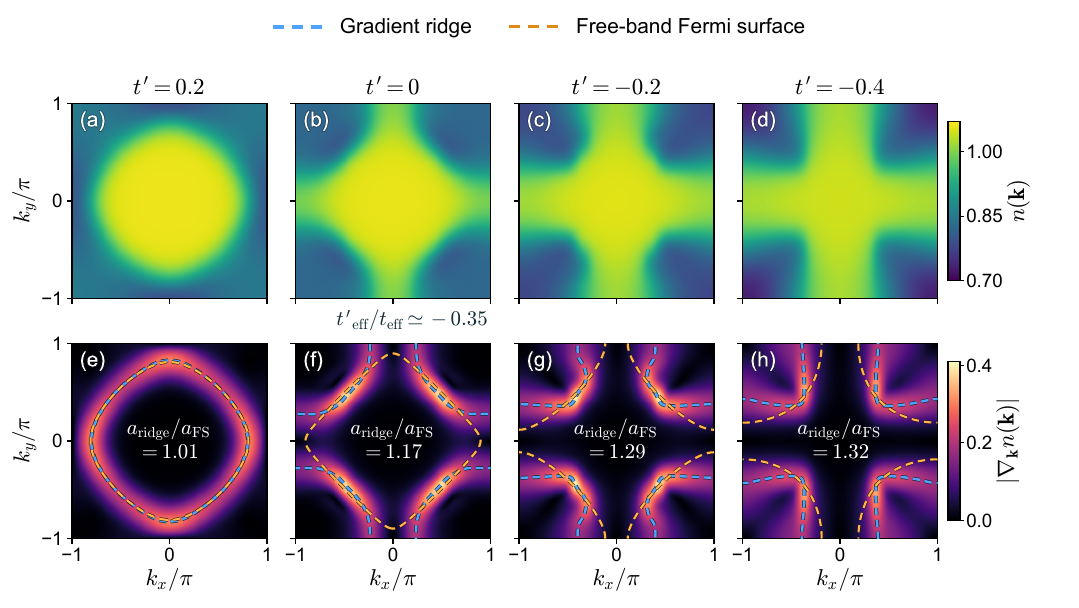}
    \caption{Electron momentum distribution on the open $8\times16$ strip at $J=0.4$, $\delta=1/16$, and $m=30000$, for $t'=0.2$, $0$, $-0.2$, and $-0.4$ (columns). Upper row (a--d): $n(\mathbf k)$ of Eq.~\eqref{eq:dmrg_single_particle_momentum}; lower row (e--h): $|\nabla_{\mathbf k}n(\mathbf k)|$. Curves, color scales, and labels follow Fig.~\ref{fig:square_momentum_tprime}. For $t'=0$, a band with $t'_{\rm eff}/t_{\rm eff}\simeq-0.35$ fits the ridge, with the effective chemical potential also adjusted in the fit.}
    \label{fig:strip_momentum_tprime}
\end{figure*}

Figures~\ref{fig:square_momentum_tprime} and~\ref{fig:strip_momentum_tprime} follow $n(\mathbf k)$ and its ridge through the $t'$ sequences of the $8\times8$ square and the $8\times16$ strip at $J=0.4$, $\delta=1/16$. Along this sequence the contour changes from a loop around $\Gamma$ to a loop around $M$, so we trace the zero line of $\hat{\mathbf g}^{T}H_n\hat{\mathbf g}$ in both the $\Gamma$-centered and the $M$-centered zone, select the closed loop with the largest mean $|\nabla_{\mathbf k}n|$, and verify that $|\nabla_{\mathbf k}n|$ is a local maximum along $\hat{\mathbf g}$ at every point. A loop around $\Gamma$ is electron-like and bounds the occupied side; a loop around $M$ is hole-like and bounds the low-occupation side. For the strip at $t'=-0.4$ and the square at $t'=-0.2$, this procedure reproduces the ray-based areas to four digits. Over all bond dimensions available for each case, between $m=20000$ and $40000$, the ridge areas change by at most $0.0012$ of the zone. For unprojected free electrons on the same open lattices, the same extraction gives $a_{\rm free,OBC}$ between $0.453$ and $0.477$ for every $t'$, within $3.5\%$ of $a_{\rm FS}$, with degenerate Fermi-level shells filled fractionally. This finite-size and boundary offset is about five times smaller than the smallest excess of the interacting states at $t'\le0$, $16\%$.

On the eight-leg strip (Fig.~\ref{fig:strip_momentum_tprime}), the ridge evolves with $t'$ as on the $8\times8$ square of Sec.~\ref{subsec:8x8_dmrg_scan}{, where the scan extends} to $t'=-0.4$. For $t'=0.2$, the ridge-inferred Fermi surface not only satisfies Luttinger's theorem but also follows the free-fermion Fermi-surface geometry [Fig.~\ref{fig:strip_momentum_tprime}(e)]. From $t'=0$ on, however, the Fermi-surface topology becomes hole-like, and the area $a_{\rm ridge}$ enclosed by the ridge is significantly larger than $a_{\rm FS}=15/32$; the ratio $a_{\rm ridge}/a_{\rm FS}$ is $1.01$, $1.17$, $1.29$, and $1.32$ at $t'=0.2$, $0$, $-0.2$, and $-0.4$, respectively. Allowing both the chemical potential and $t'$ to be adjusted, we obtain $t'_{\rm eff}/t_{\rm eff}\simeq -0.35$ for the bare $t'=0$ (Appendix~\ref{app:ridge_effective_fit}), close to the value $-0.33$ on the square. For $t'\leq 0$, the near-nodal gradient remains sharp, while the antinodal ridge broadens [Fig.~\ref{fig:strip_momentum_tprime}(f--h)]. Because the strip geometry already distinguishes the two lattice directions, the agreement with the square shows that this reorganization does not depend on the $C_4$ symmetry of the lattice.

\subsection{\texorpdfstring{Effective curvature of the zero-$t'$ ridge}{Effective curvature of the zero-t-prime ridge}}
\label{app:ridge_effective_fit}

{For input $t'=0$, we parameterize the entire occupation-gradient ridge by the contour}
\begin{equation}
{-2(\cos k_x+\cos k_y)-4r_{\rm eff}\cos k_x\cos k_y
       =\bar\mu_{\rm eff},}
\end{equation}
where $r_{\rm eff}=t'_{\rm eff}/t_{\rm eff}$ and $\bar\mu_{\rm eff}=\mu_{\rm eff}/t_{\rm eff}$ are both adjustable. The contour determines these ratios. We sample the original ridge uniformly in arc length and minimize its mean squared shortest distance to the continuous band contour. The fit uses the full loop without symmetry averaging and takes $t''_{\rm eff}=0$. It gives $r_{\rm eff}=-0.349$ and $-0.328$ for the $8\times16$ and $8\times8$ lattices, with root-mean-square momentum distances $0.0144\pi$ and $0.00862\pi$, respectively. These are the values summarized in Figs.~\ref{fig:strip_momentum_tprime} and~\ref{fig:square_momentum_tprime}.

The corresponding band fillings are $n_{\rm fit}=1.1000$ and $1.0915$. If instead the band is constrained to the actual density $15/16$, the optimal ratios are $-0.369$ and $-0.353$, but the root-mean-square distances increase to $0.0671\pi$ and $0.0626\pi$. The negative fitted ratio is stable under this constraint; the failure to reproduce the contour position at the same density is the independent content of the area mismatch. The quoted ratios are geometric fit parameters. Their comparison with the ARPES fits of Ref.~\cite{Yoshida2006UnderlyingFS} concerns the direction of the deformation; that study uses $t''=-t'/2$.

\section{\texorpdfstring{The $8\times8$ square at $t'=0$: doping dependence and the $\delta=1/8$ benchmark}{The 8x8 square at t'=0: doping dependence and the delta=1/8 benchmark}}

None of the $C_4$-symmetric lattices in the main text shows a unidirectional stripe at the dopings studied there. To connect with the stripe order reported near $\delta=1/8$, we study the $8\times8$ lattice at $t'=0$ and $J=0.4$ at larger doping. DMRG finds that stripes form as $\delta$ approaches $1/8$ (Appendix~\ref{app:dmrg_8x8_doping_appendix}), and at $\delta=1/8$ a stripe-initialized finite PEPS reproduces the DMRG stripe at a lower variational energy (Appendix~\ref{app:hole8_benchmark}).

\subsection{\texorpdfstring{Doping dependence from DMRG}{Doping dependence from DMRG}}
\label{app:dmrg_8x8_doping_appendix}

\begin{figure*}[htbp]
    \centering
    \includegraphics[width=0.95\textwidth]{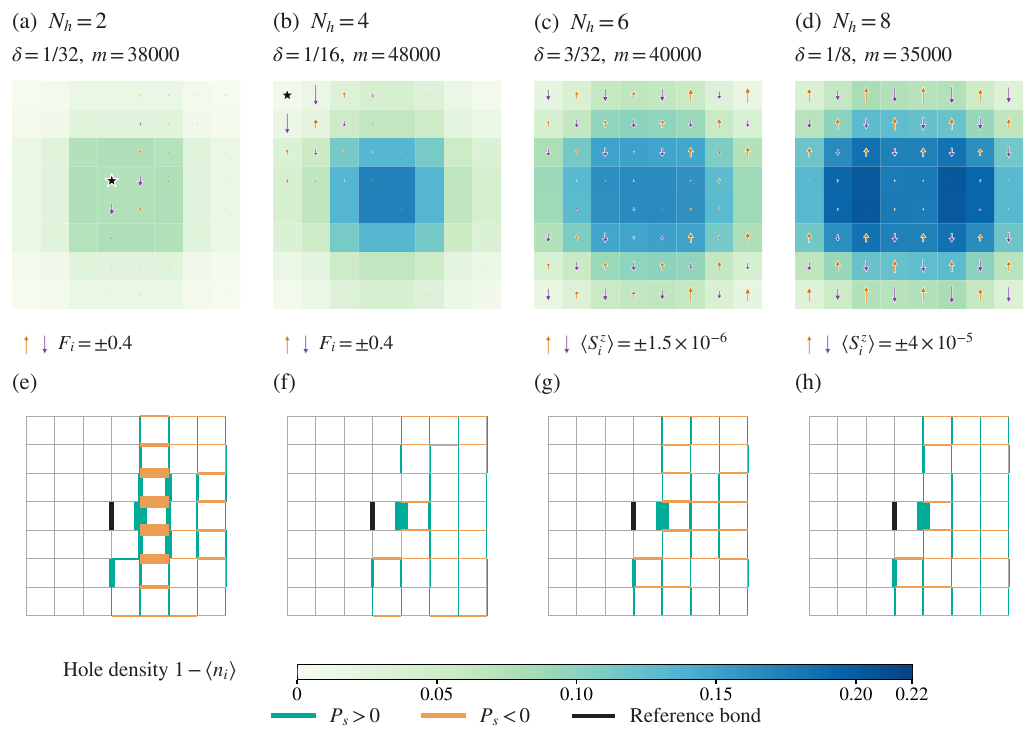}
    \caption{DMRG results on the open $8\times8$ lattice at $t'=0$ and $J=0.4$ for $N_h=2,4,6$, and $8$. Panels (a)--(d) show the hole density on a common color scale. Arrows in (a,b) show the spin correlation $F_i=\langle\mathbf S_{r_0}\cdot\mathbf S_i\rangle$ with the black-star reference sites. Arrows in (c,d) show the residual local polarization $\langle S_i^z\rangle$. Orange upward and purple downward arrows denote positive and negative values; lengths are linear in magnitude, with a common scale for (a,b) and separate scales for (c,d). Panels (e)--(h) show the singlet-pairing correlation $P_s$ relative to the black vertical bond. Teal and orange bonds denote positive and negative values, respectively. Bond widths are linear in $|P_s|$ and scaled independently in each panel.}
    \label{fig:app_dmrg_8x8_doping}
\end{figure*}

Figure~\ref{fig:app_dmrg_8x8_doping} follows the DMRG ground state from $N_h=2$ to $8$ ($\delta=1/32$ to $1/8$). For $N_h=2$ and $4$, the holes are concentrated in the center of the lattice and the density profile is nearly $C_4$ symmetric. For $N_h=6$, the hole-rich region elongates along the horizontal direction. For $N_h=8$, it extends across the full width of the lattice along the horizontal middle line, with the holes concentrated in two islands to the left and right of the center. The spin structure changes at the same doping. The local moments are tiny, $\max_i|\langle S_i^z\rangle|\simeq1.2\times10^{-8}$--$3.8\times10^{-5}$, as expected for DMRG states at fixed $S^z_{\rm tot}=0$ with nearly restored spin-rotation symmetry, so their magnitudes are not meaningful, but their sign pattern changes with doping. At $N_h=2$ and $4$ the residual moments are smaller than \(1.1\times10^{-7}\) and have irregular signs, so panels (a,b) show the spin correlations $F_i$ instead. At $N_h=6$ and $8$ the residual moments form a staggered pattern whose phase shifts by $\pi$ across the horizontal middle line, where the moments nearly vanish; for $N_h=8$ the spin correlations $F_i$ confirm this $\pi$ shift (Appendix~\ref{app:hole8_benchmark}). The hole-rich region is an antiphase domain wall of the antiferromagnetic background. At $N_h=8$ the domain wall runs across the full width, including the less doped region between the two islands, so the state is a stripe with a charge modulation along its length, breaking the $C_4$ symmetry of the Hamiltonian. This is the tendency toward stripe formation near $\delta=1/8$ reported in earlier DMRG studies~\cite{Shoushu2021Robust,YiFan2024Ground,Chen2025PNAS}. The $C_4$-symmetric patterns found at $\delta=1/16$ in the main text (Figs.~\ref{fig:8x8_t2_scan_} and~\ref{fig:benchmark}) do not conflict with the stripe order reported near $\delta=1/8$: on the same lattice, DMRG finds a stripe only as the doping approaches $1/8$.

The pairing correlations carry the $d$-wave sign structure at all four dopings, with opposite signs on horizontal and vertical bonds. At $N_h=2$, where the lattice holds a single hole pair, they are confined to a narrow column near the reference bond; for $N_h\ge4$ they spread over the displayed half of the lattice. Stripe formation at $N_h=6$ and $8$ leaves the $d$-wave sign structure intact.

\subsection{\texorpdfstring{PEPS--DMRG benchmark at $\delta=1/8$}{PEPS-DMRG benchmark at delta=1/8}}
\label{app:hole8_benchmark}

\begin{figure}[htbp]
    \centering
    \includegraphics[width=\linewidth]{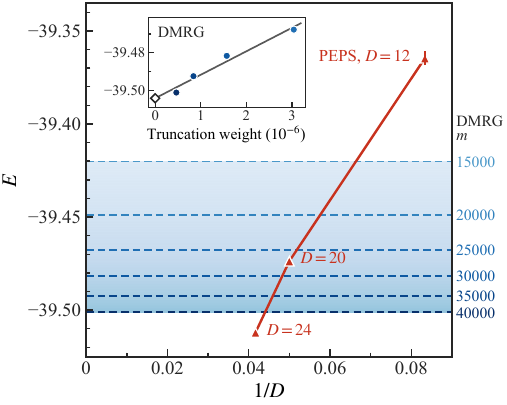}
    \caption{Energy benchmark on the open $8\times8$ lattice at $t'=0$, $J=0.4$, and $N_h=8$ ($\delta=1/8$). Red triangles: PEPS energies at $D=12$, $20$, and $24$ versus $1/D$; the $D=24$ point is the optimization-tail estimate $E=-39.512(1)$. Dashed lines: DMRG energies with $m=15000$, $20000$, $25000$, $30000$, $35000$, and $40000$ kept states. Inset: DMRG energies for $m=25000$--$40000$ versus the local truncation weight at the final recorded bond update of the last sweep, with a linear fit giving $E_{\rm ext}\simeq-39.5042$ at zero truncation weight.}
    \label{fig:benchmark_hole8_energy}
\end{figure}

At $\delta=1/8$, DMRG on the $8\times8$ lattice is not converged with the accessible numbers of kept states: its energy still decreases visibly from $m=15000$ to $40000$ (Fig.~\ref{fig:benchmark_hole8_energy}). Instead of benchmarking against a single DMRG reference as in Table~\ref{tab:benchmark_energy}, we compare the two methods as variational sequences. The PEPS tensors carry $U(1)$ particle-number symmetry and are initialized from a simple-update state prepared with a temporary stripe-pinning field (Sec.~\ref{subsec:Initialization}); the Monte Carlo sampling is restricted to the same $N_h=8$, $S^z_{\rm tot}=0$ sector as DMRG. The PEPS energy decreases steadily with $D$ and reaches $E=-39.512(1)$ at $D=24$, below the $m=40000$ DMRG energy: at this doping, finite PEPS reaches lower variational energies than the largest DMRG calculation we can perform on this lattice. The PEPS energy is even below the linear zero-truncation extrapolation of the DMRG energies, $E_{\rm ext}\simeq-39.5042$ (inset). Since the PEPS energy is an upper bound on the ground-state energy $E_0$, this implies $E_0<E_{\rm ext}$: the linear extrapolation is not reliable here, presumably because the accessible values of $m$ have not reached the regime in which the DMRG energy is linear in the truncation weight. The finite-PEPS study of Dong \emph{et al.} found a diagonal stripe on the $8\times8$ lattice at the same $J=0.4$ and $\delta=1/8$, with $E/N=-0.61212(2)$~\cite{Dong2020StableDiagonalStripes}. Our axial stripe, with $E/N=-0.61738(2)$ at $D=24$, is lower by $5.3\times10^{-3}$ per site (0.9\%), so on this lattice the stripe along a lattice axis is variationally favored over the diagonal one.

\begin{figure}[t]
    \centering
    \includegraphics[width=\linewidth]{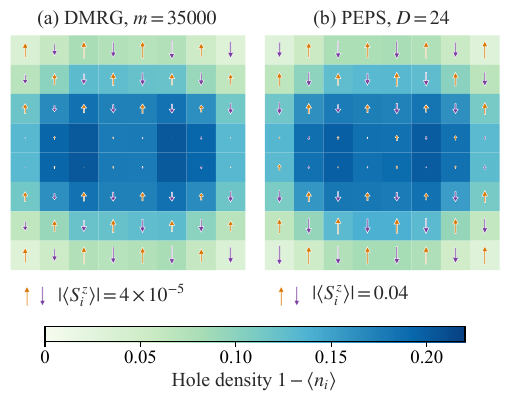}
    \caption{Hole density (color) and local spin polarization $\langle S_i^z\rangle$ (arrows) on the open $8\times8$ lattice at $t'=0$, $J=0.4$, and $N_h=8$ ($\delta=1/8$), from DMRG with $m=35000$ (left) and PEPS with $D=24$ (right). The density color scale is shared. Orange upward and purple downward arrows denote positive and negative polarization, respectively. Arrow lengths are linear in $|\langle S_i^z\rangle|$; their scales differ by a factor of $10^3$ between the panels, as indicated by the keys.}
    \label{fig:benchmark_hole8_pattern}
\end{figure}

The two states also agree in real space (Fig.~\ref{fig:benchmark_hole8_pattern}). In both, the holes concentrate in two islands, to the left and right of the center, on a band along the horizontal middle line; the density maxima of about $0.2$ sit at nearly the same positions, and the density is suppressed near the upper and lower edges. The band is an antiphase domain wall of the antiferromagnetic background. In PEPS, whose local moments reach magnitudes of order $10^{-2}$, the staggered polarization reverses sign across the band. In DMRG, the local moments are only of order $10^{-5}$ because spin-rotation symmetry is nearly restored, but the spin correlations $F_i=\langle\mathbf S_0\cdot\mathbf S_i\rangle$ measured from a corner site have opposite staggered signs on the two sides of the band, and even the residual moments outside the two central rows follow the PEPS sign pattern up to a global spin flip. Both methods find a stripe of the type reported near $\delta=1/8$ in DMRG, quantum Monte Carlo, and tensor-network studies of the $t$-$J$ and Hubbard models~\cite{SimonsCollaboration2017,Corboz2011Stripes,Corboz2014Competing,Shoushu2021Robust,YiFan2024Ground,Chen2025PNAS}; its orientation is arbitrary on the $C_4$-symmetric lattice, and a single $8\times8$ lattice cannot fix its period or filling. Because the PEPS optimization starts from a stripe-pinned state, this benchmark shows that the ansatz at $D=24$ represents a stripe accurately, not that the optimization finds one spontaneously.

\section{\texorpdfstring{DMRG on the $12\times12$ lattice at $t'=-0.2$}{DMRG on the 12x12 lattice at t'=-0.2}}
\label{app:dmrg_12x12_t2m02}

As a check on the PEPS results at $t'=-0.2$ that does not depend on the PEPS initial states, we performed a DMRG calculation on the same open $12\times12$ lattice at $J=0.4$. Because $|\psi_{\rm pPDW}\rangle$ carries more than 90\% of its weight in the $N_h=12$ sector (Fig.~\ref{fig:12x12_particle_number_distribution}), we fix $N_h=12$ ($\delta=1/12$). With $m=20000$ kept states, DMRG on a width-12 lattice is far from converged, and the grand-canonical energy of the resulting state lies above the PEPS energies [dashed line in Fig.~\ref{fig:12x12_energy_competition}(c)]. We use this calculation only to identify which real-space features already appear in an unconverged DMRG state, not as an energy benchmark.

\begin{figure}[htbp]
    \centering
    \includegraphics[width=\linewidth]{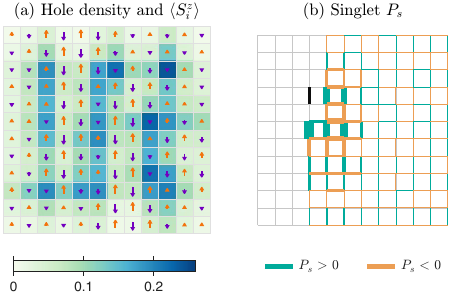}
    \caption{DMRG results on the open $12\times12$ lattice at $t'=-0.2$, $J=0.4$, and $N_h=12$ ($\delta=1/12$), with $m=20000$ kept states. (a) Hole density $\langle n_{i,h}\rangle$ with the local spin polarization $\langle S_i^z\rangle$ shown as arrows. Orange upward and purple downward arrows denote positive and negative $\langle S_i^z\rangle$, respectively; the largest displayed magnitude is $\max_i|\langle S_i^z\rangle|=0.302$. (b) Bond-resolved singlet-pairing correlation $P_s$ relative to the black reference bond. Teal and orange denote positive and negative correlations, respectively, and the bond width scales as $\sqrt{|P_s|}$; the largest displayed magnitude is $\max_b|P_s|=2.83\times10^{-4}$.}
    \label{fig:app_dmrg_12x12_t2m02}
\end{figure}

The hole density in Fig.~\ref{fig:app_dmrg_12x12_t2m02}(a) is strongly inhomogeneous but shows no regular pattern, which we attribute to the incomplete convergence. The local moments are large, $\max_i|\langle S_i^z\rangle|=0.302$, on a predominantly antiferromagnetic background: at this $m$ the DMRG state, like a finite-$D$ PEPS, has limited entanglement and breaks spin-rotation symmetry. The pair--pair correlations [Fig.~\ref{fig:app_dmrg_12x12_t2m02}(b)] do not follow the uniform $d$-wave sign structure. Instead, plaquettes whose four bonds carry the same sign appear both near the reference bond and at larger distances, as on the $8\times8$ lattice at the same $t'$ [Fig.~\ref{fig:8x8_t2_scan_}(f)].

At this unconverged $m$, the magnitude and spatial decay of $P_s$ do not establish the range of the converged pairing correlations at $t'=-0.2$. What the calculation does show, independently of the PEPS initial states, is that a variational state at $t'=-0.2$ with $N_h=12$ develops a strongly inhomogeneous charge distribution and pair correlations with a plaquette-like rather than uniform $d$-wave sign structure, in line with the PEPS ground state.

\section{\texorpdfstring{Metastable PEPS states on the $12\times12$ lattice}{Metastable PEPS states on the 12x12 lattice}}
\label{app:metastable_branches}

At each $t'$, the two initial states of Sec.~\ref{subsec:Initialization} converge to distinct local minima of the variational energy. The main text presents the lower one; the higher one is metastable within the variational manifold at the accessible $D$. {Following each minimum across $t'$ infers the features that $t'$ drives.} Below we present the two metastable states, at $t'=0$ and $t'=-0.2$.

\subsection{\texorpdfstring{Plaquette-PDW minimum: metastable at $t'=0$, ground state at $t'=-0.2$}{Plaquette-PDW minimum: metastable at t'=0, ground state at t'=-0.2}}
\label{app:metastable_t20}

\begin{figure}[htbp]
    \centering
    \includegraphics[width=\columnwidth]{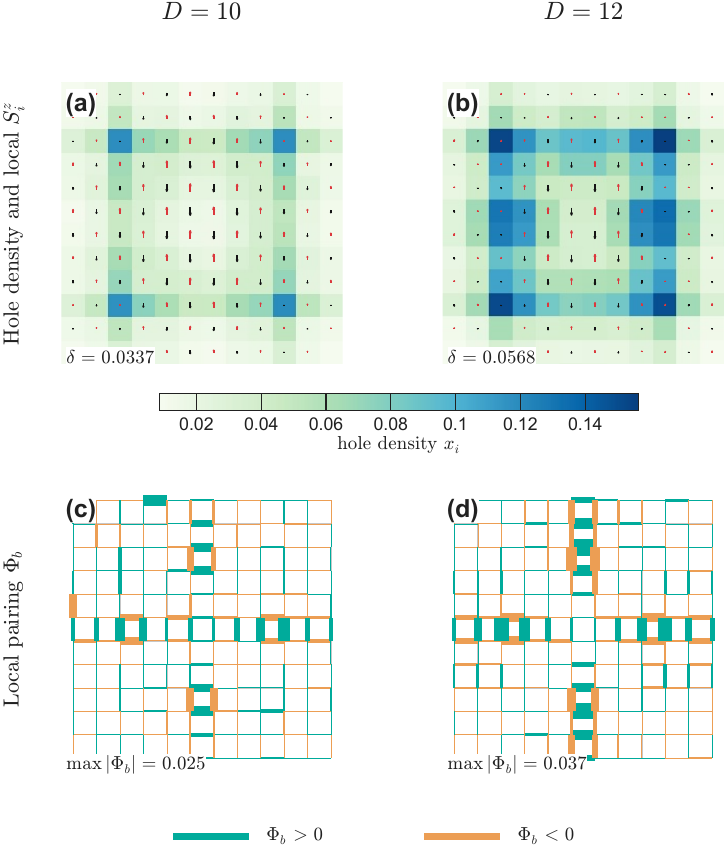}
    \caption{Hole density, local spin polarization, and singlet-pairing amplitude of the metastable state $|\psi_{\rm pPDW}\rangle$ at $t'=0$, $\mu=1.55$, and $J=0.4$ on the $12\times12$ lattice. The left and right columns show $D=10$ and $D=12$, respectively. Panels (a,b) show the hole density $\langle n_{i,h}\rangle$ together with the local spin polarization $\langle S_i^z\rangle$; the density color scale and spin-arrow scale are shared between the two panels. Panels (c,d) show the bond-resolved singlet-pairing amplitude $\Phi_b$. Teal and orange bonds denote positive and negative amplitudes, respectively, and bond width represents $|\Phi_b|$ using a common scale for both bond dimensions. The in-panel labels report the mean hole density $\delta$ and $\max_b|\Phi_b|$.}
    \label{fig:app_t20_pdw_density_sc}
\end{figure}

The plaquette-PDW local minimum at $t'=0$ already has the charge structure of the $t'=-0.2$ ground state [Fig.~\ref{fig:app_t20_pdw_density_sc}(a,b); compare Fig.~\ref{fig:12x12_density_spin_triptych}(c,f)]. The holes gather on a square frame on the third ring from the edge, with the highest density at its four corners, around a weakly doped interior with a clear antiferromagnetic pattern.
From a comparison with the $t'=-0.2$ ground state [Fig.~\ref{fig:12x12_density_spin_triptych}(c,f)], we find
that the frame belongs to this minimum and is not created by negative $t'$. At  {$\mu=1.55$ and $t'=0$}, its hole density is only $\delta=0.034$ ($0.057$) for $D=10$ ($12$), well below the $0.063$ ($0.073$) of the $d$-wave ground state with the same $\mu$ and $t'$.

The plaquette-PDW local minimum at $t'=0$ retains appreciable local pairing but loses the $2\times2$ pairing pattern of its initial state. The strongest amplitudes cluster along the central row and column [Fig.~\ref{fig:app_t20_pdw_density_sc}(c,d)]. Both horizontal and vertical amplitudes change sign across the lattice, so a uniform $d$-wave sign pattern is absent at both bond dimensions.

The charge frame persists from $D=10$ to $12$, while the mean hole density grows by about 68\% and $\max_b|\Phi_b|$ increases from $0.025$ to $0.037$, by about 49\%. By comparison, the hole densities of the $d$-wave SC ground states change by at most 16\% over the same range of $D$ (Table~\ref{tab:peps_runs}). At $D=12$, this minimum thus combines stronger local pairing with a spatially varying pairing phase.

\subsection{\texorpdfstring{$d$-wave minimum: ground state at $t'=0$, metastable at $t'=-0.2$}{d-wave minimum: ground state at t'=0, metastable at t'=-0.2}}
\label{app:t2minus_dwave_branch}

\begin{figure}[htbp]
    \centering
    \includegraphics[width=\columnwidth]{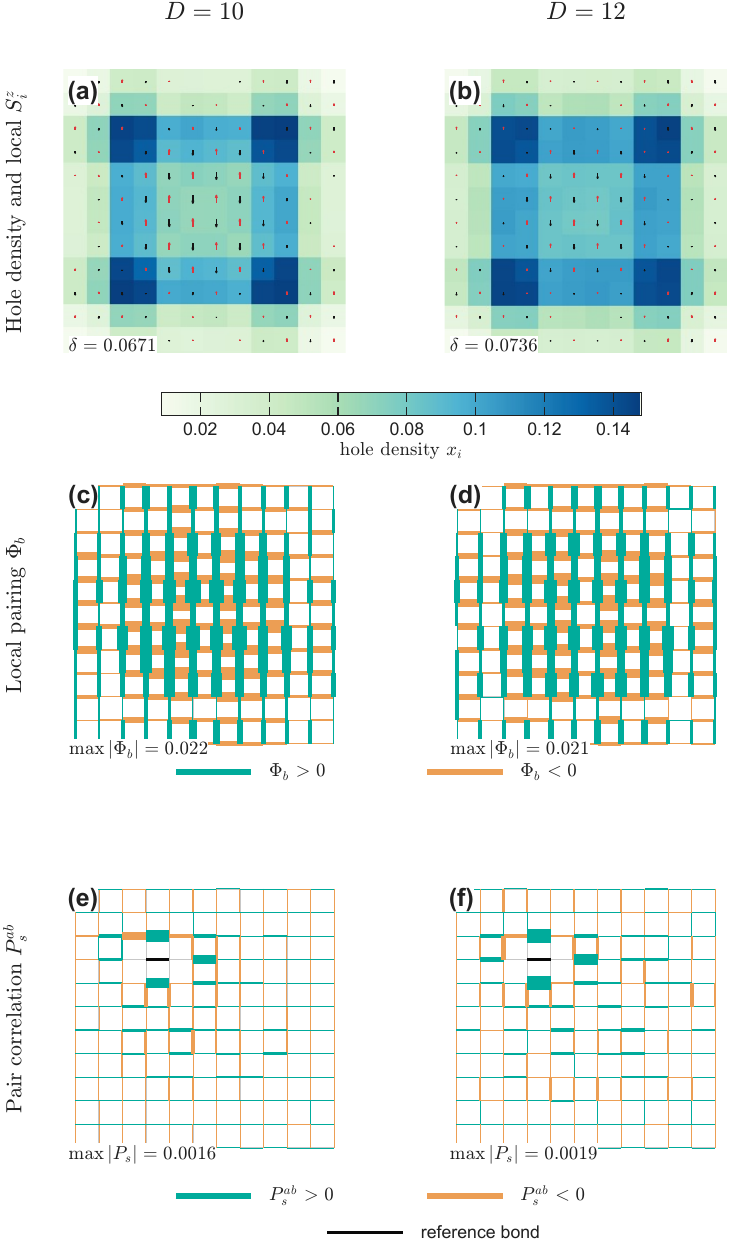}
    \caption{Real-space observables of the metastable state $|\psi_d\rangle$ at $t'=-0.2$, $\mu=1.55$, and $J=0.4$ on the $12\times12$ lattice. The left and right columns show $D=10$ and $D=12$, respectively. Panels (a,b) show the hole density $\langle n_{i,h}\rangle$ and local spin polarization $\langle S_i^z\rangle$, using shared density and spin-arrow scales. Panels (c,d) show the bond-resolved singlet-pairing amplitude $\Phi_b$, while panels (e,f) show the singlet pair-pair correlation $P_s^{ab}$ measured relative to the black reference bond. Teal and orange bonds denote positive and negative values, respectively; within each row, bond widths represent the corresponding absolute magnitudes on a common scale across $D=10$ and $D=12$. The in-panel labels report $\delta$, $\max_b|\Phi_b|$, and $\max_b|P_s^{ab}|$.}
    \label{fig:app_t2minus_dwave_density_sc}
\end{figure}

At $t'=-0.2$, the metastable $d$-wave minimum retains the $d$-wave sign on every bond [Fig.~\ref{fig:app_t2minus_dwave_density_sc}(c,d)]. Its uniform pairing component is less than half that at $t'=0$, while the total residual decreases more slowly [Table~\ref{tab:sc_summary}(b)]. Its pairing amplitude therefore has larger spatial variations relative to the uniform $d$-wave background.

The charge distribution evolves in the same direction [panels (a,b)]. The faint hole-rich spots at the corners of the ring at $t'=0$ [Fig.~\ref{fig:12x12_density_spin_triptych}(b,e)] grow into four $2\times2$ hole-rich islands at $t'=-0.2$, a $C_4$-symmetric pattern distinct from the frame of the plaquette-PDW minimum. At $t'=-0.2$ both minima are strongly charge modulated, so charge order at this $t'$ does not depend on the initial state; the two minima differ in how the holes are arranged and in the fate of the uniform pairing.

\section{\texorpdfstring{DMRG pair correlations in momentum space: eight-leg strip and Wick background}{DMRG pair correlations in momentum space: eight-leg strip and Wick background}}
\label{app:dmrg_pair_momentum}

\subsection{\texorpdfstring{Eight-leg strip at $t'=-0.4$}{Eight-leg strip at t'=-0.4}}
\label{app:strip_pair_fourier}

\begin{figure}[htbp]
    \centering
    \includegraphics[width=\columnwidth]{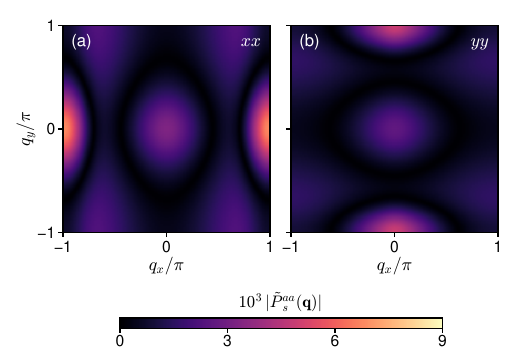}
    \caption{Orientation-resolved singlet-pair correlations $|\tilde P_s^{aa}(\mathbf q)|$ of Eq.~\eqref{eq:dmrg_strip_pair_fourier} on the open $8\times16$ strip at $t'=-0.4$, $J=0.4$, and $\delta=1/16$: (a) horizontal bonds and (b) vertical bonds. The signed correlations at $m=22000$, $24000$, $25000$, $27000$, and $30000$ are extrapolated linearly in the maximum final-sweep discarded weight to zero discarded weight before Fourier transformation. Colors give the absolute value in units of $10^{-3}$, with the normalization of Fig.~\ref{fig:pair_momentum_squares}.}
    \label{fig:app_strip_pair_fourier}
\end{figure}

{On the eight-leg strips the $d$-wave sign structure of the pair correlations survives at $t'=-0.2$ and is lost only at $t'=-0.4$ [Fig.~\ref{fig:8leg_sc_2d}], whereas on the $C_4$-symmetric squares it is lost already at $t'=-0.2$ (Sec.~\ref{sec:c4_square_clusters}): the crossover in $t'$ depends on geometry. The strip counterpart of the square states analyzed in Sec.~\ref{subsec:dmrg_strip_pair_fourier} is therefore the $8\times16$ strip at $t'=-0.4$. Its pair correlations decay faster than $r^{-2}$ ($K_{\rm sc}\simeq3.2$; Fig.~\ref{fig:8leg_sc}) and change sign across the strip on both bond orientations without an evident pattern [Fig.~\ref{fig:8leg_sc_2d}(d)]. In momentum space they show the structure found on the squares [Fig.~\ref{fig:app_strip_pair_fourier}]: $|\tilde P_s^{xx}(\mathbf q)|$ is largest at $\mathbf Q_x$ and $|\tilde P_s^{yy}(\mathbf q)|$ at $\mathbf Q_y$, $2.09$ and $1.90$ times the corresponding zero-momentum values, and excluding separations below two, three, or four lattice spacings in Eq.~\eqref{eq:dmrg_strip_pair_fourier} preserves these maxima. A geometry whose two lattice directions are inequivalent thus selects the same plaquette momenta in both bond orientations.}

\subsection{Normal Wick background}
\label{app:pair_wick_decomposition}

{A finite-momentum feature in a four-electron correlation can partly inherit the momentum structure of its single-electron contractions. To test this possibility, we compare the DMRG pair correlations with their normal Wick contribution, constructed from the same state's one-electron density matrix.}
We compare $P_s^{ab}(i,j)=\langle\Delta_a^\dagger(i)\Delta_b(j)\rangle$ with the normal Wick background built from $G^\sigma_{ij}=\langle c^\dagger_{i\sigma}c_{j\sigma}\rangle$:
\begin{equation}
\begin{aligned}
 P_{s,{\rm Wick}}^{ab}(i,j)
 &=\sum_\sigma\bigl(G^\sigma_{ij}G^{\bar\sigma}_{i'j'}
                  +G^\sigma_{ij'}G^{\bar\sigma}_{i'j}\bigr),\\
 R^{ab}(i,j)&=P_s^{ab}(i,j)-P_{s,{\rm Wick}}^{ab}(i,j).
\end{aligned}
\label{eq:pair_wick_residual}
\end{equation}
Here, $i'=i+a$ and $j'=j+b$ are the other endpoints of the two bonds, and $\bar\sigma$ is the opposite spin. The fixed particle number and $S^z$ make anomalous and spin-flip averages vanish but leave these normal contractions. We apply Eq.~\eqref{eq:dmrg_strip_pair_fourier} identically to $P_s$, $P_{s,{\rm Wick}}$, and $R$, using the same nonoverlapping bonds and $1/N_a$ normalization.

The normal Wick background does not single out the plaquette momenta. On the strip it is largest at $\mathbf q=0$, and on the square it is nearly flat between $\mathbf q=0$ and the plaquette momentum. At $\mathbf Q_x$ in the $xx$ channel and $\mathbf Q_y$ in the $yy$ channel, $|\tilde P^{aa}_{s,{\rm Wick}}|$ is $0.57$--$1.11$ times its value at $\mathbf q=0$, whereas the full correlation $|\tilde P_s^{aa}|$ is $1.8$--$2.0$ times larger there than at $\mathbf q=0$. The plaquette maxima of Figs.~\ref{fig:pair_momentum_squares}(a,e) and~\ref{fig:app_strip_pair_fourier} therefore do not originate from the single-particle contractions.

\begin{figure}[htbp]
    \centering
    \includegraphics[width=\columnwidth]{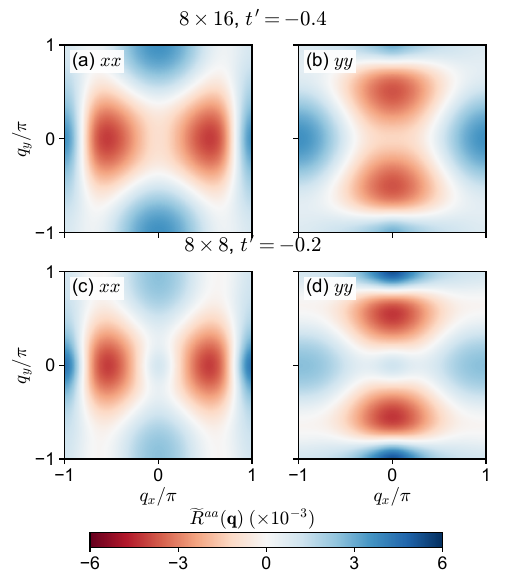}
    \caption{Signed residual of the singlet-pair correlations after subtracting the normal Wick background, Eq.~\eqref{eq:pair_wick_residual}, at $J=0.4$, $\delta=1/16$, and $m=30000$, without extrapolation. Panels (a,b) show the $8\times16$ strip at $t'=-0.4$ and (c,d) the $8\times8$ square at $t'=-0.2$; the left and right columns are the $xx$ and $yy$ channels. Colors show $\tilde R^{aa}(\mathbf q)$ on one common scale, with blue positive and red negative. The Fourier sum follows Eq.~\eqref{eq:dmrg_strip_pair_fourier}, excluding bonds that share a site.}
    \label{fig:dmrg_pair_wick_fourier}
\end{figure}

Figure~\ref{fig:dmrg_pair_wick_fourier} shows the signed Fourier residual $\tilde R^{aa}(\mathbf q)$. At the plaquette momenta, it remains positive and retains about $56$--$59\%$ of the total signed Fourier amplitude on both lattices. On the $8\times8$ square these remain the residual's largest positive and absolute values. On the $8\times16$ strip, the deepest negative extrema occur near $(\pm0.54\pi,0)$ in $xx$ and $(0,\pm0.50\pi)$ in $yy$, where the pair correlations are suppressed relative to the normal Wick background. Regions of both signs are in fact required by this nonlocal measure: excluding bonds that share a site removes every zero-displacement term from Eq.~\eqref{eq:dmrg_strip_pair_fourier}, so the Brillouin-zone average of $\tilde R^{aa}(\mathbf q)$ vanishes. The physical content lies in where the positive weight sits. Its location at the plaquette momenta establishes a substantial finite-momentum pairing component beyond normal single-particle contractions on both geometries.

\section{Real-space pairing modes from the pair-correlation matrix}
\label{app:pair_eigenmodes}

The Fourier analysis of Sec.~\ref{subsec:dmrg_strip_pair_fourier} projects the pair correlations onto one plane wave at a time, with a uniform envelope over the whole lattice. On open lattices, where the correlations decay within a few lattice spacings and the hole density is inhomogeneous, the pairing need not be organized as a single plane wave. We therefore ask which real-space bond patterns carry the weight behind the $(\pi,0)$ and $(0,\pi)$ maxima in Figs.~\ref{fig:pair_momentum_squares}(a,e) and~\ref{fig:app_strip_pair_fourier}: what their internal bond form factor is, how their amplitude is distributed in space, and how the pair phases in different regions are related. To answer this without presupposing any of the three, we diagonalize the full bond--bond pair-correlation matrix.

We regard each nearest-neighbor bond $b$ as the set of its two sites and, following Wietek~\cite{Wietek2022Fragmented}, retain only the correlations between bonds that share no site:
\begin{equation}
    \widetilde C_{bb'}=
    \begin{cases}
        \langle\Delta_b^\dagger\Delta_{b'}\rangle,& b\cap b'=\varnothing,\\
        0,&\text{otherwise}.
    \end{cases}
    \label{eq:nonlocal_pair_matrix}
\end{equation}
This removes the local density and spin contributions that dominate overlapping bond pairs. All horizontal--horizontal, vertical--vertical, and mixed horizontal--vertical elements enter with their signs. The eigenvectors of $\widetilde C$,
\begin{equation}
    \sum_{b'}\widetilde C_{bb'}v_n(b')
      =\widetilde\lambda_n v_n(b),\qquad
      \sum_b|v_n(b)|^2=1,
    \label{eq:nonlocal_pair_modes}
\end{equation}
with $\widetilde\lambda_1\ge\widetilde\lambda_2\ge\cdots$, are pair wave functions defined on the bonds; neither a plane-wave envelope nor a bond form factor is imposed. Each eigenvector defines a pair operator $B_n=\sum_b v_n(b)\Delta_b$, and $\widetilde\lambda_n$ is the part of $\langle B_n^\dagger B_n\rangle$ that comes from bonds sharing no site. It thus measures how strongly pairs in the pattern $v_n$ are correlated across the lattice. In a phase-coherent superconductor, one eigenvalue grows in proportion to the number of bonds and its eigenvector extends over the whole system~\cite{Yang1962ODLRO}; several eigenvalues of comparable size indicate that the pair correlations are shared among distinct spatial patterns~\cite{Wietek2022Fragmented}. Here we use the spectrum on finite lattices to resolve this spatial organization. Because we use the unnormalized operator of Eq.~\eqref{eq: pairing correlation}, our eigenvalues are twice those obtained with the normalized operator of Ref.~\cite{Wietek2022Fragmented}; $\widetilde C$ is not divided by the number of bonds. With its diagonal removed, $\widetilde C$ is traceless and has negative as well as positive eigenvalues; we focus on the largest positive ones.

\begin{figure*}[t]
    \centering
    \includegraphics[width=0.98\textwidth]{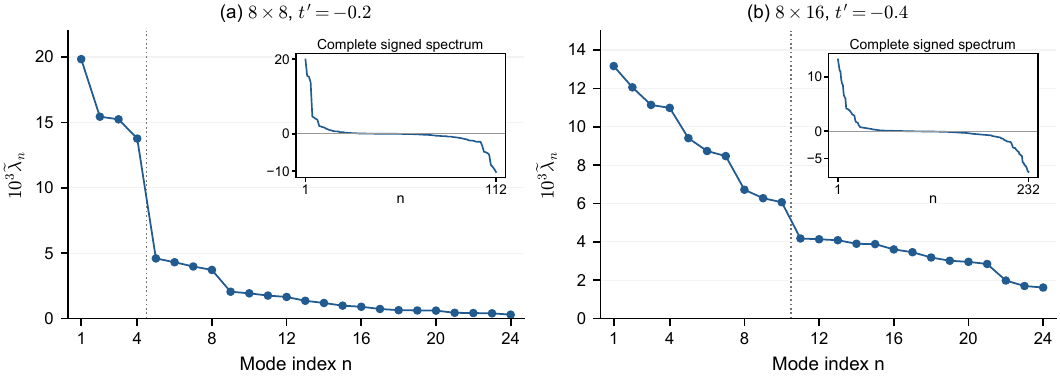}
    \caption{Eigenvalues of the nonlocal pair-correlation matrix $\widetilde C$, Eq.~\eqref{eq:nonlocal_pair_matrix}, for (a) the open $8\times8$ square at $t'=-0.2$ and (b) the open $8\times16$ strip at $t'=-0.4$, both at $J=0.4$ and $\delta=1/16$. Main panels show the 24 largest eigenvalues; insets show the complete spectra of the $112\times112$ and $232\times232$ matrices, including the negative eigenvalues. Dotted lines mark the drop after mode 4 on the square and after mode 10 on the strip. Panel (a) uses $m=30000$. In (b), the matrix elements at $m=22000$, $24000$, $25000$, $27000$, and $30000$ are extrapolated linearly in the maximum final-sweep discarded weight before diagonalization, as in Fig.~\ref{fig:app_strip_pair_fourier}.}
    \label{fig:nonlocal_pair_spectrum}
\end{figure*}

\begin{figure*}[t]
    \centering
    \includegraphics[width=0.98\textwidth]{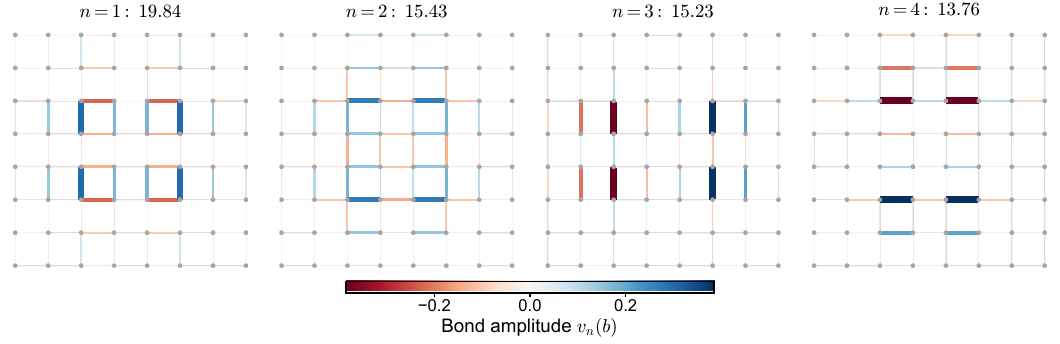}
    \caption{Four leading pairing modes on the $8\times8$ square of Fig.~\ref{fig:nonlocal_pair_spectrum}(a). Each panel gives the mode index $n$ and $10^3\widetilde\lambda_n$. Blue (red) bonds denote positive (negative) components of the normalized eigenvector $v_n(b)$, and the bond width is proportional to $|v_n(b)|$. All panels share one amplitude scale; the overall sign of each mode is arbitrary. Mode 1: $d$-wave sign structure with strong plaquettes on a $2\times2$ superlattice. Mode 2: the P2 plaquette pattern. Modes 3 and 4: vertical- and horizontal-bond patterns with odd envelopes, related by a $90^\circ$ rotation.}
    \label{fig:nonlocal_pair_modes_8x8}
\end{figure*}

On the $8\times8$ square, four eigenvalues stand well above the rest [Fig.~\ref{fig:nonlocal_pair_spectrum}(a)]: $10^3\widetilde\lambda_{1,\ldots,4}=19.84$, $15.43$, $15.23$, and $13.76$, followed by $10^3\widetilde\lambda_5=4.59$. Figure~\ref{fig:nonlocal_pair_modes_8x8} shows the four leading modes. The first two reproduce, without any input from the PEPS calculations, the two ingredients of Eq.~\eqref{eq:plaquette_modulation}. Mode~1 has the $d$-wave sign structure throughout the interior, with opposite signs on horizontal and vertical bonds, but its amplitude is strongly modulated with period two: the weight is concentrated on four strong plaquettes that form a $2\times2$ superlattice, while the bonds connecting them nearly vanish. This is a uniform $d$-wave component with a large plaquette admixture, the same structure as the total pair field of the PEPS state at $t'=0$ (Sec.~\ref{subsubsec:12x12_plaquette}). Mode~2 is the plaquette pattern itself, Eq.~\eqref{eq:plaquette_pdw}. Its sign alternates between neighboring horizontal bonds along $x$ and between neighboring vertical bonds along $y$, so that plaquettes whose four bonds share one sign alternate in sign on a checkerboard, as in the P2 state of Ref.~\cite{Zheng2025CompetingPDW}. Modes~3 and~4 are related by a $90^\circ$ rotation. Mode~3 resides on vertical bonds, and its envelope changes sign between the left and right halves of the square, but it also shows strong modulation inside each {domain}; mode~4 is the corresponding horizontal-bond pattern, with opposite signs in the lower and upper halves. {The small difference between $\widetilde\lambda_3$ and $\widetilde\lambda_4$ presumably reflects the finite number of kept states in DMRG.} Because these envelopes are odd, a uniform-envelope plane wave at $\mathbf Q_y$ or $\mathbf Q_x$ has zero overlap with them, even though their eigenvalues are nearly as large as that of mode~2. The eigenmode analysis thus recovers substantial pairing weight that is invisible at any single wave vector.

\begin{figure*}[t]
    \centering
    \includegraphics[width=0.98\textwidth]{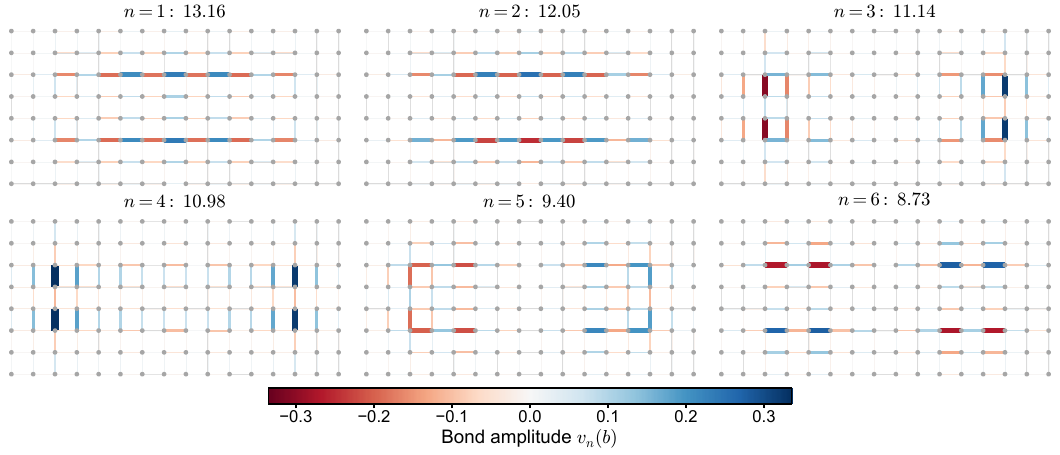}
    \caption{First {six} pairing modes on the $8\times16$ strip of Fig.~\ref{fig:nonlocal_pair_spectrum}(b), displayed as in Fig.~\ref{fig:nonlocal_pair_modes_8x8}. Each panel gives $n$ and $10^3\widetilde\lambda_n$, and all panels share one amplitude scale. Modes 1 and 2: horizontal pairing that alternates in sign along $x$ on two rows, with equal and opposite relative phases. Modes 3 and 4: vertical pairing near the two ends of the strip, with opposite and equal relative phases.}
    \label{fig:nonlocal_pair_modes_8x16}
\end{figure*}

{On the $8\times16$ strip, the pairing weight is spread over more modes [Fig.~\ref{fig:nonlocal_pair_spectrum}(b)]. Figure~\ref{fig:nonlocal_pair_modes_8x16} shows the first six modes. Modes~1 and~2 are almost purely horizontal. These modes carry the $(\pi,0)$ maximum of $|\tilde P_s^{xx}(\mathbf q)|$.  Modes~3 and~4 are predominantly vertical and localized near the two short ends of the strip, with opposite and equal phases, respectively. {Modes~5 and~6 add further nodes to the spatial envelope but still show the $(\pi,0)$ sign structure within each domain. Such envelope-modulated pairing weight is not resolved by the plane-wave analysis.}}

The two lattices therefore show that the plaquette maxima in momentum space are carried by a few spatially localized pairing modes rather than by a single extended plane wave. On the square, the leading modes are the uniform $d$-wave with a strong plaquette modulation and the P2 plaquette pattern. On the strip, period-two pairing is confined to two rows whose relative phase is only weakly fixed. A phase-coherent superconductor would instead concentrate the weight in one extended mode with a single relative phase across the system. The strip thus gives a direct real-space view of the state discussed in Sec.~\ref{sec:fluctuating_pdw}: plaquette pair correlations that are well developed locally but only weakly phase-coherent between different regions of the lattice.

\clearpage
\onecolumngrid
\section{Summary of DMRG bond dimension and truncation error}
\label{app:dmrg_runs}
Table~\ref{tab:dmrg_runs}
{lists the bond dimension and truncation error for the DMRG calculations on each lattice size. Unless indicated otherwise, we quote as truncation error $\epsilon$ the largest discarded weight during the last complete sweep at the listed $m$.}

{In the table, the reference $m$ specifies the state used for the hole density and spin correlations.
{For the $8\times16$ pairing correlations , we}
fit each sign-stable bond correlation to $P_s(\epsilon)=P_s(0)+a\epsilon+b\epsilon^2$ at all four values of $t'$.

\begin{table}[!ht]
    \caption{Number of kept states $m$ and truncation error $\epsilon$ of the DMRG calculations in this work.}
    \label{tab:dmrg_runs}
    \centering
    \small
    \setlength{\tabcolsep}{7pt}
    \renewcommand{\arraystretch}{1.1}
    \begin{tabular}{lccrrccl}
    \toprule
    Lattice & $\delta$ & $J/t$ & $t'/t$ &
    \shortstack[r]{{Reference $m$}} &
    \shortstack[c]{{$m$ range for extrapolation}} &
    $\epsilon$ & Shown in \\
    \midrule
    $6\times32$ & 1/16 & 0.4 & $0.2$ & 30000 & -- & $4.23\times10^{-6}$ & Figs.~\ref{fig:6leg_density}--\ref{fig:6leg_sc_2d}, \ref{fig:app_6leg_spin_decay} \\
                &    &     & $0$   & 45000 & -- & $3.44\times10^{-6}$ & \\
                &    &     & $-0.2$ & 32000 & -- & $4.68\times10^{-6}$ & \\
                &    &     & $-0.5$ & 30000 & -- & $6.43\times10^{-6}$ & \\
    \addlinespace
    $6\times16$ & 1/16 & 0.4 & $0.2$ & 43000 & -- & $2.96\times10^{-6}$ & Figs.~\ref{fig:app_dmrg_6x16}, \ref{fig:app_dmrg_6x16_pairing} \\
                &   &     & $0$   & 46000 & -- & $2.79\times10^{-6}$ & \\
                &   &     & $-0.2$ & 38000 & -- & $2.26\times10^{-6}$ & \\
                &   &     & $-0.4$ & 25000 & -- & $6.80\times10^{-6}$ & \\
    \addlinespace
    $8\times16$ & 1/16 & 0.4 & $0.2$ & 30000 & 15000--30000 & $3.29\times10^{-5}$ & Figs.~\ref{fig:8leg_density}--\ref{fig:8leg_sc_2d} \\
                &   &     & $0$   & 30000 & 14000--30000 & $3.43\times10^{-5}$ & \\
                &   &     & $-0.2$ & 33000 & 30000--43000 & $2.67\times10^{-5}$ & \\
                &   &     & $-0.4$ & 30000 & 20000--30000 & $2.31\times10^{-5}$ & \\
    \addlinespace
    $6\times6$ & 1/18 & 0.3 & $0$ & 20000 & -- & $1.96\times10^{-7}$ & Table~\ref{tab:benchmark_energy}, Fig.~\ref{fig:benchmark} \\
    \addlinespace
    $8\times8$ & 1/32 & 0.4 & $0$ & 38000 & -- & $2.44\times10^{-6}$ & Table~\ref{tab:benchmark_energy}, Fig.~\ref{fig:app_dmrg_8x8_doping} \\
               & 1/16 & 0.4 & $0$ & 40000 & -- & $3.15\times10^{-5}$ & Table~\ref{tab:benchmark_energy}, Figs.~\ref{fig:8x8_t2_scan_}, \ref{fig:square_momentum_tprime} \\
               &   &     & $0$ & 48000 & -- & $2.47\times10^{-5}$ & Fig.~\ref{fig:app_dmrg_8x8_doping} \\
               &   &     & $0.2$ & 35000 & -- & $3.39\times10^{-5}$ & Figs.~\ref{fig:8x8_t2_scan_}, \ref{fig:square_momentum_tprime} \\
               &   &     & $-0.2$ & 30000 & -- & $2.44\times10^{-5}$ & Figs.~\ref{fig:8x8_t2_scan_}, \ref{fig:square_momentum_tprime} \\
               & 3/32 & 0.4 & $0$ & 40000 & -- & $9.32\times10^{-5}$ & Fig.~\ref{fig:app_dmrg_8x8_doping} \\
               & 1/8 & 0.4 & $0$ & 35000 & 15000--40000 & $1.04\times10^{-4}$ & Figs.~\ref{fig:app_dmrg_8x8_doping}, \ref{fig:benchmark_hole8_energy}, \ref{fig:benchmark_hole8_pattern} \\
    \addlinespace
    $12\times12$ & 1/12 & 0.4 & $-0.2$ & 20000 & -- & $2.18\times10^{-4}$ & Figs.~\ref{fig:12x12_energy_competition}(c), \ref{fig:app_dmrg_12x12_t2m02} \\
    \bottomrule
    \end{tabular}
    \par\smallskip
\end{table}
\twocolumngrid

\clearpage
\bibliography{apssamp}

\end{document}